\documentclass[a4paper,11pt]{article}

\pdfoutput=1

\usepackage{jcappub}
\usepackage[T1]{fontenc}
\usepackage{lmodern}
\usepackage{graphicx}
\usepackage{amsmath,amssymb,amsfonts}
\usepackage{bm}
\usepackage{slashed}
\usepackage{booktabs}
\usepackage{textcomp}

\title{Optical Images of the Braneworld Black Hole Surrounded by an Optically Thin Accretion Disk}

\author[a]{Wen-Hao Deng,}
\author[b,1]{Sen Guo,\note{Corresponding author.}}
\author[a]{Qing-Quan Jiang,}
\author[c]{Kai Lin,}
\author[b]{Pei Wang,}

\affiliation[a]{School of Physics and Astronomy, China West Normal University, Nanchong 637009, People's Republic of China}
\affiliation[b]{College of Physics and Electronic Engineering, Chongqing Normal University, Chongqing 401331, People's Republic of China}
\affiliation[c]{Universidade Federal de Campina Grande, Campina Grande, PB, Brazil}

\emailAdd{sguophys@126.com}

\begin{document}
\maketitle
\begin{abstract}
This work examines the observational signatures of rotating black holes with tidal charge in the Randall--Sundrum braneworld scenario. Combining an elliptic-integral-based analytical treatment with numerical ray tracing, we characterize photon motion around braneworld black holes in detail. For small observer inclinations, secondary images remain embedded inside the primary emission rings. As the inclination becomes larger, the primary and secondary images gradually separate and produce a strongly asymmetric image morphology. We show that the image asymmetry and the deformation of the inner shadow are jointly controlled by the black hole spin, the tidal charge, and the observer inclination. To analyze the frequency shifts across the accretion disk, we extend the emitting region from the ISCO down to the event horizon by including the plunging flow. The results indicate that the observer inclination is the dominant factor governing the frequency-shift distribution. In addition, we reconstruct braneworld black hole images using a fisheye-lens ray-tracing model. The optical morphology and brightness distribution show a clear dependence on the observing frequency, especially when comparing 230~GHz with 86~GHz. We also contrast the brightness distributions of prograde and retrograde disks, finding that both the total intensity and the peak intensity at 86~GHz are higher than those at 230~GHz. For negative values of the tidal charge, we further investigate the corresponding frequency-shift behavior and 230~GHz intensity profiles, which may provide useful theoretical guidance for future studies of extra-dimensional gravity models.
\end{abstract}

\keywords{Braneworld black hole, optically thin accretion disk, inner shadow, black hole image}

\section{Introduction}

\par
General relativity is one of the central pillars of modern physics and provides the basic theoretical framework for black hole physics and cosmology. Within this theory, black holes first appeared as particular solutions of Einstein's field equations, long before they were established as astrophysical objects \cite{1}.
A black hole represents a distinctive spacetime region whose outer boundary is a null hypersurface, namely the event horizon. Many characteristic properties of black holes are therefore encoded in the behavior of this horizon. In the early stage of the subject, research was mainly theoretical because direct observational evidence was still lacking. As observational facilities and analysis techniques have improved, an increasing number of measurements have provided both direct and indirect evidence for black holes in the observable universe \cite{2}.

\par
A major observational milestone was reached in 2019, when the Event Horizon Telescope collaboration released the first horizon-scale image of the shadow of the black hole in M87 \cite{3}. This result supplied direct empirical support for the predictions of GR in the strong-field regime. The subsequent polarized image of M87*, published by the EHT collaboration in 2021, revealed a highly complex magnetic-field structure close to the black hole and provided important clues about the launching of relativistic jets \cite{4}. In 2022, the EHT further reported the image of the supermassive black hole at the Galactic center, Sagittarius A* (Sgr A*), marking the first direct visualization of a black hole in the Milky Way. Together with stellar-orbit measurements, these observations strongly support the broad applicability of general relativity to astrophysical black holes \cite{5,6}. Such developments have opened a new stage in strong-gravity studies and have made physically reliable shadow models, especially those incorporating realistic astrophysical environments, an important task in gravitational physics \cite{7}.

 \par  
Polarimetric observations of M87* by the Event Horizon Telescope also supplied strong evidence that magnetic fields are present in the vicinity of black holes. In realistic astrophysical systems, black holes are generally expected to be surrounded by magnetized plasma; the magnetar detected near Sgr A* provides a representative example of such an environment \cite{8}. In addition, the size and circularity of the M87* shadow are closely related to the inference of black hole spin and are often used to test the applicability of the Kerr description \cite{9,10}. These observations therefore deepen our understanding of strong-gravity phenomena and underline the importance of gravitational lensing in black hole imaging.

\par 
The black hole shadow is associated with the innermost unstable photon orbits and appears as a central dark depression, whereas the neighboring bright ring is produced by photons that are strongly lensed before reaching the observer \cite{11}. The shadow and photon ring therefore serve as important probes of black hole parameters and provide useful constraints on gravity in the strong-field regime \cite{12}.
At the same time, the detailed appearance of the photon ring is shaped not only by gravitational lensing but also by radiation from the accretion flow. The geometry and emissivity of the disk play a decisive role in determining the observed brightness pattern.

\par
The intrinsic properties of the shadow are mainly controlled by the underlying spacetime geometry, while the surrounding luminous structures are strongly affected by the emission model of the accretion disk. Earlier theoretical work has classified black hole shadows and photon rings in detail and has shown that small-scale disk structures can leave visible imprints on the image \cite{13}. The ``critical curve'', associated with the photon orbit, forms a basic element in shadow studies and has motivated a large body of later work. Supermassive black holes are generally believed to accrete hot, magnetized plasma and to produce bright disk emission. In the observational bands relevant to black hole imaging, thermal synchrotron radiation from electrons in the disk is usually considered a plausible source of the detected signal \cite{14}.

\par 
Numerical images of Kerr black holes surrounded by optically thin accretion disks usually show two characteristic features: a dark central region and a narrow ring-like photon structure \cite{15}. Based on this general picture, Wang studied Kerr--de Sitter black holes and discussed how the cosmological constant modifies the imaging properties \cite{16}. Hou considered Kerr--Melvin black holes with thin accretion disks and demonstrated that the inner shadow and critical curve can be used to diagnose the magnetic-field strength \cite{12}. Guo further analyzed the optical appearances of Kerr--Newman and Kerr--Sen black holes, focusing on the effects of spin, charge, and observer inclination \cite{5,17}. Yang investigated the rotating Ghosh--Kumar black hole and later carried out a systematic study of rotating charged black holes in the Kalb--Ramond gravitational framework \cite{18,19}.

\par
Among the theoretical extensions of GR, braneworld models have received
considerable attention because of their relevance to cosmology, particle
physics, and string theory. In this framework, our observable universe
is regarded as a four-dimensional timelike hypersurface embedded in a
higher-dimensional spacetime. Extra dimensions can affect both particle
physics and cosmology, and their possible imprints on black hole images
are therefore worth studying \cite{20}. Ref.~\cite{21} analyzed the
shadow of rotating black holes in the Randall--Sundrum braneworld
scenario, and later work introduced a cosmological constant to extend
this analysis \cite{22}. Gravitational lensing and retrolensing in
braneworld black hole spacetimes have also been examined \cite{23}.
More recently, the M87* image has been used to constrain the parameters
of rotating braneworld black holes \cite{24}. Although substantial
theoretical progress has been made, previous investigations have mainly
focused on vacuum shadow boundaries, shadow sizes, deviations from
circularity, or unresolved spectral observables. The resolved radiative
appearance of a rotating braneworld black hole surrounded by an
optically thin accretion flow remains comparatively less explored.

\par
From an astronomical perspective, the assumption that black holes are
strictly electrically neutral is not universally guaranteed. Small
electric charges may arise through the selective accretion of charged
particles or through the Wald induction mechanism when a rotating black
hole is immersed in an external magnetic field
\cite{39,40}. Although the realistic electric charge
expected for Sgr~A* is many orders of magnitude below the extremal value
and is therefore insufficient to modify the background geometry
appreciably, it may still affect charged-particle motion, the ISCO, and
the surrounding plasma emission \cite{40}. In addition,
primordial black holes residing in virialized dark-matter halos have
been predicted to acquire a net negative electric charge because
electrons are lighter and may be accreted more efficiently than protons
\cite{41}. These studies indicate that charge-related degrees of
freedom need not vanish identically in realistic astrophysical
environments.

\par
The tidal charge considered in the present work is, however,
fundamentally different from an ordinary electromagnetic charge. It is
not generated by the selective accretion of electrons or protons, but
encodes the nonlocal gravitational influence of the higher-dimensional
bulk projected onto the four-dimensional brane
\cite{42,25}. In the Kerr--Newman geometry, the electric
charge enters the metric through $Q^2$, and hence positive and negative
electric charges produce the same spacetime geometry for neutral
particles and photons. By contrast, the tidal charge $q$ enters the
braneworld metric linearly, so that its sign directly affects the event
horizon, photon orbits, the ISCO, and the resulting black hole image.
Consequently, the negative-$q$ sector should be interpreted as a
sign-dependent gravitational modification rather than as an ordinary
electrically negative black hole.

\par
The negative-tidal-charge sector also has a direct astronomical
motivation. Comparisons between theoretical accretion-disk spectra and
the optical luminosities of quasars in a related higher-dimensional
model have reported a model-dependent preference for a negative
effective charge parameter \cite{43}. Moreover, analyses of
the shadow properties of M87* and Sgr~A* have constrained large negative
tidal charges but have not completely excluded a limited negative
parameter range \cite{44,45}. These results do not
constitute evidence for the detection of a negative tidal charge, but
they motivate a more detailed investigation of its observable
consequences in realistic radiative environments.

\par
The novelty of the present analysis is therefore not merely the
replacement of the Kerr metric by a rotating braneworld metric in an
existing ray-tracing framework. Previous braneworld studies have mainly
concentrated on vacuum shadow geometries, shadow diameters,
circularities, or unresolved continuum spectra. In contrast, we connect
the tidal charge to disk-dependent and frequency-dependent imaging
observables, including the inner shadow, higher-order images, frequency
shifts, and intensity distributions. We also provide quantitative
comparisons with the Kerr limit $q=0$, including the changes in the
shadow eccentricity, characteristic frequency shifts, and image
intensities. These results provide theoretical imaging templates for
identifying the effects of a negative effective tidal charge and for
distinguishing them from ordinary electromagnetic-charge effects.

\par
To address this issue, we study photon trajectories around braneworld black holes surrounded by optically thin accretion disks. By combining general relativity with ray-tracing methods, we focus on the observable signatures of this spacetime. Specifically, we calculate photon orbits, examine the influence of gravitational lensing and particle dynamics, and analyze frequency-shift and intensity variations at 230~GHz and 86~GHz. In addition to improving our understanding of braneworld black hole images, this study provides a theoretical framework for exploring how the tidal charge affects photon trajectories, frequency shifts, and intensity distributions in an optically thin disk model.

\par
The remainder of this paper is arranged as follows. Section \ref{sec:2} summarizes the basic properties of braneworld black holes and uses a semi-analytic approach to calculate photon deflection and lensing-ring formation. Section \ref{sec:3} derives the frequency-shift and intensity distributions and presents synthetic images at different observing frequencies for both prograde and retrograde accretion disks. Section \ref{sec:4} gives the main conclusions.

\section{Ray Tracing of a Spinning Braneworld Black Hole}
\label{sec:2}
This section provides a concise review of the braneworld black hole solution and describes the ray-tracing approach using the trajectories of test particles in this spacetime \cite{25,26}.

\begin{eqnarray}
\label{1}
ds^2 &=& -\left(1 - \frac{2Mr - q}{\Sigma}\right) dt^2
     - 2 \frac{a(2Mr - q) \sin^2\theta}{\Sigma} dt d\phi
     + \frac{\Sigma}{\Delta} dr^2 \nonumber\\
&& + \Sigma d\theta^2 
     + \left(r^2 + a^2 + \frac{2Mr - q}{\Sigma} a^2 \sin^2\theta\right)
     \sin^2\theta d\phi^2 .
\end{eqnarray}
where the metric functions are defined as

\begin{equation}
\label{2}
\Delta = r^2 - 2Mr + a^2 + q,
\end{equation}

\begin{equation}
\label{3}
\Sigma = r^2 + a^2 \cos^2{\theta}.
\end{equation}

\par
The parameter $M$ is the black hole mass, and $a = J/M$ denotes the spin parameter. The parameter $q$ is the tidal charge, which represents non-local gravitational effects produced by the higher-dimensional bulk spacetime. In the braneworld picture, this parameter should not be interpreted as an electric charge located on the brane. Rather, the bulk tidal field modifies the effective gravitational field on the brane in a form similar to a charge contribution. Different from the squared electric charge $Q^{2}$ in the Kerr--Newman solution, $q$ is not required to be positive; several studies have argued that negative tidal charge may even be more natural in braneworld models~\cite{20,25,26}.

The metric is obtained by assuming that the induced four-dimensional brane geometry has the Kerr--Schild form, although whether this brane metric can be embedded as an exact solution of the full higher-dimensional bulk equations is still unclear. The standard Kerr spacetime is recovered for $q = 0$, while the non-rotating limit $a = 0$ gives the static braneworld black hole solution discussed in Ref.~\cite{27}. The event horizon follows from the condition $\Delta = 0$~\cite{21}:

\begin{equation}
\label{4}
 r_{H} = M \pm \sqrt{M^{2} - (a^{2} + q)}.
\end{equation}

\par
For simplicity, we use geometric units and set the black hole mass to $M = 1$, so that all relevant quantities are dimensionless. The event horizon radius is denoted by $r_{H}$. From the viewpoint of imaging, photons that terminate on the event horizon create a central dim region, which can be identified as the black hole inner shadow. The existence of an event horizon requires $q \leq M^{2} - a^{2}$, and the equality corresponds to an extremal black hole with $r_{H} = M$. For positive tidal charge, the allowed spin is restricted by $a \leq \sqrt{M^{2}-q}$, which reduces to the usual Kerr bound $a \leq M$ when $q = 0$. For negative tidal charge, however, the standard bound can be relaxed, allowing an extremal black hole with $a > M$.

\begin{figure}[htbp]
\centering
\includegraphics[width=2.8cm]{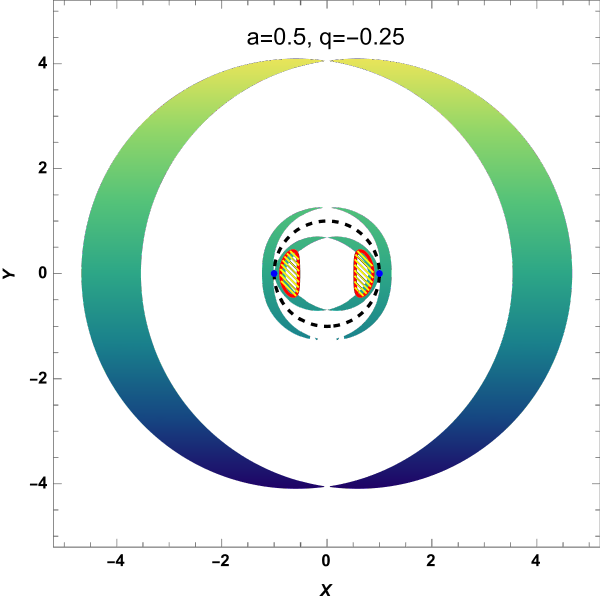}
\includegraphics[width=2.8cm]{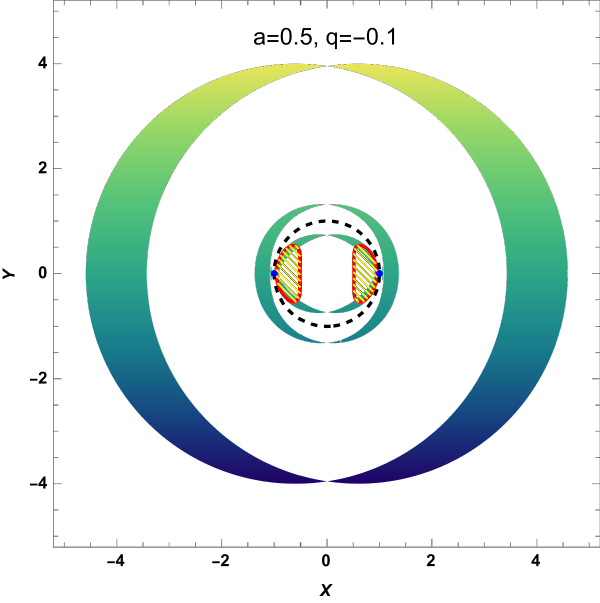}
\includegraphics[width=2.8cm]{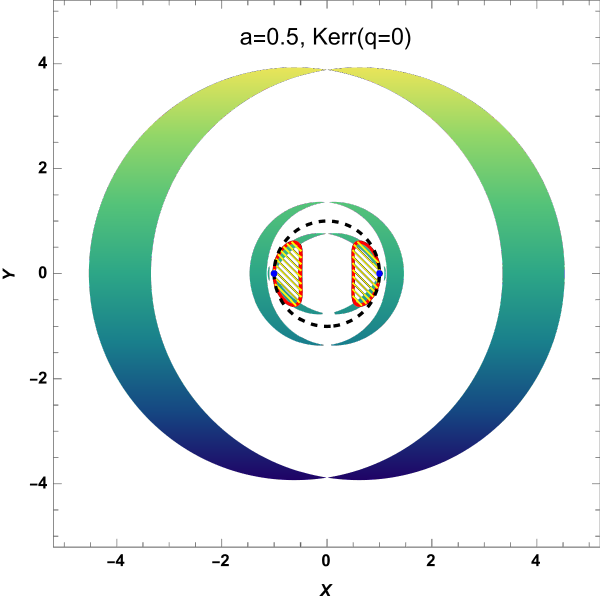}
\includegraphics[width=2.8cm]{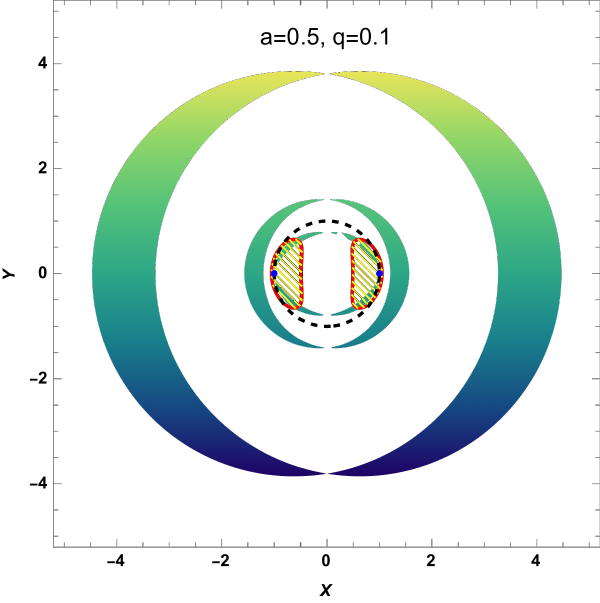}
\includegraphics[width=2.8cm]{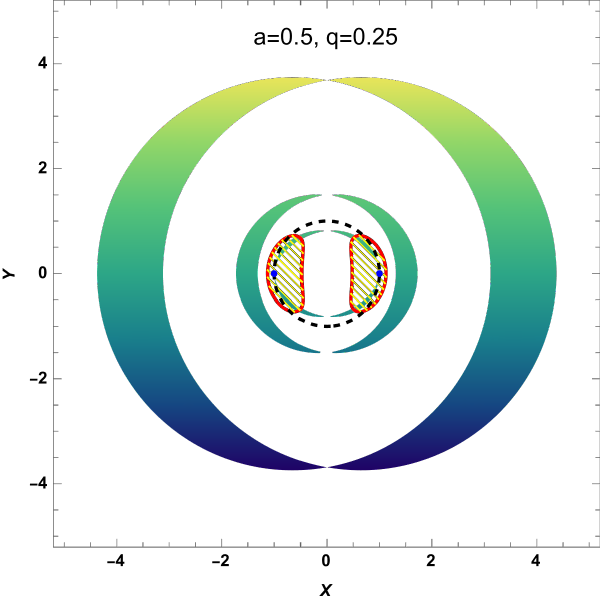}
\vspace{2mm}
\caption{The photon shell and photon-orbit structures of the braneworld black hole are illustrated. In all panels, the spin parameter is fixed at $a=0.5$, while the tidal charge $q$ varies from left to right as $q=-0.25$, $-0.1$, $0$, $0.1$, and $0.25$, with $q=0$ corresponding to the Kerr solution. In all cases, the black hole mass is set to $M=1$.}
\label{fig:1}
\end{figure}

\par
To clarify the physical differences between braneworld black holes and the Kerr spacetime, we adopt the method of Refs.~\cite{28,29} and examine the photon regions for several tidal-charge values. The Kerr limit is recovered when $q = 0$. Figure~1 displays the corresponding photon regions in the $(r,\theta)$ plane. In each panel, the green shaded area marks unstable photon orbits, the yellow hatched domain represents causality-violating regions, and the black dashed curve denotes the event horizon. The radial direction is plotted over the full spacetime domain. For clarity, the radial coordinate has been scaled as $M \exp(r/M)$ for $r < 0$ and to $r + M$ for $r > 0$, so that the throat at $r = 0$ appears as the black dashed circle.

\par
The figure gives a direct view of how the tidal charge changes photon motion for a fixed spin. Because $q$ may be either positive or negative in the braneworld scenario, we use the representative interval $-0.25 \leq q \leq 0.25$ to illustrate its effect. With increasing tidal charge, the outer edge of the photon region moves inward, whereas the domain of bound photon orbits becomes larger. The internal arrangement of photon trajectories also changes visibly. The case $q = 0$ corresponds to the Kerr black hole, and the comparison between negative and positive tidal charge emphasizes the characteristic behavior of the braneworld solution.

\subsection{Photon Trajectories near a Braneworld Black Hole}
To investigate photon trajectories around a braneworld black hole, examining the dynamical behavior of particles within the corresponding spacetime is of critical importance. The trajectories of photons in curved spacetime are prescribed by the geodesic equation, reflecting the effect of spacetime curvature \cite{30}:
  
\begin{equation}    
\label{5}
\frac{d^{2}x^{\mu}}{d\lambda^{2}}+ \Gamma^{\mu}_{\alpha\beta}\frac{dx^{\alpha}}{d\lambda}\frac{dx^{\beta}}{d\lambda}=0,
\end{equation}
    
Here, $\lambda$ is the affine parameter, and $\Gamma^{\mu}_{\alpha\beta}$ denotes the Christoffel symbol. Once appropriate initial conditions are specified, the photon path can in principle be obtained from this equation, although direct integration is usually inconvenient in practice.

\par
For this reason, we use the Hamilton--Jacobi formalism to treat the geodesic motion. In the present spacetime, the Hamilton--Jacobi equation is written as \cite{31}:

{
\begin{equation}
\label{6}
 \frac{\partial S}{\partial\lambda}+\frac{1}{2}g^{\mu\nu}\frac{\partial S}{\partial x^{\mu}}\frac{\partial S}{\partial x^{\nu}}=0,
\end{equation}}

Because of the spacetime symmetries, photon orbits can be described using conserved quantities. For an axisymmetric rotating black hole, three constants of motion are sufficient for determining the photon trajectories: $p_{t}=-E$, $p_{\phi}=L$, and $\mathcal{C}=p_{\theta}^{2}-\cos^{2}\theta\bigl(a^{2}E^{2}-L^{2}\csc^{2}\theta\bigr)$, where $E$ is the conserved energy, $L$ is the conserved angular momentum, and $\mathcal{C}$ is the Carter constant. Introducing the dimensionless impact parameters, namely the normalized angular momentum $\xi$ and the rescaled Carter constant $\eta$, gives

\begin{equation}
\label{7}
 \xi=\frac{L}{E},\qquad \eta=\frac{\mathcal{C}}{E^{2}}
\end{equation}

 \par
 
For braneworld black holes, the photon four-momentum $p^{\mu}$ along the trajectory is written as

\begin{equation}
\label{8}
\frac{\Sigma}{E} p^{r} = \pm_{r} \sqrt{\mathcal{R}(r)},
\end{equation}

\begin{equation}
\label{9}
\frac{\Sigma}{E} p^{\theta} = \pm_{\theta} \sqrt{\Theta(\theta)},
\end{equation}

\begin{equation}
\label{10}
\frac{\Sigma}{E} p^{\phi} = \frac{\xi}{\sin^{2}{\theta}} + \frac{a(a^{2}+r^{2}-a\xi)}{\Delta} - a,
\end{equation}

\begin{equation}
\label{11}
\frac{\Sigma}{E} p^{t} = a\xi - a^{2}\sin^{2}\theta + \frac{a^{4} + 2a^{2}r^{2} + r^{4} - a\xi(a^{2}+r^{2})}{\Delta}.
\end{equation}

where the radial potential $\mathcal{R}(r)$ and angular potential $\Theta(\theta)$ are given by

\begin{equation}
\label{12}
\mathcal{R}(r) = -\Delta\bigl[(\xi - a)^{2} + \eta\bigr] + (a^{2} + r^{2} - a\xi)^{2},
\end{equation}

\begin{equation}
\label{13}
\Theta(\theta) = a^{2}\cos^{2}\theta + \eta - \xi^{2}\cot^{2}\theta.
\end{equation}

\par
The symbols $\pm_{r}$ and $\pm_{\theta}$ give the signs of $p^{r}$ and $p^{\theta}$, respectively. The radial and angular turning points are located at the values of $r$ and $\theta$ for which $\mathcal{R}(r)$ and $\Theta(\theta)$ vanish. We consider a photon emitted from $(t_{s}, r_{s}, \theta_{s}, \phi_{s})$ and received by a distant observer at $(t_{o}, \infty, \theta_{o}, \phi_{o})$. To describe the trajectory, we introduce the Mino time $\tau$.

The equations of motion written in Mino time take the form \cite{32}

\begin{equation}
\label{14}
    \frac{dx^{\mu}}{d\tau}\equiv\frac{\Sigma}{E}p^{\mu}
\end{equation}

\par
Using Eqs.~(\ref{8})--(\ref{11}), the equations of motion can be recast into the following integral form \cite{32}:

\begin{equation}
\label{15}
\Delta t = t_{o} - t_{s} = I_{t} + a^{2} G_{t},
\end{equation}

\begin{equation}
\label{16}
\tau = I_{r} = G_{\theta},
\end{equation}

\begin{equation}
\label{17}
\Delta \phi = \phi_{o} - \phi_{s} = I_{\phi} + \xi G_{\phi},
\end{equation}

\par
The detailed derivation is given in Appendix~A. The integrals $I_{t}$, $I_{r}$, and $I_{\phi}$ represent path integrals between the source and observer coordinates, $x_{s}^{\mu}$ and $x_{o}^{\mu}$, respectively. For angular motion with $\eta > 0$, the geodesic is described by the polar turning points $\theta_{\pm}$, which are symmetric with respect to the equatorial plane~\cite{33}:

\begin{equation}
\label{18}
\theta_{\pm} = \arccos\!\left( \mp \sqrt{ \omega_{+} } \right),
\end{equation}

where  

{
\begin{equation}
\label{19}
\omega_{\pm} = \frac{1}{2} - \frac{ \eta + \xi^{2} }{ 2 a^{2} } 
\pm \sqrt{ \frac{ \eta }{ a^{2} } + \frac{1}{4} \left( 1 - \frac{ \eta + \xi^{2} }{ a^{2} } \right)^{2} }.
\end{equation}}

To avoid coordinate singularities at the poles, we restrict $\theta$ to $0 < \theta < \pi$ and rewrite the angular potential in terms of $\omega = \cos^{2}\theta$. The four roots of $\Theta(\theta)$ are then $\theta = \arccos(\pm\sqrt{\omega_{\pm}})$.

Special cases occur only when $\omega_{+}=0$, $\omega_{-}=0$, or $\omega_{+}=\omega_{-}$. These conditions divide the $(\xi,\eta)$ plane into distinct regions. Within each region, the angular potential preserves the same qualitative structure, including the number of real roots and the sign of the potential near them. Therefore, the behavior of each region can be effectively characterized by examining a representative point inside it.

\par
In this formulation, the positive angular integral can be expressed by elliptic integrals and classified by the number of turning points $z$ along the trajectory. Following Refs.~\cite{32,33}, we obtain

\begin{equation}
\label{20}
G_{\theta} = \frac{1}{ a \sqrt{ -\omega_{-} } } \left[ 2 z K \pm_{s} F_{s} \mp_{o} F_{o} \right],
\end{equation}

\begin{equation}
\label{21}
G_{\phi} = \frac{1}{ a \sqrt{ -\omega_{-} } } \left[ 2 z \Pi \pm_{s} \Pi_{s} \mp_{o} \Pi_{o} \right],
\end{equation}

\begin{equation}
\label{22}
G_{t} = - \frac{ 2 \omega_{+} }{ a \sqrt{ -\omega_{-} } } \left[ 2 z E' \pm_{s} E'_{s} \mp_{o} E'_{o} \right].
\end{equation}

\par

Here, $K$, $\Pi$, and $E'$ are the complete elliptic integrals of the first, third, and second kinds, respectively. The quantities $K_{i}$, $\Pi_{i}$, and $E'_{i}$ denote the corresponding incomplete elliptic integrals.

\par
In the braneworld black hole spacetime, the tidal charge modifies the effective radial potential directly. By analyzing the roots of this potential, the conserved photon impact parameters, $\xi$ and $\eta$, can be obtained explicitly from the simultaneous solution of Eqs.~(\ref{2}), (\ref{3}), and (\ref{12}), yielding:

\begin{equation}
\label{23}
\mathcal{R}(r) = r^{4} + A r^{2} + B r + D,
\end{equation}

where

\begin{equation}
\label{24}
A = a^{2} - \eta - \xi^{2},
\end{equation}

\begin{equation}
\label{25}
B = 2 M (\eta + \xi^{2} + a^{2} - 2 a \xi),
\end{equation}

\begin{equation}
\label{26}
D = - a^{2} \eta - q (\eta + \xi^{2} + a^{2} - 2 a \xi),
\end{equation}

From Eq.~(\ref{23}), the four roots of the radial potential satisfy $r_{1}+r_{2}+r_{3}+r_{4}=0$ and can be expressed as

\begin{equation}
\label{27}
r_{1} = -j - \sqrt{-\frac{A}{2} - j^{2} + \frac{B}{4 j}},
\end{equation}

\begin{equation}
\label{28}
r_{2} = -j + \sqrt{-\frac{A}{2} - j^{2} + \frac{B}{4 j}},
\end{equation}

\begin{equation}
\label{29}
r_{3} = j - \sqrt{-\frac{A}{2} - j^{2} - \frac{B}{4 j}},
\end{equation}

\begin{equation}
\label{30}
r_{4} = j + \sqrt{-\frac{A}{2} - j^{2} - \frac{B}{4 j}},
\end{equation}

where

\begin{equation}
\label{31}
j = \sqrt{\frac{\varpi_{+} + \varpi_{-} - \frac{A}{3}}{2}},
\end{equation}

\begin{equation}
\label{32}
\varpi_{\pm} = \sqrt[3]{-\frac{A}{6}(\frac{A^{2}}{36} - D) - \frac{B^{2}}{16} 
\pm \sqrt{\frac{A^{2}}{36} - \frac{D^{3}}{3} - \frac{A}{6}(\frac{A^{2}}{36} - D) - \frac{B^{2}}{16}}}.
\end{equation}

\par

For an observer at spatial infinity, a photon trajectory can have at most one radial turning point outside the event horizon. If $r_{4}$ is real and lies strictly outside the horizon, the turning point is located at $r_{4}$; otherwise, the photon falls directly into the black hole. The radial integral can therefore be written in the generalized form~\cite{33}:

\begin{equation}
\label{33}
I_t \sim \int_{r_s}^{r_o} dr \sim \int_{r_s}^{r_o} \cdots + 2 \omega \int_{r_t}^{r_s} dr \cdots.
\end{equation}

In this way, the radial integration takes the specific form:

\begin{equation}
\label{34}
I_r = \int_{r_s}^{\infty} \frac{dr}{\sqrt{\mathcal{R}(r)}} + 2 \omega \int_{r_4}^{r_s} \frac{dr}{\sqrt{\mathcal{R}(r)}}.
\end{equation}
where $\omega$ is defined in Eq.~(\ref{18}) and Eq.~(\ref{19}).
Depending on the light source, the total radial integral should consider two distinct cases \cite{33}:

\begin{equation}
\label{35}
I_r^{\rm total} =
\left\{
\begin{array}{ll}
\displaystyle 2\int_{r_s}^{\infty}
\frac{dr}{\sqrt{\mathcal{R}(r)}},
& r_{+} < r_4 \in \mathbb{R}, \\[1em]
\displaystyle \int_{r_{+}}^{\infty}
\frac{dr}{\sqrt{\mathcal{R}(r)}},
& {\rm otherwise}.
\end{array}
\right.
\end{equation}

\par
The root $r_{4}$ represents a physical radial turning point only when it is real and outside the event horizon. Following the analytical method of Ref.~\cite{33}, the inverse path integral is obtained by exchanging the observer and source positions. This reverses the direction of integration along the photon path and changes $I_{r}$ to $-I_{r}$. The source radius $r_{s}$ can then be written explicitly as

\begin{equation}
\label{36}
r_s =
\frac{
r_4 r_{31}
- r_3 r_{41}\,{\rm sn}^2\left(\frac{1}{2}\sqrt{r_{31}r_{42}}\,I_r\right)
- \mathcal{F}_o\left|\frac{r_{32}r_{41}}{r_{31}r_{42}}\right|
}{
r_{31}
- r_{41}\,{\rm sn}^2\left(\frac{1}{2}\sqrt{r_{31}r_{42}}\,I_r\right)
- \mathcal{F}_o\left|\frac{r_{32}r_{41}}{r_{31}r_{42}}\right|
}.
\end{equation}

Here, $r_{ij}=r_{i}-r_{j}$ denotes the difference between the corresponding roots, and the auxiliary function $\mathcal{F}_{o}$ is defined by~\cite{33}

{
\begin{equation}
\label{37}
\mathcal{F}_o =
F(\arcsin \sqrt{
\frac{r_{31}}{r_{41}}} |
\frac{r_{32}r_{41}}{r_{31}r_{42}}
).
\end{equation}}

\par
As in Kerr spacetime, the radial and angular parts of photon geodesics in a braneworld black hole can be written analytically in terms of elliptic integrals. Because photons may wind around the black hole more than once, we introduce the fractional orbital number

\begin{equation}
\label{38}
n = \frac{G_\theta}{2 \int_{\theta_-}^{\theta_+} d\theta \Theta(\theta)^{-{\frac{1}{2}}}} 
  = \frac{a \sqrt{-\omega_-}}{4K} I_r. 
\end{equation}

By combining Eqs.~(\ref{20}) and ~(\ref{38}), the orbital fraction \( n \) can be related to the number of turning points \( z \) through a derived expression \cite{32,33}:

\begin{equation}
\label{39}
n = \frac{1}{2} z \pm_o \frac{1}{4} \left[ (-1)^z \frac{F_s}{K} \mp_o \frac{F_o}{K} \right]. 
\end{equation}

\par
For a given pair of impact parameters $\xi$ and $\eta$, the radial potential in Eq.~(\ref{12}) has four roots, as listed in Eqs.~(\ref{27})--(\ref{30}). The real roots correspond to radial turning points. Similar to the Schwarzschild case, the limiting values of the impact parameters determine the photon-ring location. In the rotating case, the radial potential can develop a double root at a critical radius $\tilde{r}$. The corresponding critical impact parameters, $\tilde{\xi}$ and $\tilde{\eta}$, are fixed by $\mathcal{R}(\tilde{r}) = \mathcal{R}'(\tilde{r}) = 0$ and can be written as~\cite{20}

{
\begin{equation}
\label{40}
\tilde{\xi} = a+\frac{\left[(2a^2+ 2q - 3M\tilde{r} + \tilde{r}^2)\tilde{r}\right]}{a(M - \tilde{r})},
\end{equation}}

\begin{equation}
\label{41}
\tilde{\eta} = -\frac{\tilde{r}^2 \left[4a^2q - 4a^2M\tilde{r} + (2q + \tilde{r}(\tilde{r} - 3M))^2\right]}{a^2(M - \tilde{r})^2}. 
\end{equation}

\par
From Eq.~(\ref{41}), $\tilde{\eta}$ is a quartic function of $\tilde{r}$ for a braneworld black hole. The usual trigonometric transformations used in simpler cases are therefore not directly applicable. We solve this equation numerically to determine the critical curve accurately. It should also be noted that $\eta > 0$ for any geodesic that crosses the equatorial plane. The photon-shell boundaries are fixed by circular photon orbits confined to the equatorial plane, for which $\tilde{\eta} = 0$. After simplifying Eq.~(\ref{41}), one obtains

\begin{equation}
\label{42}
-\tilde{r}^4 + 6M\tilde{r}^3 - (9M^2 + 4q)\tilde{r}^2 + (4a^2M + 12Mq)\tilde{r} - 4(a^2q + q^2) = 0. 
\end{equation}

\par
The radial potential $\mathcal{R}(r)$ can be written as a fourth-degree polynomial, 
$\mathcal{R}(r) = (r-\tilde{r}_{1})(r-\tilde{r}_{2})(r-\tilde{r}_{3})(r-\tilde{r}_{4})$.
The roots are ordered as $\tilde{r}_{1}<\tilde{r}_{2}<\tilde{r}_{3}<\tilde{r}_{4}$ and satisfy the constraint $\tilde{r}_{1}+\tilde{r}_{2}+\tilde{r}_{3}+\tilde{r}_{4}=0$.

\par
Figure~2 displays the dependence of $\tilde{\eta}$ on $\tilde{r}$ and shows four roots ordered as $\tilde{r}_{1}<\tilde{r}_{2}<r_{H}<\tilde{r}_{3}<\tilde{r}_{4}$. Since the event horizon is at $r_{H}$, only the two roots outside the horizon, $\tilde{r}_{3}$ and $\tilde{r}_{4}$, are physically relevant. We therefore label them as $r_{-}$ and $r_{+}$, respectively.

\begin{figure}[htbp]
\centering
\includegraphics[width=13cm]{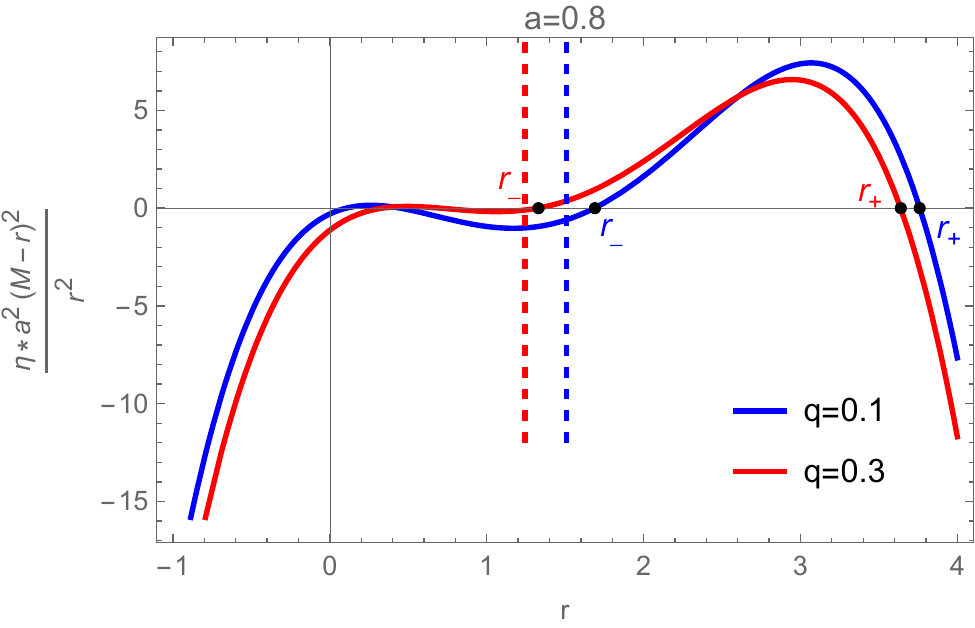}

\vspace{2mm}
\caption{The functional dependence of $\tilde{\eta}$ on $\tilde{r}$ is shown. The dashed vertical line marks the event horizon radius $r_H$, while the solid curves represent the behavior of $\tilde{\eta}$ as a function of $\tilde{r}$ for different values of the tidal charge $q$. The curves are plotted for the parameter set $a=0.8$ and $M=1$, with the blue and red curves corresponding to $q=0.1$ and $q=0.3$, respectively.}
\label{fig:2}
\end{figure}

\subsection{Lensing Bands and Photon Orbits in Braneworld Spacetime}
We consider an observer positioned at spatial infinity. On the observer's image plane, the apparent position of a photon is described by the orthogonal celestial coordinates $(\alpha, \beta)$. This configuration effectively maps the observer's celestial sphere onto a two-dimensional Cartesian plane, where the coordinates scale with the trigonometric functions of the photon's incidence angles. For a braneworld black hole, where photon trajectories are characterized by two constants of motion, these celestial coordinates $\alpha$ and $\beta$ are defined as \cite{32,33}:

\begin{equation}
\label{43}
\alpha = -\frac{\xi}{\sin\theta_0},
\end{equation}

\begin{equation}
\label{44}
\beta = \pm_0 \sqrt{\Theta(\theta)} = \pm_0 \sqrt{\eta + a^2 \cos^2\theta_0 - \xi^2 \cot^2\theta_0}.
\end{equation}
Note that, in the above expressions, the parameter $\theta_{0}$ denotes the observer inclination angle. Based on the analytical relations derived above, the critical impact parameters $\xi(\tilde{r})$ and $\tilde{\eta}(\tilde{r})$ can be explicitly determined. By projecting these parameters onto the celestial coordinates $\tilde{\alpha}(\tilde{r})$ and $\tilde{\beta}(\tilde{r})$, one can naturally characterize the photon ring on the observer's screen for a braneworld black hole surrounded by an optically thin accretion disk.

\begin{figure}[htbp]
\centering
\includegraphics[width=4.5cm]{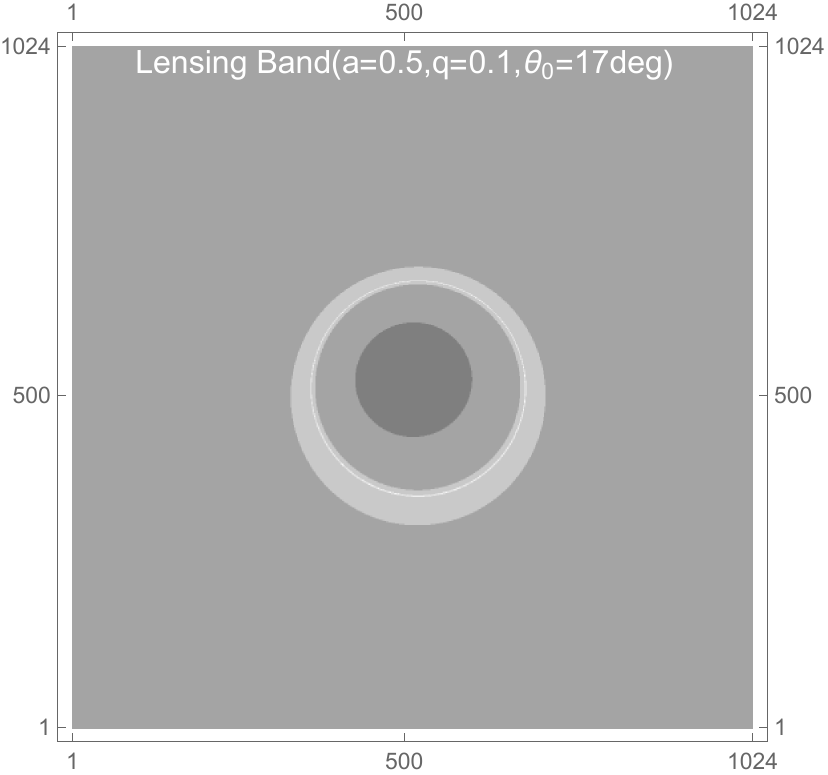}
\includegraphics[width=4.5cm]{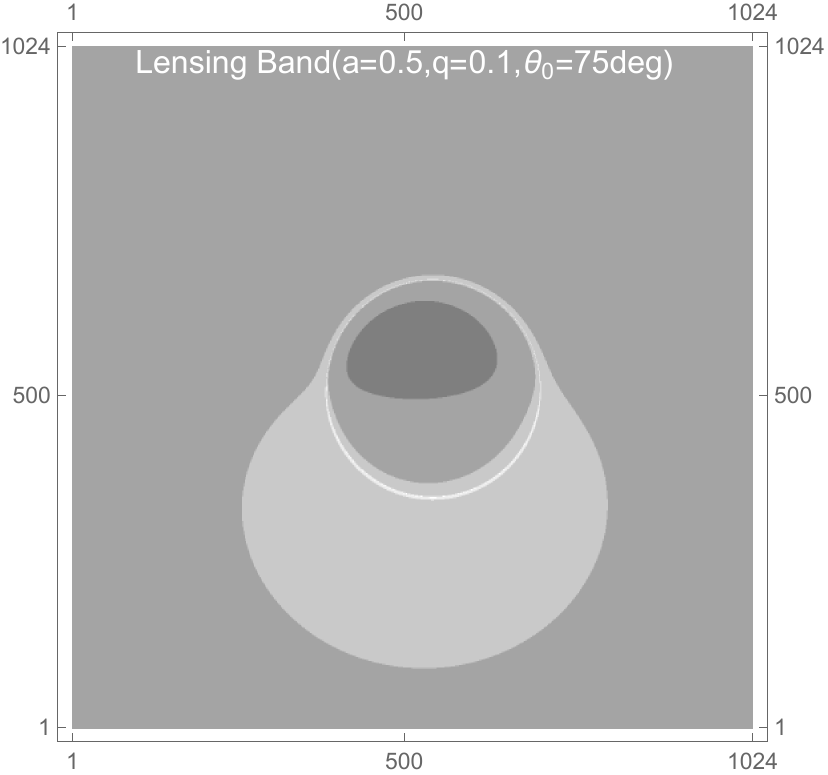}
\includegraphics[width=4.5cm]{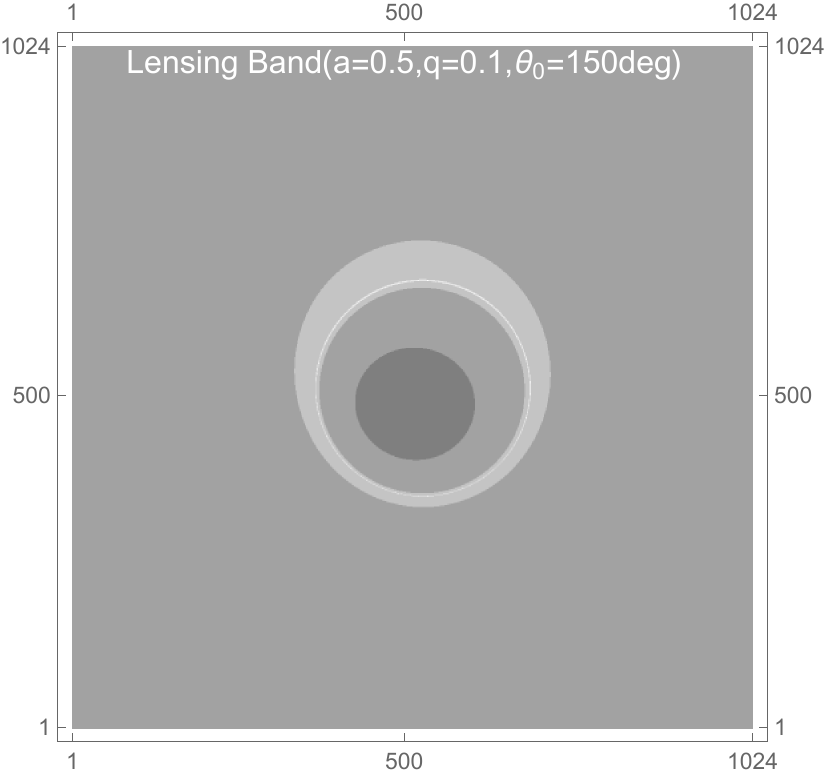}
\includegraphics[width=4.5cm]{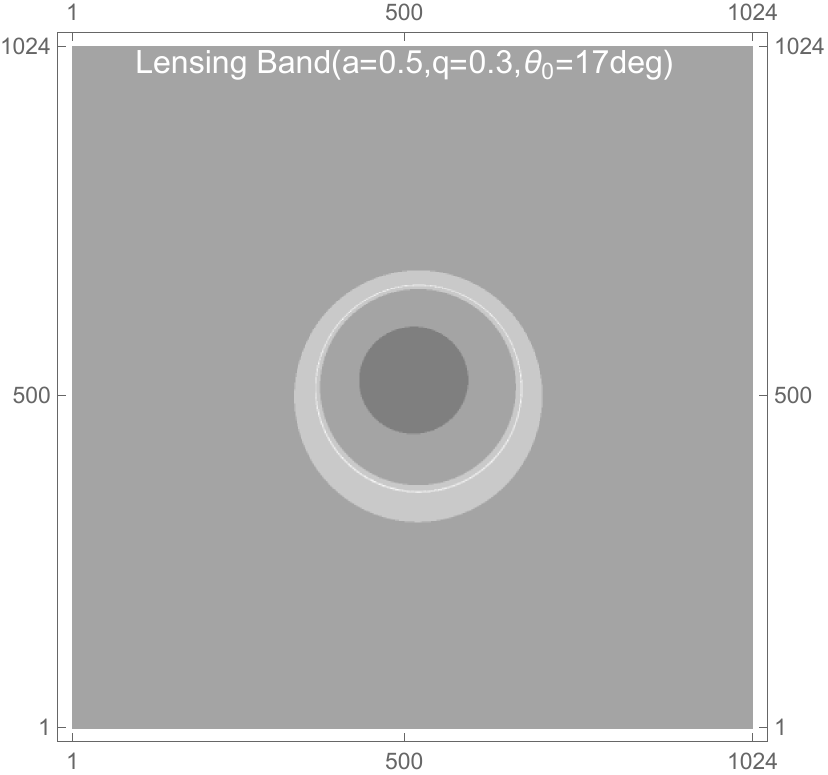}
\includegraphics[width=4.5cm]{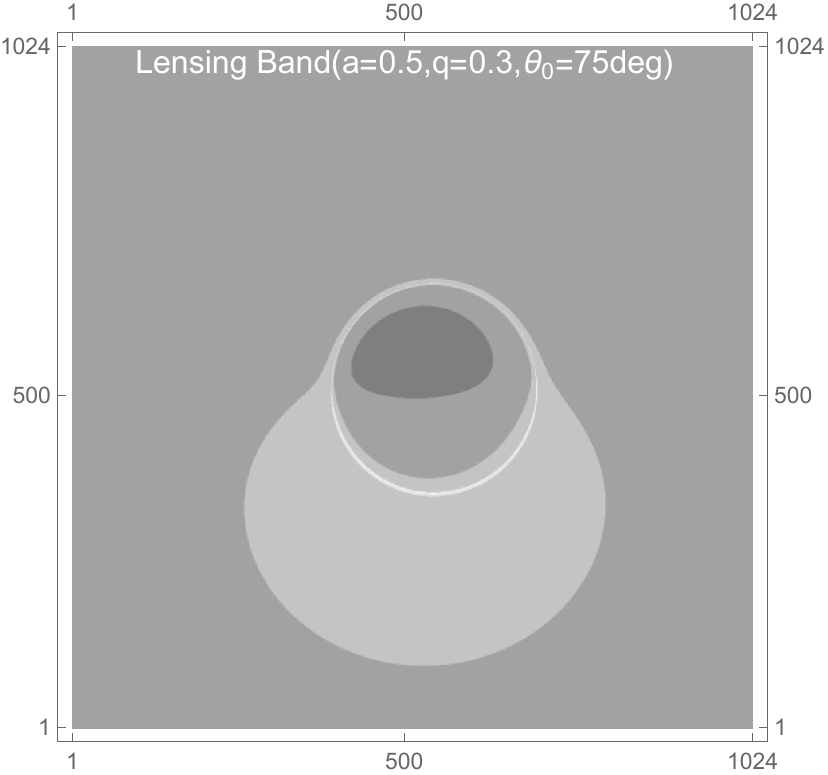}
\includegraphics[width=4.5cm]{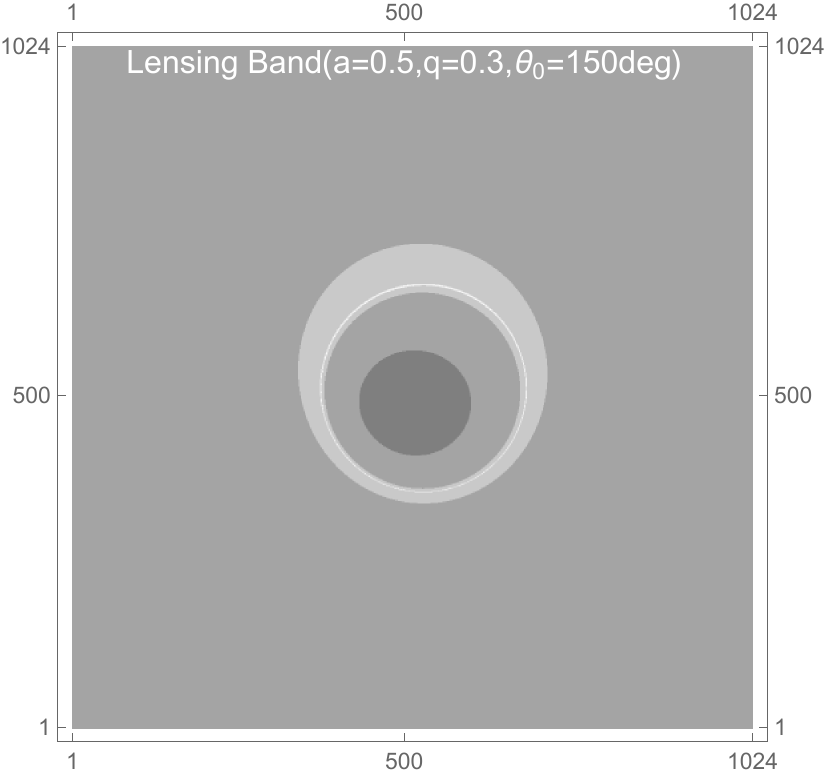}
\includegraphics[width=4.5cm]{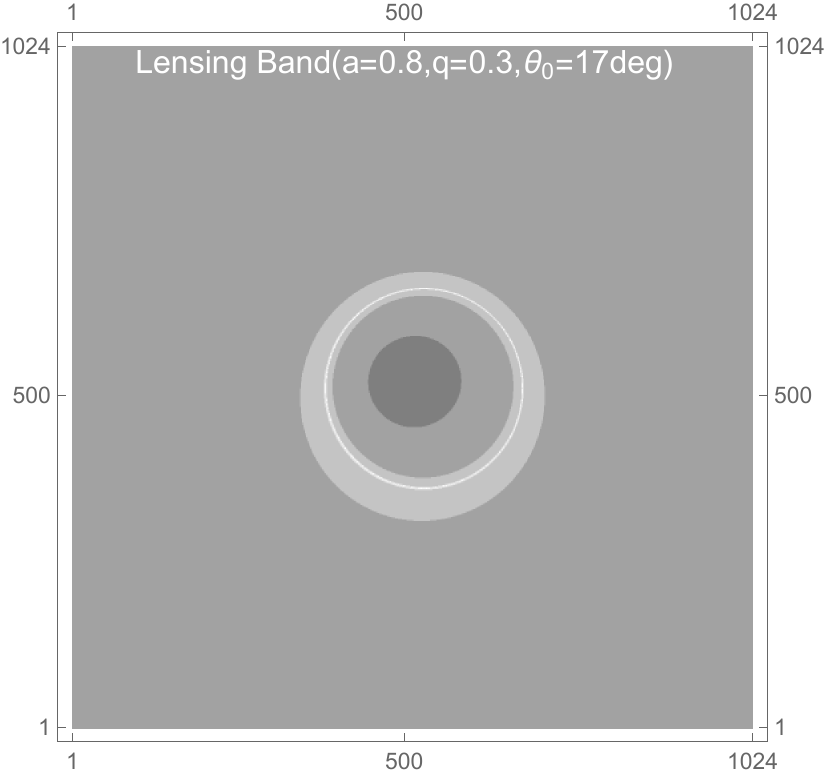}
\includegraphics[width=4.5cm]{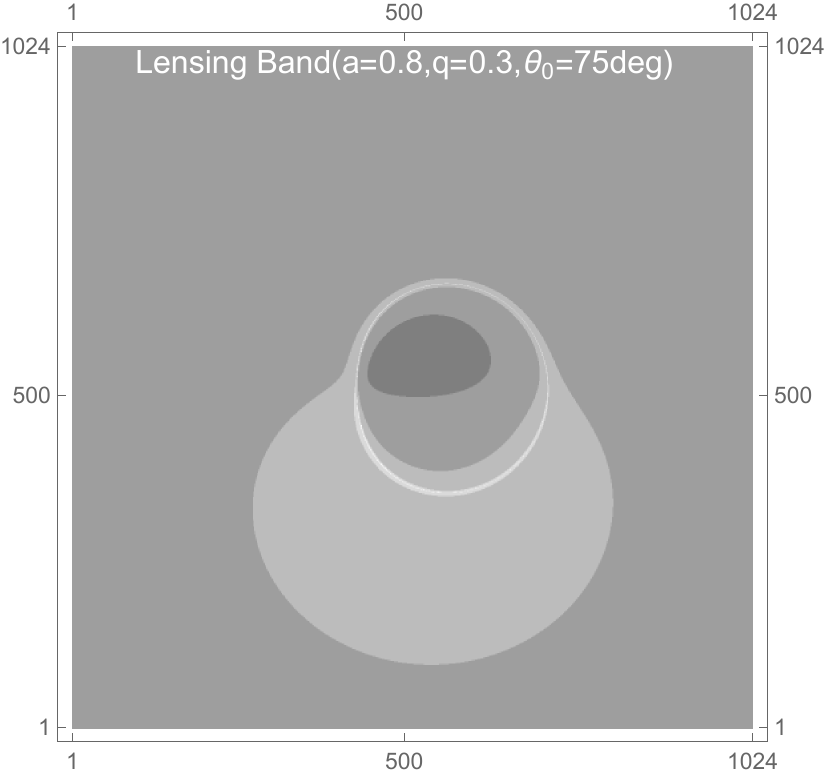}
\includegraphics[width=4.5cm]{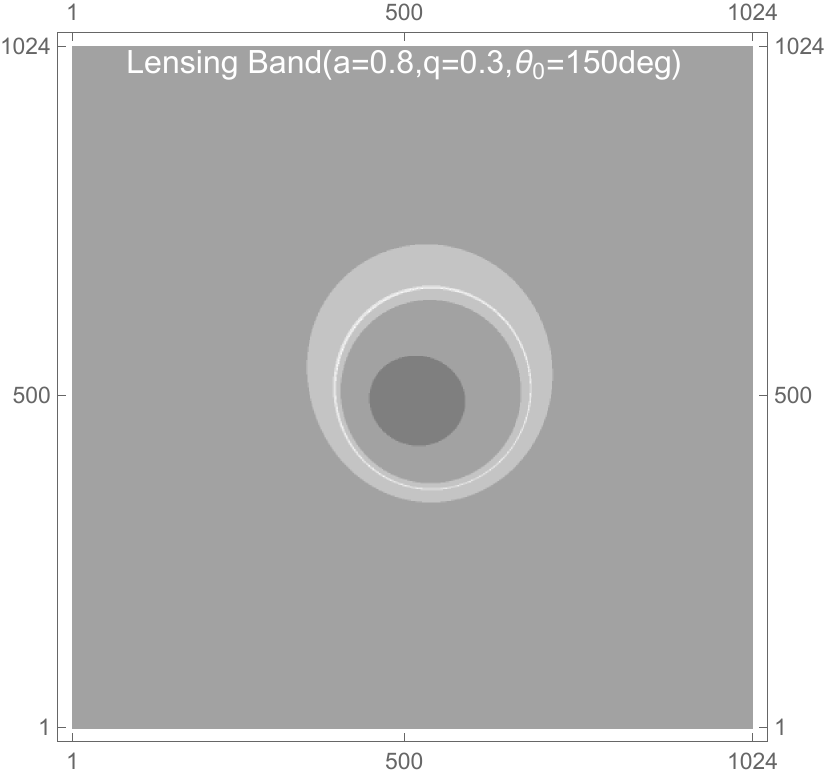}
\includegraphics[width=4.5cm]{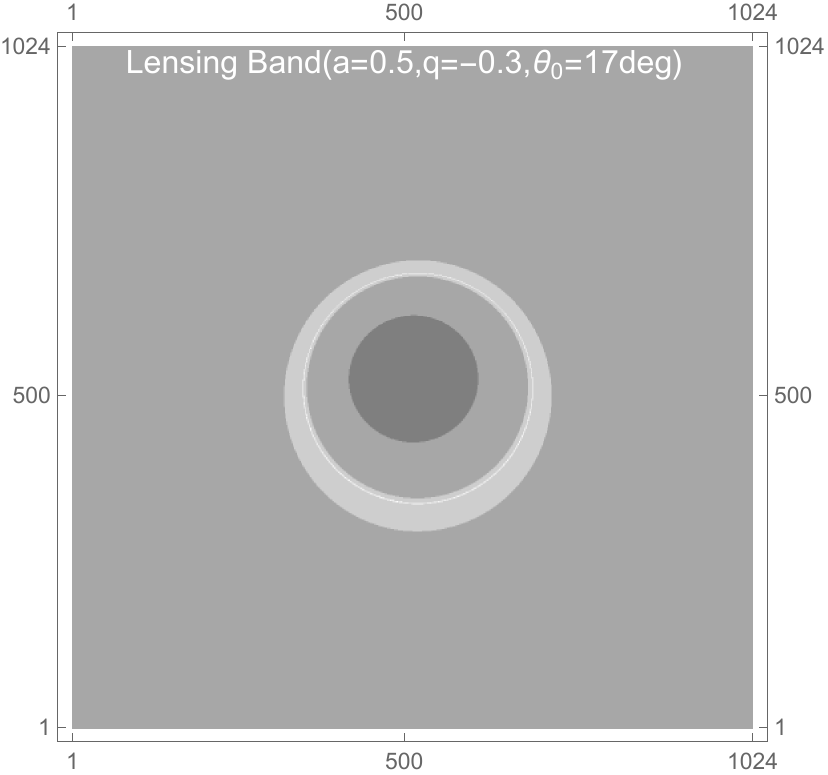}
\includegraphics[width=4.5cm]{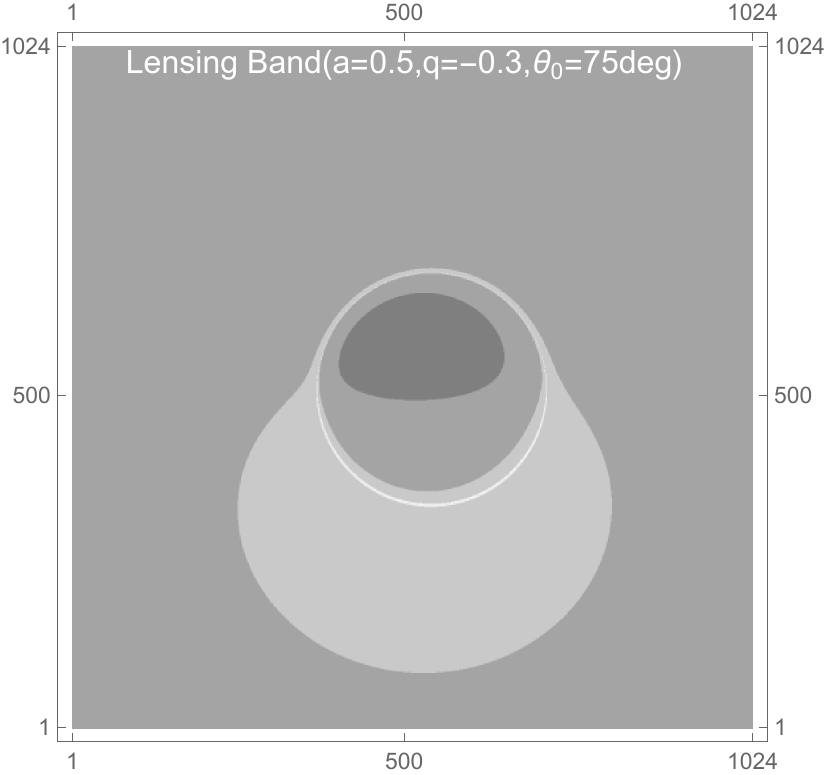}
\includegraphics[width=4.5cm]{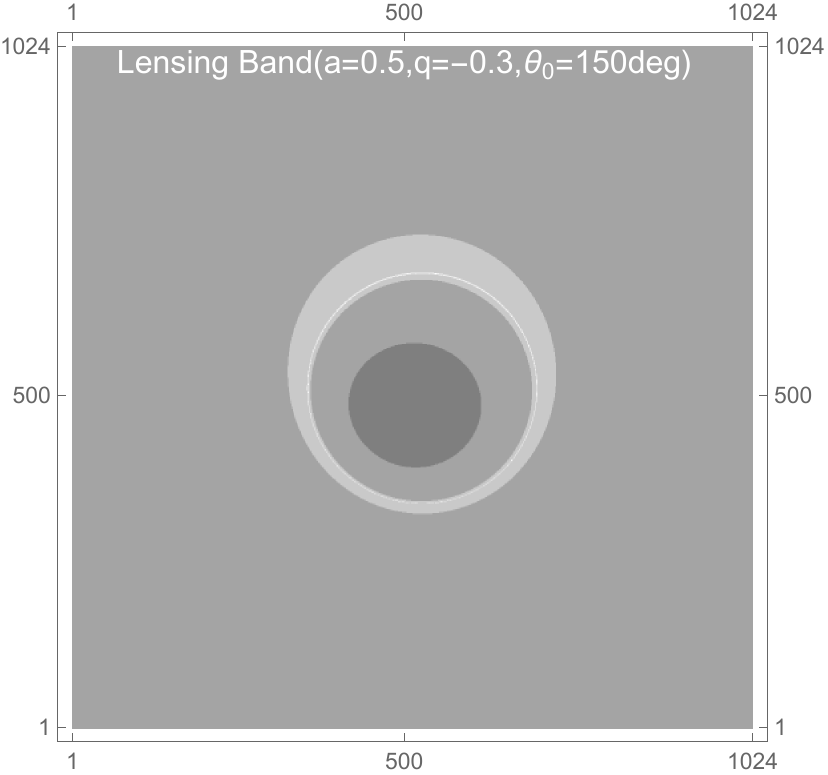}

\vspace{2mm}
\caption{Photon lensing bands around a braneworld black hole for different combinations of the spin parameter $a$, tidal charge $q$, and observer inclination angle $\theta_{0}$. The four parameter sets considered are $a=0.5$, $q=-0.3$; $a=0.5$, $q=0.1$; $a=0.5$, $q=0.3$; and $a=0.8$, $q=0.3$. The corresponding images are shown for inclination angles of $17^\circ$, $75^\circ$, and $150^\circ$. The dark gray and light gray regions denote light rays intersecting the equatorial plane once and twice, respectively. The white line marks the critical curve, and the innermost region represents the inner shadow of the braneworld black hole. In all cases, the black hole mass is fixed at $M=1$.}
\label{fig:3}
\end{figure}

\par
Figure~3 presents the photon lensing bands of the braneworld black hole for different observer inclinations, with each image computed at a resolution of $1024 \times 1024$ pixels. The solid white contour denotes the critical photon curve. We compare three inclination angles, $\theta_{0}=17^\circ$, $75^\circ$, and $150^\circ$. The case $\theta_{0}=17^\circ$ is close to the inclination inferred for M87$^{*}$ from EHT observations. The dark and light gray regions mark light rays that cross the equatorial plane once and twice, respectively, corresponding to the lensing ring and photon ring. The central dark region is the inner shadow of the braneworld black hole.

Comparison of the panels shows that, as the viewing inclination $\theta_{0}$ increases from the polar direction toward the equatorial plane, namely within $0<\theta_{0}<\pi/2$, the critical curve becomes more asymmetric and is gradually stretched toward the southwest direction on the observer's screen. This asymmetric deformation mainly arises from frame dragging generated by black hole rotation, and it becomes stronger as the line of sight moves closer to the equatorial plane. A larger spin further enhances frame dragging and therefore increases the distortion of the critical curve. When the inclination passes the equatorial plane and approaches larger values, the orientation of the critical curve is almost reversed. This behavior can be interpreted through the intrinsic D-shaped profile of the critical curve for a rotating black hole: changing the viewing direction from $\theta_{0}$ to $180^\circ-\theta_{0}$ approximately produces a mirror-symmetric mapping on the image plane.

\par
We now consider photon orbits around the braneworld black hole, mainly focusing on sources on the equatorial plane. For an observer with inclination $\theta_{0}\neq 0$, the first incomplete elliptic integral $F_{s}$ vanishes. We then obtain

\begin{equation}
\label{45}
\sqrt{-w_-}\, a I_r + {\rm sign}(\beta) F_o = 2 z K
\end{equation}
In this expression, $F_{o}$ denotes the incomplete elliptic integral of the first kind, while $K$ denotes the corresponding complete elliptic integral. This equation establishes the functional dependence of the image coordinates $(\alpha,\beta)$ on the radial source position $r_{s}$ and the source order $z$.

\section{Image of the Braneworld Black Hole}
\label{sec:3}
We now examine the observational appearance of a braneworld black hole illuminated by an optically thin accretion disk. Following the standard relativistic disk model, we assume that the accreting material moves along quasi-circular geodesics in the equatorial plane. The inner edge of the disk is taken to be located at the innermost stable circular orbit (ISCO). Physically, the ISCO marks the transition between stable and unstable circular orbits. Under radial perturbations, particles crossing this threshold lose orbital stability, rapidly move inward, and eventually plunge into the black hole.

\par
As test particles move inward toward the ISCO of a braneworld black hole, their orbital velocities become larger. To include these relativistic kinematic effects, we extend the emitting region of the accretion disk inward toward the event horizon. In this disk model, the innermost emitting edge is placed close to the event horizon, while the main body of the disk extends outside the ISCO. This treatment requires two additional considerations:

\begin{enumerate}
    \item The gravitational redshift and radiative-transfer effects near the ISCO must be re-evaluated carefully.
    \item Ray tracing must be carried out over an extended radial interval.
\end{enumerate}

\par
{
For a massive particle in the equatorial plane, we define the conserved
specific energy and angular momentum as
$\mathcal{E}=-u_t$ and $\mathcal{L}=u_\phi$, respectively. As derived
in Appendix~B, the inward radial component of the four-velocity can be
written as
\begin{equation}
u^r
=
-\sqrt{
-\frac{
V(r,\mathcal{E},\mathcal{L})
}{
g_{rr}
}
}.
\end{equation}
Here, the negative sign selects the inward-moving branch, and the
effective potential is defined by
\begin{equation}
V(r,\mathcal{E},\mathcal{L})
=
\left[
1
+g^{tt}\mathcal{E}^{2}
-2g^{t\phi}\mathcal{E}\mathcal{L}
+g^{\phi\phi}\mathcal{L}^{2}
\right]_{\theta=\pi/2}.
\end{equation}}

\par
Using Eq.~(47), we describe the radial motion of test particles through the effective
potential. Figure~4 shows the radial profiles of $-V_{\rm eff}$ for
different spin parameters $a$ and tidal charges $q$. At fixed spin,
increasing $q$ lowers the maximum of $-V_{\rm eff}$ and shifts its
location toward a larger radius. A similar behavior is observed when
the spin is increased at fixed tidal charge. The differences among the
curves are most pronounced close to the black hole, whereas they
gradually diminish at larger radii.

\begin{figure}[htbp]
\centering
\includegraphics[width=13cm]{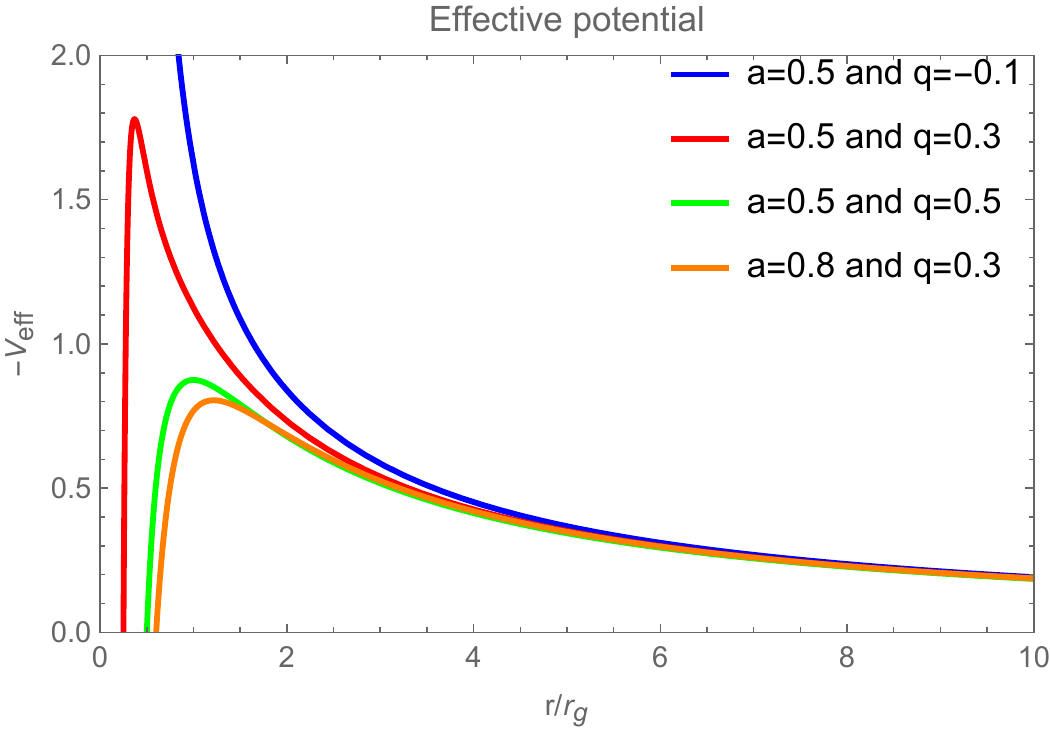}

\vspace{2mm}
\caption{Profiles of the effective potential versus radius for test particles around the braneworld black hole. Four parameter sets are depicted: $a=0.5$, $q=-0.1$; $a=0.5$, $q=0.3$; $a=0.5$, $q=0.5$; and $a=0.8$, $q=0.3$, illustrating the influence of black hole spin and tidal charge. The peak of the effective potential shifts outward and decreases in amplitude as either parameter increases. The image adopts the convention $M=1$ for the black hole mass.}
\label{fig:4}
\end{figure}

\par
{
For a circular equatorial orbit, the radial velocity vanishes. The ISCO
is determined by the simultaneous marginal-stability conditions
$V=0$, $\partial_rV=0$, and $\partial_r^2V=0$, where the radial
derivatives are evaluated while keeping $\mathcal{E}$ and
$\mathcal{L}$ fixed. The corresponding physical root outside the event
horizon is denoted by $r_{\rm ISCO}$.

Inside the ISCO, stable circular motion is no longer possible. Following
the Cunningham-type prescription adopted in this work, the conserved
quantities of the plunging matter are fixed at their ISCO values,
$\mathcal{E}=\mathcal{E}_{\rm ISCO}$ and
$\mathcal{L}=\mathcal{L}_{\rm ISCO}$. Equation~(46) then becomes
\begin{equation}
u^r=-\sqrt{-\frac{V\left(r,\mathcal{E}_{\rm ISCO},
\mathcal{L}_{\rm ISCO}
\right)}{g_{rr}}},\qquad r_H\leq r<r_{\rm ISCO}.
\end{equation}
The explicit expressions for $r_{\rm ISCO}$,
$\mathcal{E}_{\rm ISCO}$, $\mathcal{L}_{\rm ISCO}$, and the complete
plunging four-velocity
$u^\mu=(u^t,u^r,0,u^\phi)$
are presented in Appendix~C. The upper and lower signs used there
correspond consistently to the prograde and retrograde branches,
respectively.
}

\subsection{Zero-Angular-Momentum Observer}
To model a fisheye camera and define the observer's local reference frame rigorously, we introduce a standard orthonormal tetrad that provides a local basis at each point in spacetime~\cite{34}. In the braneworld black hole spacetime, the observer's orthonormal tetrad can be written as~\cite{34,35}

\begin{equation}
\label{49}
e_t = \delta \, \partial_t + \chi \, \partial_\phi, \quad 
e_r = \frac{1}{\sqrt{g_{rr}}} \, \partial_r, 
\end{equation}

\begin{equation}
\label{50}
e_\theta = \frac{1}{\sqrt{g_{\theta\theta}}} \, \partial_\theta, \quad 
e_\phi = \frac{1}{\sqrt{g_{\phi\phi}}\partial_\phi}  , 
\end{equation}

where

\begin{equation}
\label{51}
\delta = \sqrt{\frac{g_{\phi\phi}}{g_{t\phi}^2 - g_{tt} g_{\phi\phi}}}, 
\quad 
\chi = -\frac{g_{t\phi}}{g_{\phi\phi}} 
\sqrt{\frac{g_{\phi\phi}}{g_{t\phi}^2 - g_{tt} g_{\phi\phi}}}. 
\end{equation}

\par
In the ZAMO frame, photon trajectories are reversible, so photons arriving at the observer can be traced backward to their emission points. The photon four-momentum measured in the local tetrad frame can then be transformed into the coordinate basis. Thus, the four-momentum $p^{\mu}$ is given by

\begin{equation} 
\label{52}
p^\mu = \eta^{\mu\nu} e^\xi_\nu k_\xi ,
\end{equation}

The explicit components of the four-momentum are defined as \cite{16,33}:

\begin{equation} 
\label{53}
P^t = E \left( \delta - \chi \xi \right), \quad
P^r = E \, \frac{1}{\sqrt{g_{rr}}} \, \frac{\pm_r \sqrt{R(r)}}{\Delta_r}, 
\end{equation}

\begin{equation}
\label{54}
P^\theta = E \, \frac{1}{\sqrt{g_{\theta\theta}}} \, \frac{\pm_\theta \sqrt{\Theta(\theta)}}{\Delta_\theta}, \quad
P^\phi = E \, \frac{\xi}{\sqrt{g_{\phi\phi}}}.
\end{equation}

\begin{figure}[htbp]
\centering
\includegraphics[width=3.5cm]{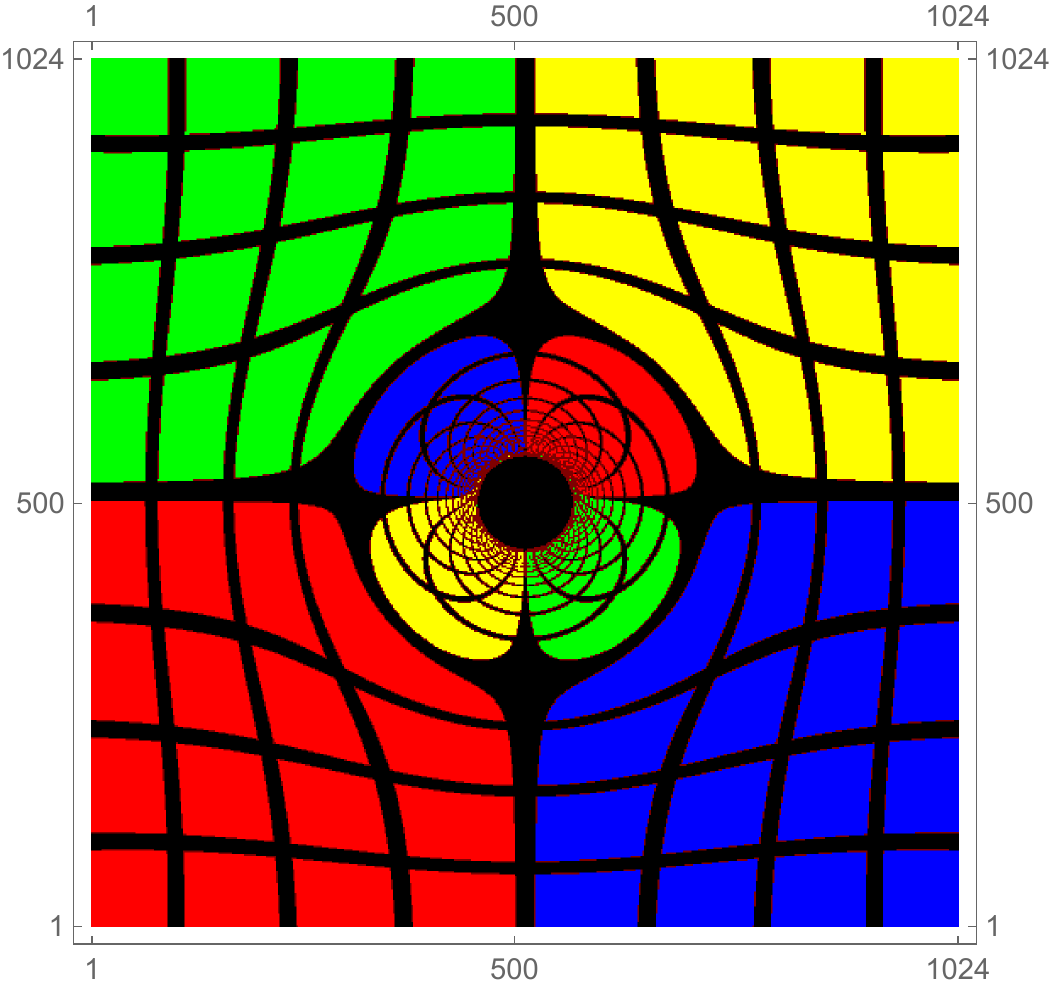}
\includegraphics[width=3.5cm]{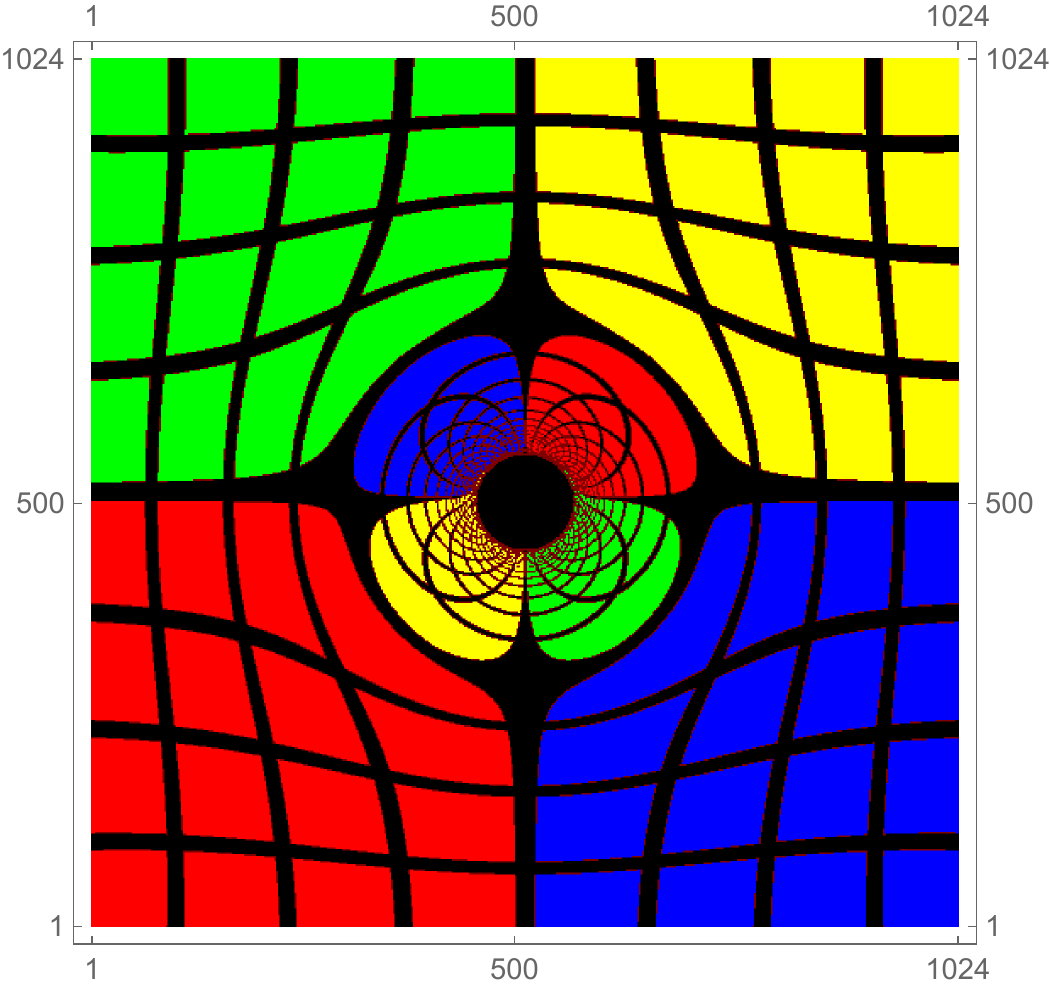}
\includegraphics[width=3.5cm]{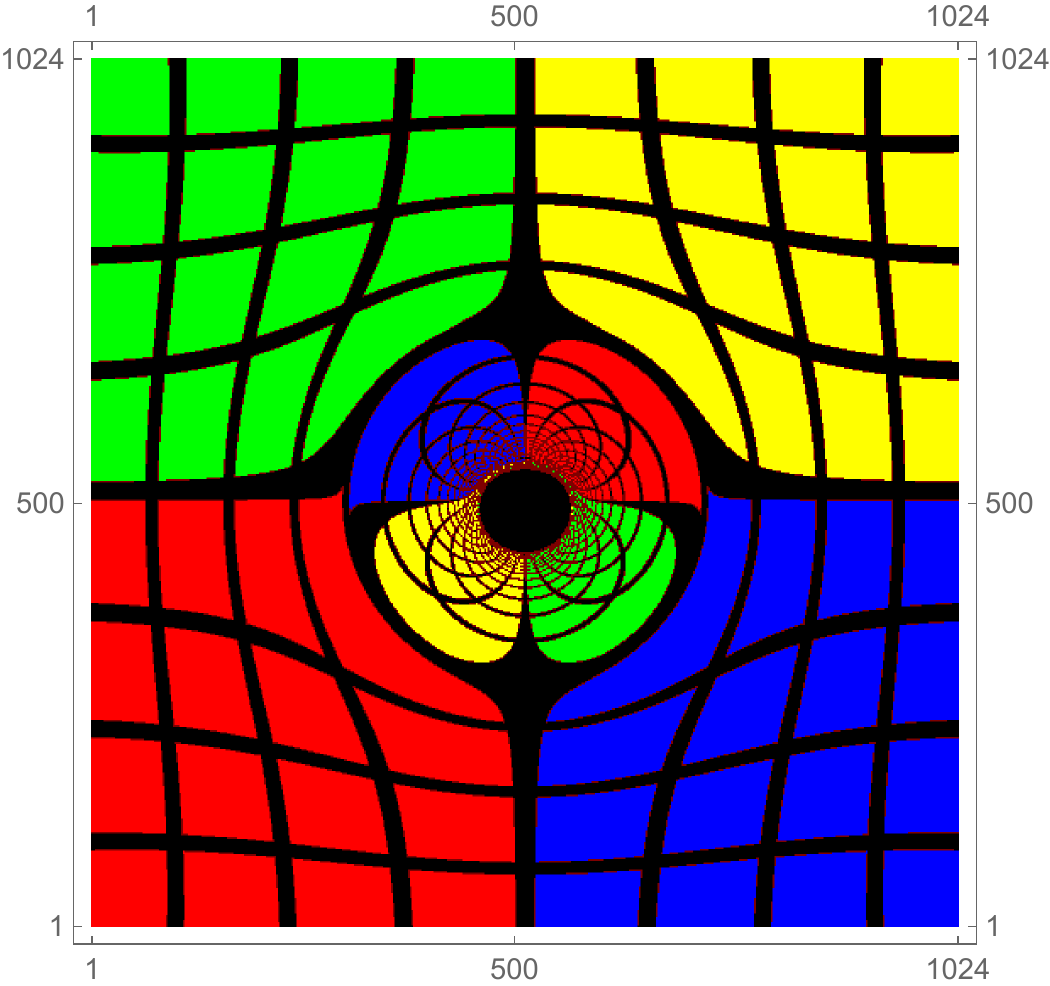}
\includegraphics[width=3.5cm]{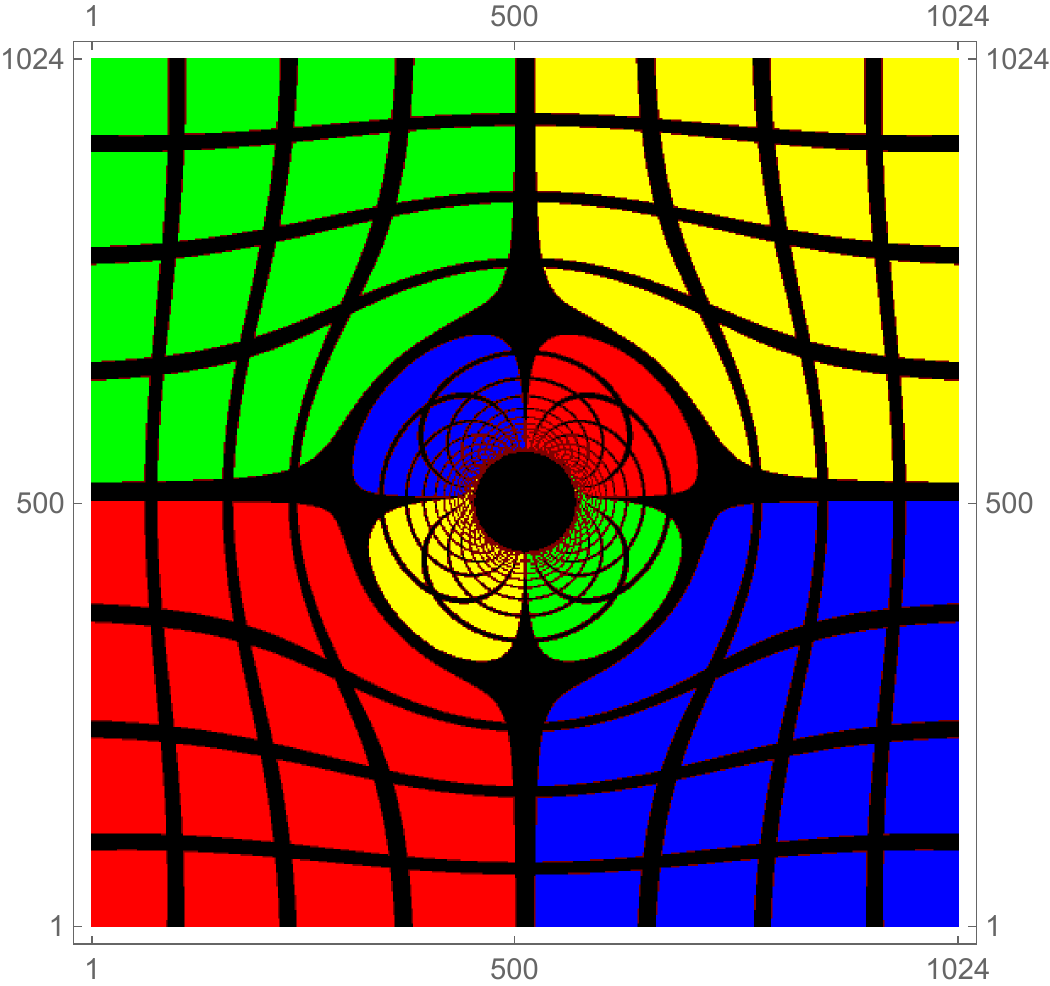}

\vspace{2mm}
\caption{The shadow cast by the braneworld black hole is studied via numerical ray tracing. From left to right, the parameter sets are $a=0.5$, $q=-0.3$; $a=0.5$, $q=0(Kerr)$; $a=0.5$, $q=0.3$; and $a=0.8$, $q=0.3$. All images adopt the convention $M=1$ for the black hole mass, and the observer inclination angle is fixed at $17^\circ$.}
\label{fig:5}
\end{figure}

\par
To show more clearly how the braneworld black hole inner shadow changes, the corresponding regions in Fig.~5 are enlarged in Fig.~6. The inner shadow is progressively distorted as the tidal charge and spin increase. Its morphology is especially more sensitive to the spin parameter than to the tidal charge.

\begin{figure}[htbp]
\centering
\includegraphics[width=3.5cm]{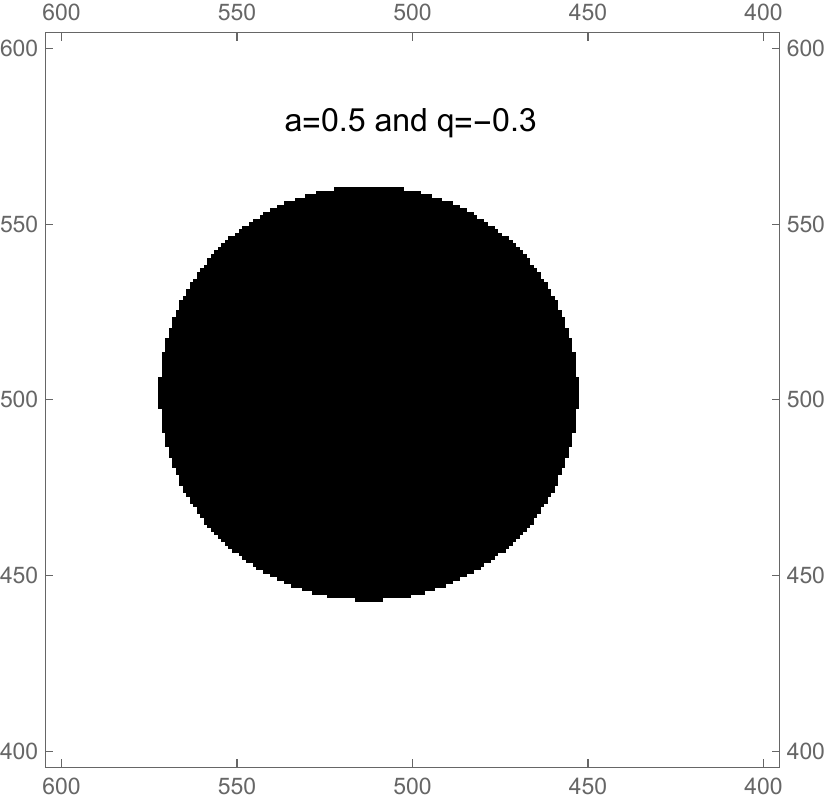}
\includegraphics[width=3.5cm]{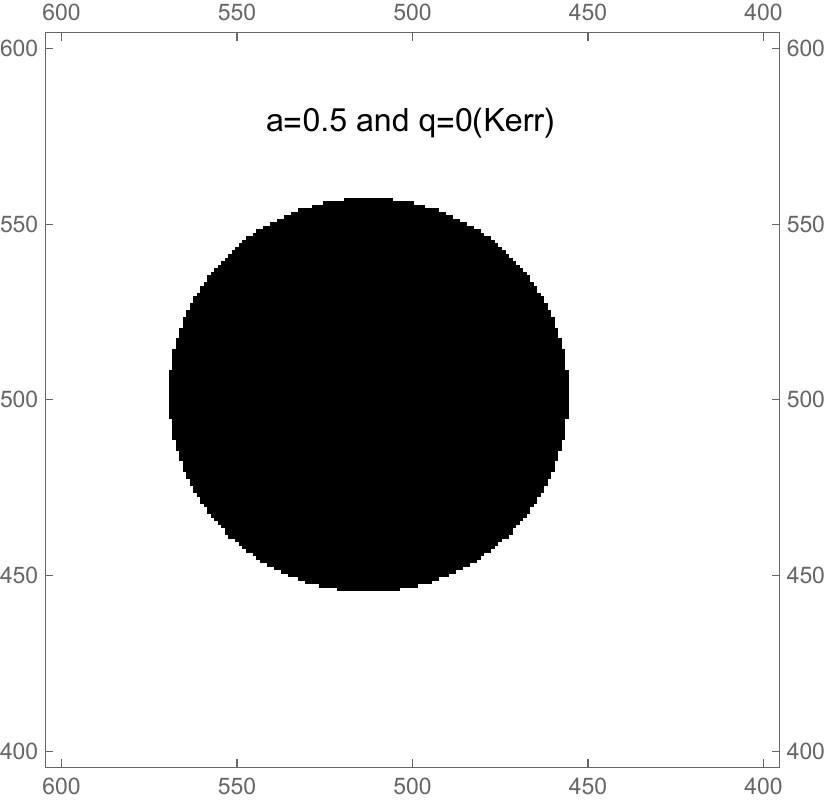}
\includegraphics[width=3.5cm]{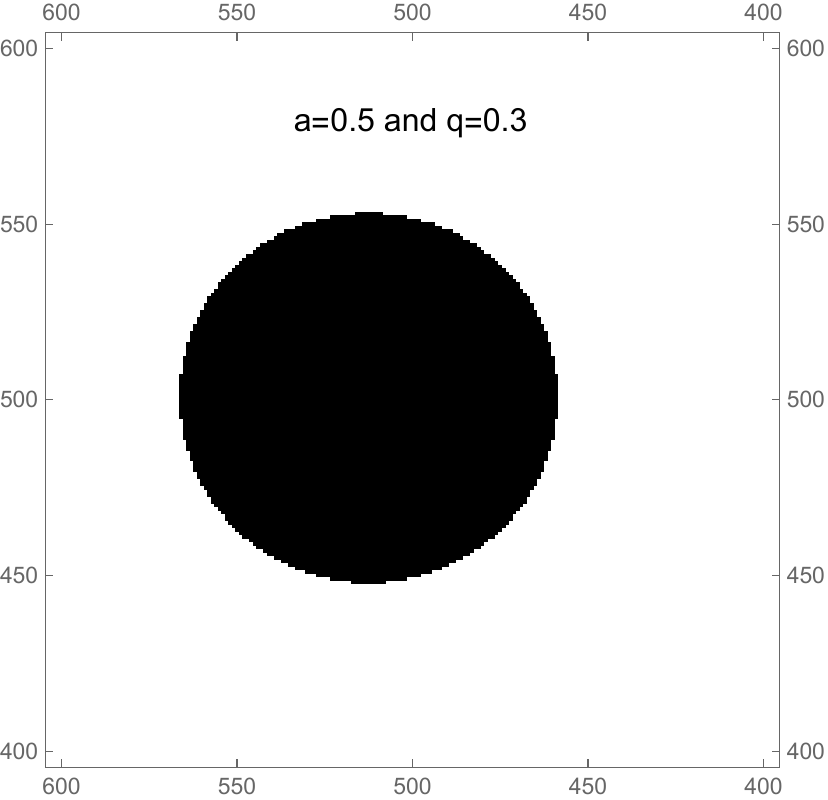}
\includegraphics[width=3.5cm]{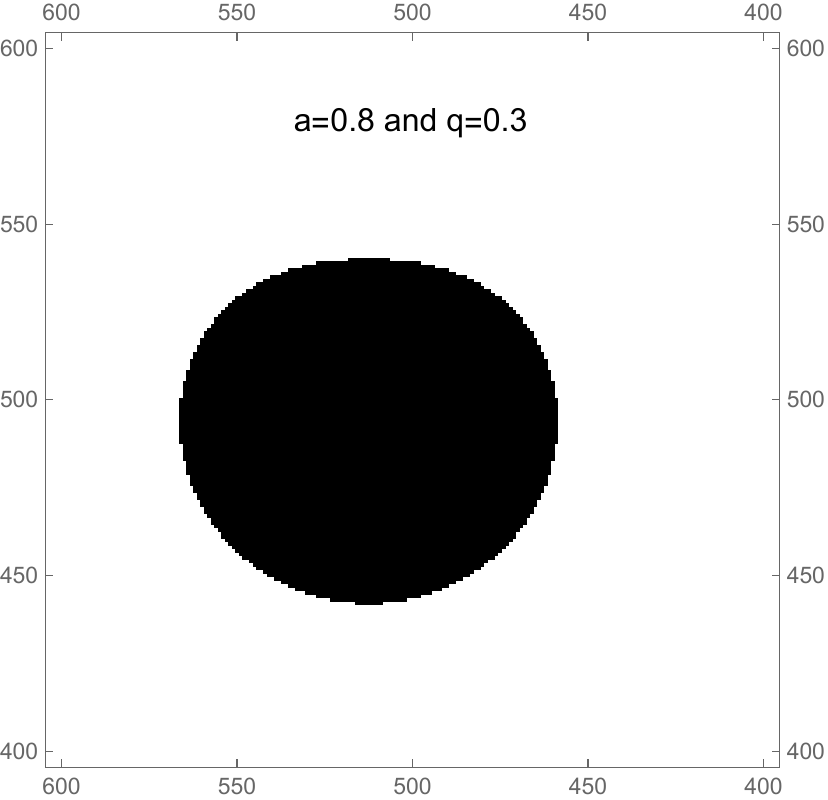}

\vspace{2mm}
\caption{The inner shadow of the braneworld black hole is studied using four parameter sets, shown from left to right: $a=0.5$, $q=-0.3$; $a=0.5$, $q=0(Kerr)$; $a=0.5$, $q=0.3$; and $a=0.8$, $q=0.3$. All images adopt the convention $M=1$ for the black hole mass, and the observer inclination angle is fixed at $17^\circ$.}
\label{fig:6}
\end{figure}

\par
Figure~7 shows the inner shadow of the braneworld black hole for different observer inclinations, tidal charges, and spin parameters. For a controlled comparison, four inner-shadow images are superposed in each panel, leading to the following trends:

\begin{itemize}
    \item As the observer inclination increases from $0^\circ$ to $90^\circ$, the inner shadow becomes progressively flattened. For inclination angles larger than $90^\circ$, its deformation gradually decreases.

    \item As the tidal charge increases, the effective gravitational field is modified, causing the inner shadow to shrink.

    \item Similarly, as the spin parameter increases, the black hole rotation strengthens the frame-dragging effect, thereby reducing the size of the inner shadow.

    \item Across all observer inclination angles, the deformation of the inner shadow is more sensitive to variations in the spin parameter than to changes in the tidal charge.
\end{itemize}

\begin{figure}[htbp]
\centering
\includegraphics[width=4.5cm]{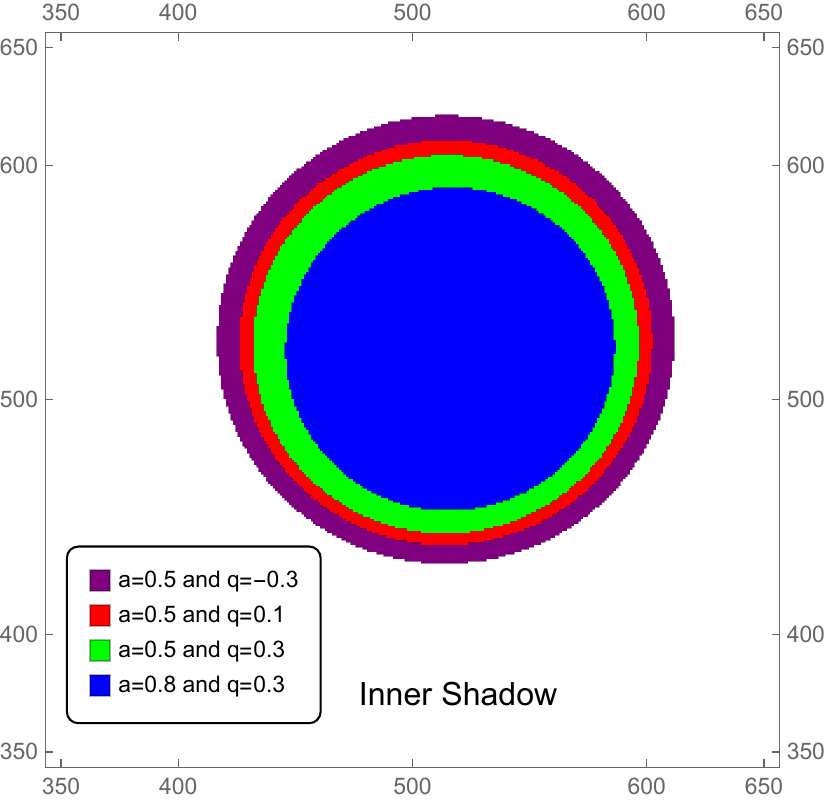}
\includegraphics[width=4.5cm]{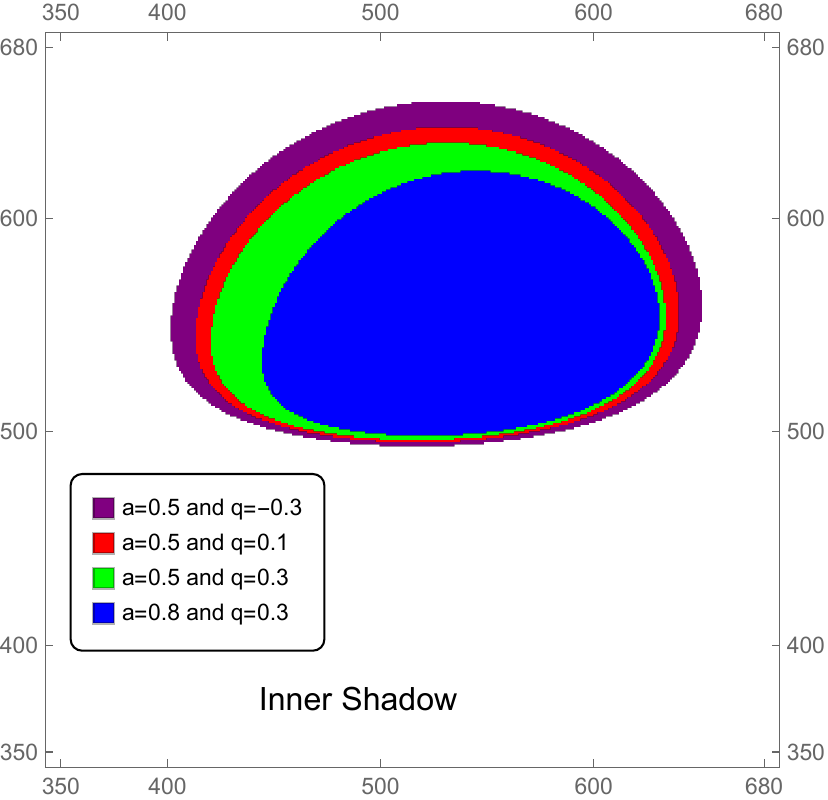}
\includegraphics[width=4.5cm]{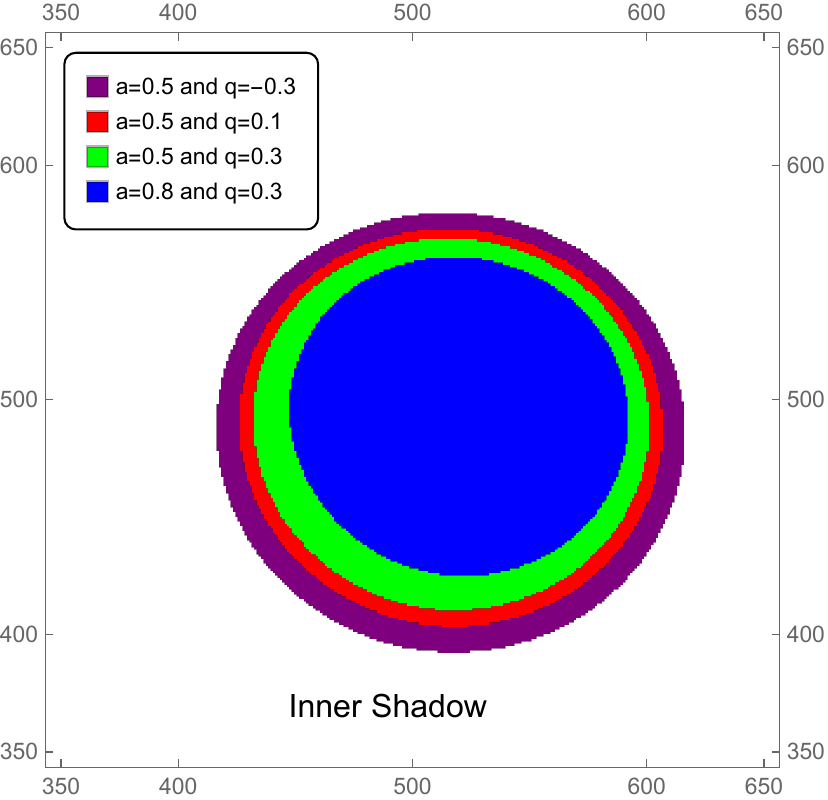}
\vspace{2mm}
\caption{A comparison of the black hole inner shadow is presented. Four parameter sets are selected: $a=0.5,\,q=-0.3$, $a=0.5,\,q=0.1$, $a=0.5,\,q=0.3$, and $a=0.8,\,q=0.3$. These cases are analyzed at three different observer inclination angles. From left to right, the columns correspond to $\theta_{0}=17^\circ$, $75^\circ$, and $150^\circ$. In all panels, the black hole mass is set to $M=1$.}
\label{fig:7}
\end{figure}

\begin{table}[htbp]

\centering
\caption{\label{tab:0}
Comparison of the effective shadow eccentricity between Kerr and rotating braneworld black holes at the observer inclination angle $\theta_0=17^\circ$.
The Kerr case is recovered by setting $q=0$. The quantity
$\Delta e_{\rm sh}=e_{\rm sh}(a,q)-e_{\rm sh}(a,0)$ measures the additional global deformation induced by the braneworld tidal charge at fixed spin.}
\begin{tabular}{l c c c c}
\hline
Model & $a$ & $q$ & $e_{\rm sh}$ & $\Delta e_{\rm sh}$ \\
\hline
Kerr       & 0.5 & 0      & 0.0533767 & 0 \\
Braneworld & 0.5 & $-0.3$ & 0.0457269 & $-0.0076498$ \\
Braneworld & 0.5 & $-0.1$ & 0.0504692 & $-0.0029075$ \\
Braneworld & 0.5 & 0.1    & 0.0567748 & 0.0033981 \\
Braneworld & 0.5 & 0.3    & 0.0657453 & 0.0123686 \\
\hline
Kerr       & 0.8 & 0      & 0.102891  & 0 \\
Braneworld & 0.8 & $-0.3$ & 0.0836967 & $-0.0191943$ \\
Braneworld & 0.8 & $-0.1$ & 0.0952338 & $-0.0076572$ \\
Braneworld & 0.8 & 0.1    & 0.112556  & 0.009665 \\
Braneworld & 0.8 & 0.3    & 0.143703  & 0.040812 \\
\hline
\end{tabular}
\end{table}

Table~\ref{tab:0} compares the effective shadow eccentricity of Kerr and rotating braneworld black holes at $\theta_0=17^\circ$. The Kerr limit is recovered when $q=0$. For fixed spin, a negative tidal charge decreases the effective shadow eccentricity, while a positive tidal charge increases it. This indicates that the tidal charge can modify the global deformation of the black hole shadow. Moreover, the variation of $\Delta e_{\rm sh}$ becomes more significant for the larger spin case, showing that the deformation effect is enhanced when the black hole rotates faster.

\subsection{Intensity and Frequency Shift}
When analyzing photons emitted from the accretion disk and reaching the observer's image plane, it is essential to account for variations in the observed intensity caused by scattering, absorption, Doppler boosting, and gravitational redshift. Under these assumptions, the radiative transfer equation governs the intensity measured by the observer~\cite{36}:

\begin{equation} 
\label{55}
\frac{d}{d\lambda}\left(\frac{I_\nu}{\nu^3}\right) = \frac{J_\nu - \kappa_\nu I_\nu}{\nu^2},
\end{equation}

Here, $\lambda$ is the affine parameter, while $I_{\nu}$, $J_{\nu}$, and $\kappa_{\nu}$ are the specific intensity, emission coefficient, and absorption coefficient at frequency $\nu$, respectively. In vacuum, $J_{\nu}$ and $\kappa_{\nu}$ vanish, so $I_{\nu}/\nu^{3}$ is conserved along the geodesic. Because the accretion disk is assumed to be geometrically thin and optically thin, the intensity profile on the observer's image plane is obtained by integrating photon trajectories. The observed intensity is therefore~\cite{12}

\begin{equation}
\label{56}
I_{\nu_0} =
\sum_{m=1}^{N_{\rm max}}
\left(\frac{\nu_0}{\nu_m}\right)^3
J_m \,
\frac{1 - e^{-\kappa_m \tau_{m-1}}}{\kappa_m}\, .
\end{equation}

Here, $\nu_{0}$ is the photon frequency on the observer's image plane, $\nu_{m}$ is the frequency measured in the disk comoving frame, and $\tau_{k}$ is the optical depth for photons emitted at point $k$. After taking into account the physical properties of the accretion disk, Eq.~(\ref{56}) can be simplified to

\begin{equation} 
\label{57}
I_{\nu_0} = \sum_{m=1}^{N_{max}} f_m \, g^3(r_m) \, J_{model}(r_m),
\end{equation}
where $f_{m}$ denotes a weighting factor, $g(r_{m})$ is the frequency-shift factor, and $J_{model}(r_{m})$ represents the emissivity model evaluated at the emission radius $r_{m}$.

In what follows, we consider radiation emitted from the equatorial plane. For each image-plane coordinate $(\alpha,\beta)$, $N_{\max}(\alpha,\beta)$ gives the number of intersections between the photon geodesic and the equatorial plane. For $N_{\max}=1$, the geodesic crosses the equatorial plane only once and produces a direct image of the equatorial emission. For $N_{\max}=2,3,4,\ldots$, multiple intersections occur and produce lensed and higher-order images.

\par
The radius $r_{m}(\alpha,\beta)$ is the radial position at which the geodesic associated with $(\alpha,\beta)$ crosses the equatorial plane. Therefore, the image order is fixed by the number of equatorial-plane intersections, rather than by the number of radial or angular turning points.

In Eq.~(\ref{57}), $J_{model}(r_{m})$ is the emissivity evaluated at the emission radius $r_{m}$ on the equatorial plane, $g$ is the combined gravitational and Doppler frequency-shift factor, and $f_{m}$ is a correction factor that adjusts the luminosity contribution from higher-order photon-ring images~\cite{32}. For the geometrically thin and optically thin disk considered here, we adopt $f_{m}=1.5$ throughout the analysis. The emissivity profile $J_{model}(r)$ is modeled as~\cite{37}

\begin{equation} 
\label{58}
\log[J_{model}(r)] = A \, \log\!\left(\frac{r}{r_H}\right) + B \left[\log\!\left(\frac{r}{r_H}\right)\right]^2, 
\end{equation}
where $A$ and $B$ are model-dependent fitting parameters, and $r_{H}$ denotes the event horizon radius of the black hole.

\par
For 230~GHz, the corresponding observing wavelength for M87$^{*}$ and Sgr~A$^{*}$ is 1.3~mm. We adopt $A=-2$ and $B=-\frac{1}{2}$ at this frequency. For 86~GHz, we set $A=0$ and $B=-\frac{3}{4}$~\cite{37}. With this choice, the luminosity of the braneworld black hole is calculated in the thin-disk model.
\begin{figure}[htbp]
\centering
\includegraphics[width=13cm]{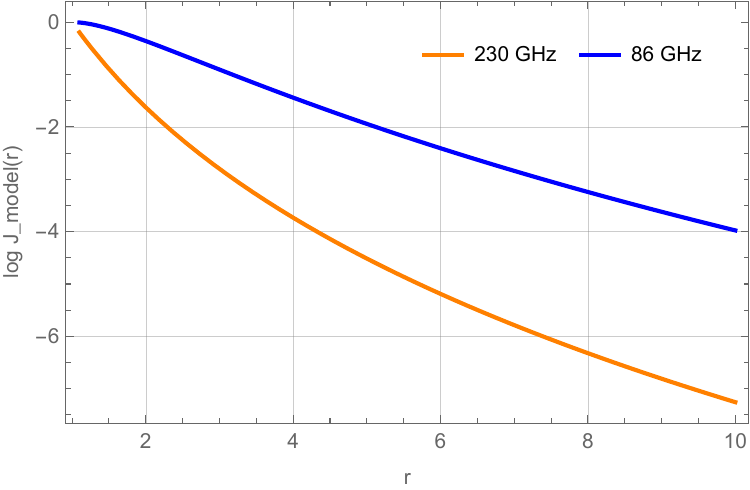}

\vspace{2mm}
\caption{Comparison of the accretion-disk emissivity profiles at 230~GHz and 86~GHz, with the event horizon radius fixed at $r_{H}=1$. In all cases, the black hole mass is set to $M=1$.} \label{fig:emissivity_profiles}
\label{fig:8}
\end{figure}

The frequency-shift factor is defined by $g=\nu_{0}/\nu_{m}$. The disk model used here extends previous models by allowing radiation from the disk inner edge to reach the event horizon, and thus enables a more detailed analysis of frequency-shift effects. Outside the ISCO, the accretion flow remains on circular orbits with angular velocity $\Omega_{m}(r)=\left.\frac{u^{\phi}}{u^{t}}\right|_{r=r_{m}}$.

The parameter $\xi$ is given in Eq.~(\ref{7}), while $\epsilon$ measures the ratio of the observed energy to the conserved energy along the geodesic~\cite{12}:

\begin{equation}
\label{59}
\epsilon = \frac{E_0}{\mathcal{E}} = \frac{P_t}{k_t} = \delta \,(1 + \xi \chi).
\end{equation}

\par
The quantities $\chi$ and $\delta$ are defined in Eq.~(\ref{51}). Since the spacetime is asymptotically flat, one has $\epsilon=1$. For an emission radius $r_{m}$ outside the ISCO, the frequency-shift factor therefore becomes~\cite{12}

\begin{equation}
\label{60}
g_m = \frac{\epsilon}{\zeta \,(1 - \xi \Omega_m)}, 
\quad r_m > r_{ISCO}, 
\end{equation}

where

\begin{equation}
\label{61}
\zeta = \sqrt{-\frac{1}{g_{tt} + 2 g_{t\phi} \Omega_m + g_{\phi\phi} \Omega_m^2}} \,\bigg|_{r = r_m}.
\end{equation}

As noted above, the ISCO fixes the radial location of the accretion disk inner boundary. Inside the ISCO, the flow enters the plunging domain and develops a radial velocity $u^{r}_{c}$. In this region, the frequency-shift factor is

{
\begin{equation}
\label{62}
g_m = \frac{\epsilon}{- \frac{u^r_c k_r}{k_t} + g^{tt} \mathcal{E}_{\rm ISCO} - \xi g^{t\phi} \mathcal{E}_{\rm ISCO} - g^{t\phi} \mathcal{L}_{\rm ISCO}+ \xi g^{\phi\phi} \mathcal{L}_{\rm ISCO}}.
\end{equation}}

\par
For a clearer presentation of the frequency-shift distribution, the observer radius is set to $r_{o}=100R_{s}$, where $R_{s}$ is the Schwarzschild radius, and the image-plane angle is chosen as $\phi=\pi/10$. Both prograde and retrograde disk rotations are included. Photons are separated according to whether they are emitted along or opposite to the black hole rotation. This distinction is important because the disk rotation direction strongly affects both the observed frequency-shift and the photon trajectories near the braneworld black hole.

\begin{figure}[htbp]
\centering
\includegraphics[width=15cm]{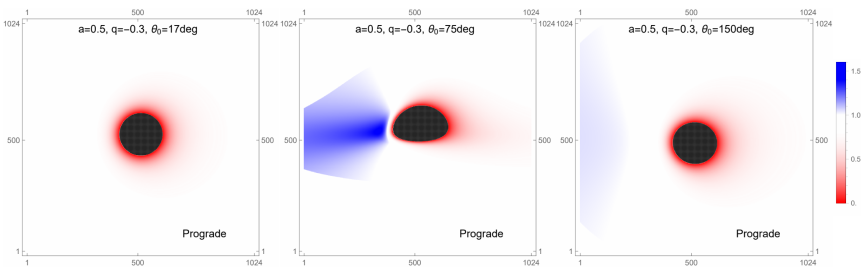}
\vspace{2mm}
\includegraphics[width=15cm]{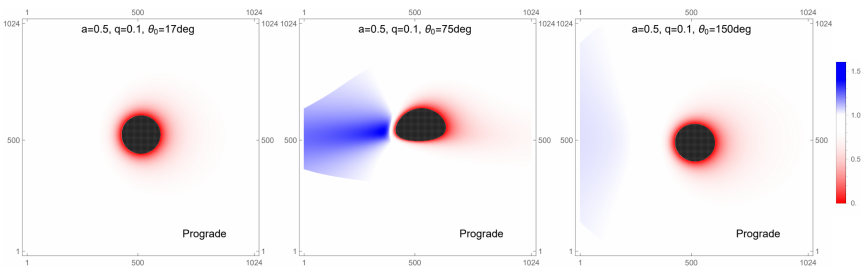}
\vspace{2mm}
\includegraphics[width=15cm]{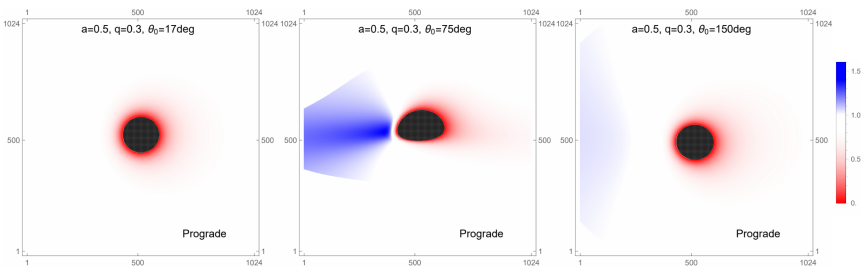}
\vspace{2mm}
\includegraphics[width=15cm]{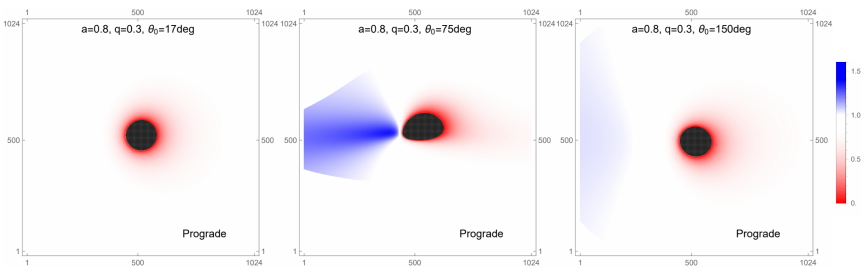}
\vspace{2mm}

\caption{First-order frequency-shift images for prograde accretion disks. The panels display the frequency-shift distribution for different black hole parameter sets and observer inclination angles. The four parameter sets are $a=0.5$, $q=-0.3$; $a=0.5$, $q=0.1$; $a=0.5$, $q=0.3$; and $a=0.8$, $q=0.3$. For each case, the frequency-shift distribution is shown at observer inclination angles of $17^\circ$, $75^\circ$, and $150^\circ$. In all panels, the black hole mass is fixed at $M=1$.}
\label{fig:9}
\end{figure}

\begin{figure}[htbp]
\centering
\includegraphics[width=15cm]{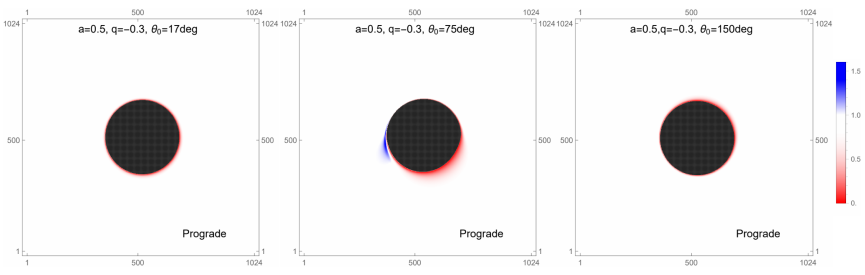}
\vspace{2mm}
\includegraphics[width=15cm]{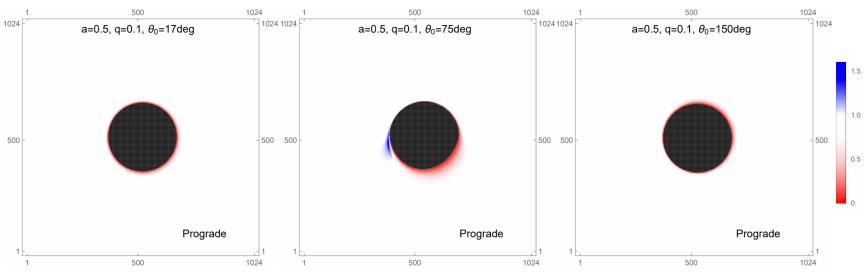}
\vspace{2mm}
\includegraphics[width=15cm]{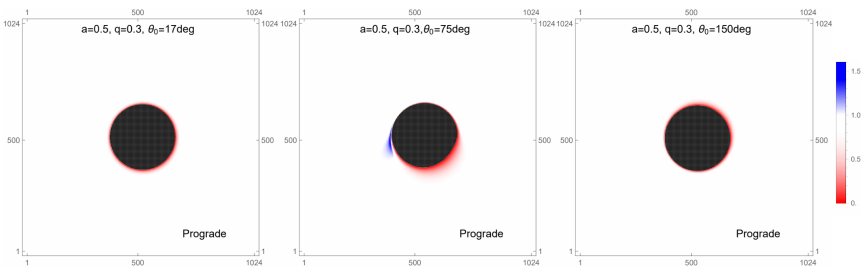}
\vspace{2mm}
\includegraphics[width=15cm]{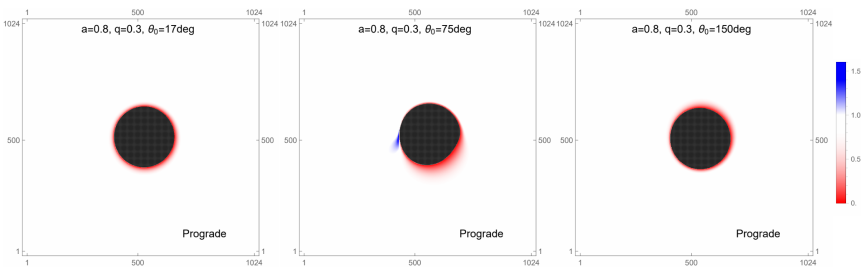}
\vspace{2mm}

\caption{Second-order frequency-shift images for prograde accretion disks. The panels display the frequency-shift distribution for different black hole parameter sets and observer inclination angles. The four parameter sets are $a=0.5$, $q=-0.3$; $a=0.5$, $q=0.1$; $a=0.5$, $q=0.3$; and $a=0.8$, $q=0.3$. For each case, the frequency-shift distribution is shown at observer inclination angles of $17^\circ$, $75^\circ$, and $150^\circ$. In all panels, the black hole mass is fixed at $M=1$.}
\label{fig:10}
\end{figure}

\begin{figure}[htbp]
\centering
\includegraphics[width=15cm]{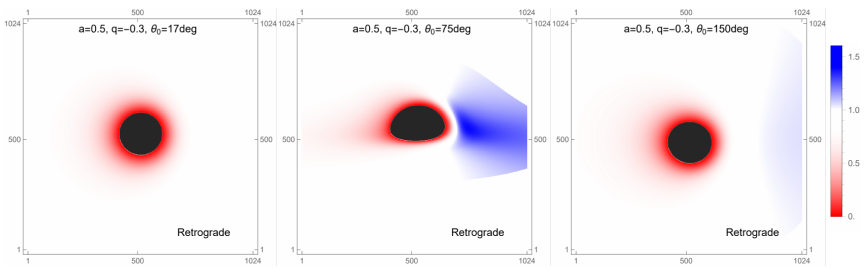}
\vspace{2mm}
\includegraphics[width=15cm]{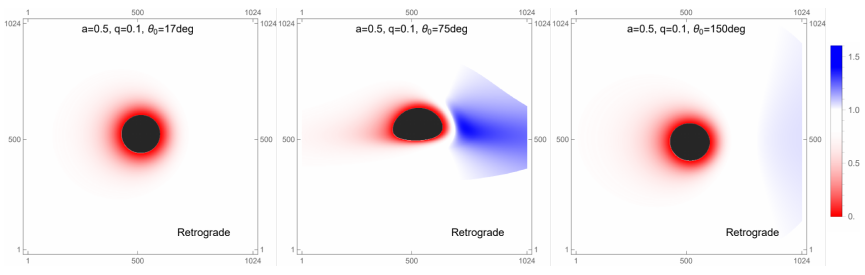}
\vspace{2mm}
\includegraphics[width=15cm]{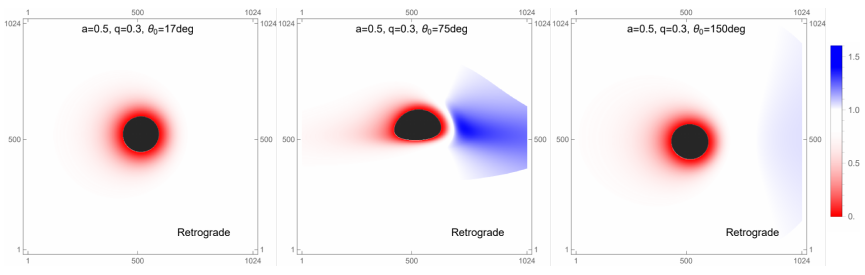}
\vspace{2mm}
\includegraphics[width=15cm]{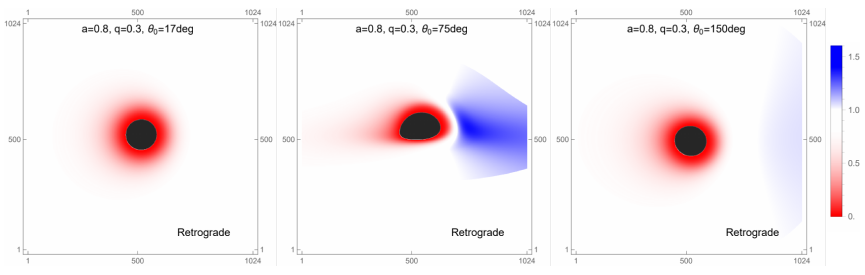}
\vspace{2mm}

\caption{First-order frequency-shift images for retrograde accretion disks. The panels display the frequency-shift distribution for different black hole parameter sets and observer inclination angles. The four parameter sets are $a=0.5$, $q=-0.3$; $a=0.5$, $q=0.1$; $a=0.5$, $q=0.3$; and $a=0.8$, $q=0.3$. For each case, the frequency-shift distribution is shown at observer inclination angles of $17^\circ$, $75^\circ$, and $150^\circ$. In all panels, the black hole mass is fixed at $M=1$.}
\label{fig:11}
\end{figure}

\begin{figure}[htbp]
\centering
\includegraphics[width=15cm]{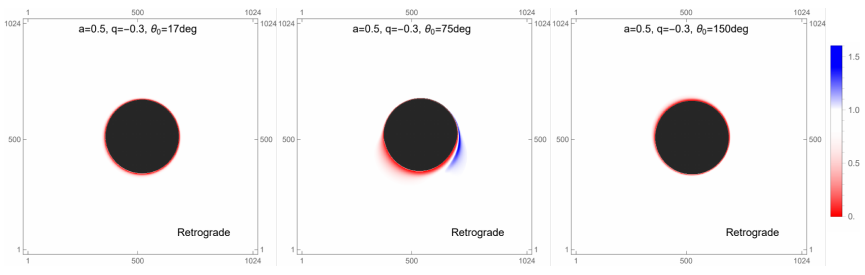}
\vspace{2mm}
\includegraphics[width=15cm]{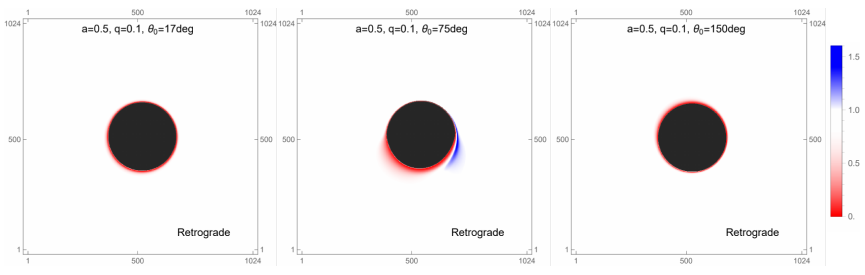}
\vspace{2mm}
\includegraphics[width=15cm]{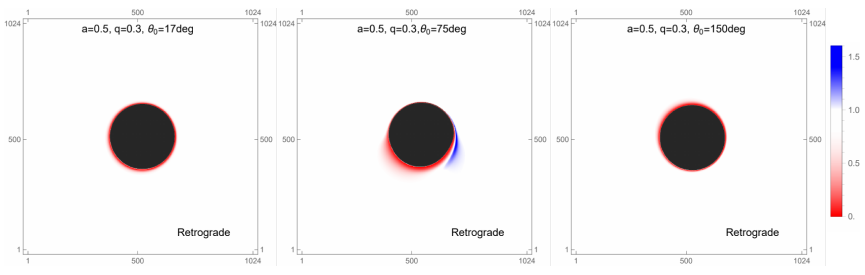}
\vspace{2mm}
\includegraphics[width=15cm]{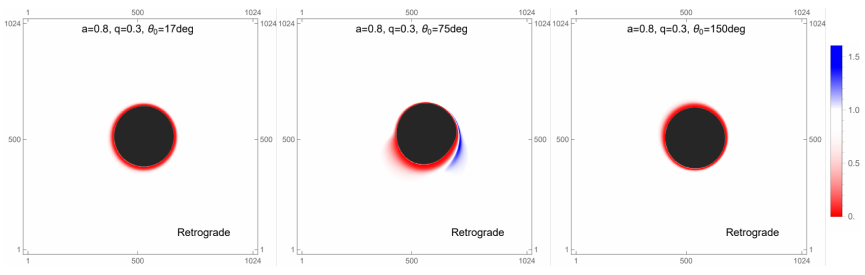}
\vspace{2mm}

\caption{Second-order frequency-shift images for retrograde accretion disks. The panels display the frequency-shift distribution for different black hole parameter sets and observer inclination angles. The four parameter sets are $a=0.5$, $q=-0.3$; $a=0.5$, $q=0.1$; $a=0.5$, $q=0.3$; and $a=0.8$, $q=0.3$. For each case, the frequency-shift distribution is shown at observer inclination angles of $17^\circ$, $75^\circ$, and $150^\circ$. In all panels, the black hole mass is fixed at $M=1$.}
\label{fig:12}
\end{figure}

\par
Here, ``first-order'' and ``second-order'' refer to the number of
intersections between a photon trajectory and the equatorial emitting
plane, rather than to a perturbative expansion of the frequency-shift factor.
The first-order map mainly represents the direct image, while the
second-order map corresponds to the lensed image. Their comparison
illustrates how stronger gravitational lensing modifies the observed
frequency-shift distribution.

\par
Figures~9--12 display the first- and second-order frequency-shift maps and show the kinematic imprint of the optically thin disk in both prograde and retrograde configurations. As the observer inclination $\theta_0$ changes from $0^\circ$ toward $90^\circ$, the redshifted area becomes smaller, while the blueshifted region expands. When $\theta_0$ increases further from $90^\circ$ to $180^\circ$, the trend is reversed: the redshifted region grows again and the blueshifted area contracts.

This trend is mainly governed by the competition between kinematic Doppler boosting and gravitational redshift. When $\theta_0$ is close to $0^\circ$ or $180^\circ$, corresponding to a face-on view, the disk rotational velocity is nearly perpendicular to the line of sight. The line-of-sight velocity component is then very small, so the observed emission is dominated by gravitational redshift and transverse Doppler redshift, and the observable region remains strongly redshifted. When $\theta_0$ approaches $90^\circ$, corresponding to an edge-on view, the projected rotational velocity becomes maximal. On the side of the disk moving toward the observer, the approaching velocity produces a strong Doppler blueshift that can overcome the gravitational redshift and enlarge the blueshifted region on the image plane. As $\theta_0$ continues toward $180^\circ$, the line-of-sight velocity component decreases, the Doppler blueshift weakens, and gravitational redshift again becomes dominant.

\par
The same figures also demonstrate how the frequency-shift pattern responds to changes in the tidal charge and spin. Increasing the spin generally moves the ISCO inward, so that radiation can be emitted from regions closer to the event horizon, where gravitational redshift is stronger. A larger tidal charge produces a similar tendency and can also enhance the gravitational redshift. Even so, the location and amplitude of the peak blueshift change only weakly with $a$ and $q$, especially when compared with the strong dependence on the observer inclination. Thus, the overall frequency-shift distribution is controlled mainly by the viewing angle, while the black hole spin and tidal charge provide secondary corrections.

\begin{table}[h]
\caption{Maximum observed frequency-shift factor of braneworld black holes with varying parameters. The row with $q=0$ corresponds to the Kerr limit.}
\label{tab1}
\centering
\setlength{\tabcolsep}{4pt}
\begin{tabular}{@{}ccrrrr@{}}
\toprule
$a$ & $q$ & \multicolumn{2}{c}{Prograde} & \multicolumn{2}{c}{Retrograde} \\
\cmidrule(lr){3-4}\cmidrule(lr){5-6}
 & & $\theta_0=75^\circ$ & $\theta_0=150^\circ$ & $\theta_0=75^\circ$ & $\theta_0=150^\circ$ \\
\midrule
0.5 & $-0.3$ & 1.53636 & 1.04895 & 1.48647 & 1.04801 \\
0.5 & 0      & 1.53347 & 1.04889 & 1.48111 & 1.04800 \\
0.5 & 0.1    & 1.53439 & 1.04886 & 1.47871 & 1.04799 \\
0.5 & 0.3    & 1.52399 & 1.04881 & 1.47662 & 1.04797 \\
0.8 & 0.3    & 1.51033 & 1.04911 & 1.46542 & 1.04766 \\
\bottomrule
\end{tabular}
\end{table}

\par
Table~\ref{tab1} lists the maximum observed frequency-shift factors for rotating braneworld black holes with different spin parameters $a$, tidal charges $q$, disk rotation directions, and observer inclination angles. The row with $q=0$ corresponds to the Kerr limit. According to the definition $g=\nu_{\rm obs}/\nu_{\rm em}$, values larger than unity indicate blueshift. The results show that the prograde disk generally reaches a larger maximum frequency-shift factor than the retrograde disk, especially for $\theta_0=75^\circ$, while the dependence on $q$ is relatively weak for $\theta_0=150^\circ$.

\begin{figure}[htbp]
\centering
\includegraphics[width=13cm]{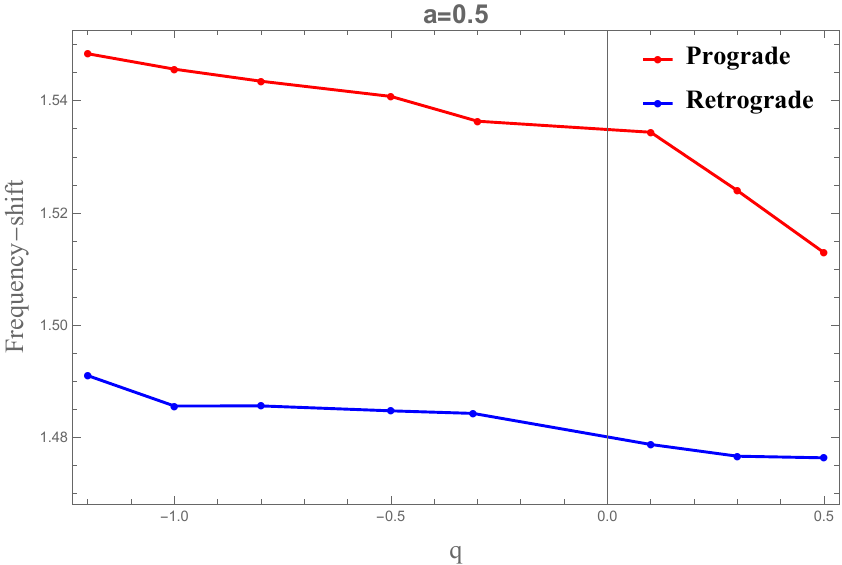}

\vspace{2mm}
\caption{Blueshift profiles for a fixed spin parameter $a=0.5$, with the tidal charge $q$ varying over the range $[-1.2,\,0.5]$. In all panels, the black hole mass is set to $M=1$.}
\label{fig:13}
\end{figure}

\begin{figure}[htbp]
\centering
\includegraphics[width=13cm]{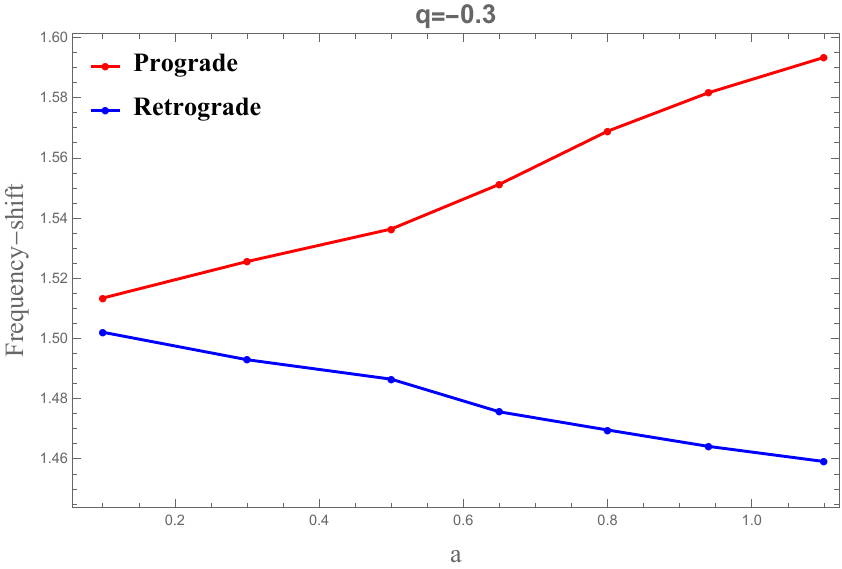}

\vspace{2mm}
\caption{Blueshift profiles for a fixed tidal charge $q=-0.3$, with the spin parameter $a$ varying over the range $[0.1,\,1.1]$. In all panels, the black hole mass is set to $M=1$.}
\label{fig:14}
\end{figure}
Figures~13 and 14 show how the frequency-shift factor varies with the tidal charge and spin parameter. At fixed spin, the maximum frequency-shift factor decreases slightly as the tidal charge increases in both the prograde and retrograde cases, although the variation is small. At fixed tidal charge, it increases with spin for prograde motion but decreases for retrograde motion. Overall, the frequency-shift factor depends only weakly on the tidal charge and spin parameter, whereas the observer inclination remains the dominant factor.

\subsection{Image of the Braneworld Black Hole within an Optically Thin Disk}
\label{sec:image-thin-disk}
We now simulate the apparent image of the braneworld black hole. Using Eq.~(\ref{56}) together with a fisheye-camera ray-tracing technique, we generate synthetic images of a braneworld black hole illuminated by an optically thin accretion disk. For an observing frequency of 230~GHz, we adopt the phenomenological parameters $A=-2$ and $B=-1/2$, which specify the radial emissivity profile as follows:

\begin{equation}
\label{63}
\log[J_{model}(r)] = -2 \, \log\!\left(\frac{r}{r_H}\right) - \frac{1}{2}\left[\log\!\left(\frac{r}{r_H}\right)\right]^2,
\end{equation}

\par
Figures~15 and 16 present the intensity profiles of the accretion disk along the x- and y-axes. The resulting trends are summarized as follows.

Along the x-axis, disk emission shows clear dependencies on the tidal charge, spin parameter, and observer inclination:

\begin{itemize}
    \item \textbf{Impact of the tidal charge:} Regardless of the disk rotation direction, the tidal charge produces a consistent effect. In both prograde and retrograde configurations, the intensity peak increases as the tidal charge increases.

    \item \textbf{Impact of the spin parameter:} In contrast to the tidal charge, the spin parameter affects the prograde and retrograde configurations in a similar manner, although the effect is more pronounced in the retrograde case. Increasing the spin parameter reduces the intensity peak, with stronger attenuation observed for retrograde accretion. Moreover, larger spin values lead to an inward contraction of the central emitting region.

    \item \textbf{Impact of the observer inclination:} The observer inclination angle produces an effect similar to that of the spin parameter. Regardless of the disk rotation direction, larger inclination angles progressively reduce the intensity peak and simultaneously drive an inward contraction of the central emitting region.
\end{itemize}

Along the y-axis, the emission characteristics differ significantly from those along the x-axis and show different dependencies on the tidal charge, spin parameter, and observer inclination angle:

\begin{itemize}
    \item \textbf{Impact of the tidal charge:} Regardless of the disk rotation direction, the peak intensity decreases steadily as the tidal charge increases. This suppressive effect appears in both prograde and retrograde configurations.

    \item \textbf{Impact of the spin parameter:} Unlike the tidal charge, the effect of the spin parameter depends on the accretion state. For fixed tidal charge and observer inclination angle, increasing the spin parameter slightly enhances the peak intensity in the prograde case. In the retrograde case, however, the peak intensity decreases sharply with increasing spin and is accompanied by a pronounced inward contraction of the central emitting region.

    \item \textbf{Impact of the observer inclination:} The observer inclination angle shows a trend opposite to that found along the x-axis. Regardless of the disk rotation direction, larger inclination angles significantly enhance the peak intensity, indicating a stronger angular dependence than that observed along the x-axis.
\end{itemize}

\begin{figure}[htbp]
\centering
\includegraphics[width=4.5cm,height=3.5cm]{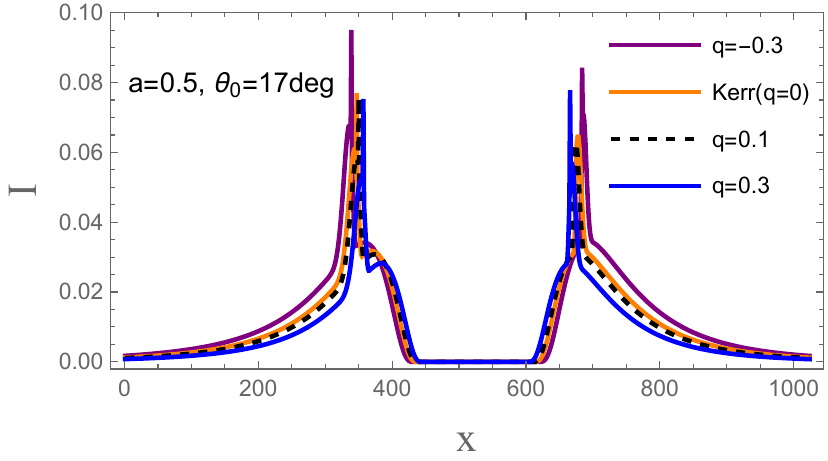}
\vspace{2mm}
\includegraphics[width=4.5cm,height=3.5cm]{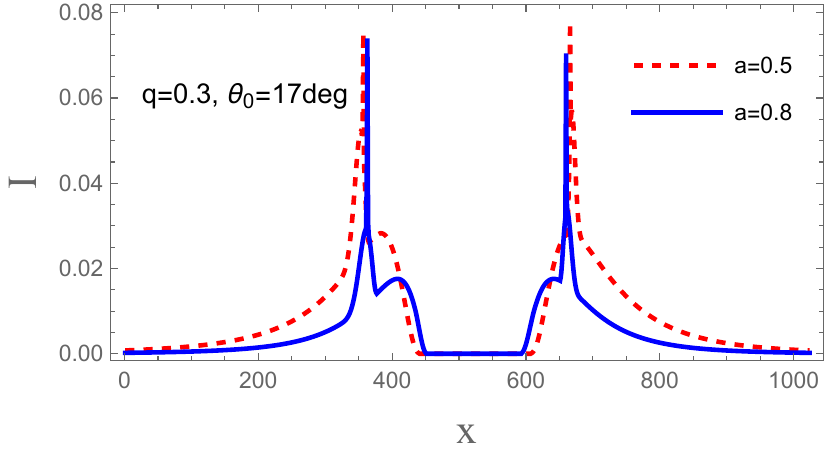}
\vspace{2mm}
\includegraphics[width=4.5cm,height=3.5cm]{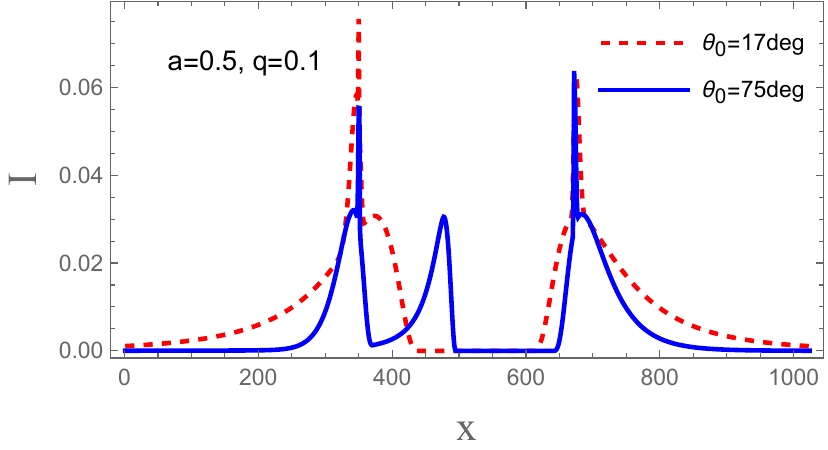}
\vspace{2mm}
\includegraphics[width=4.5cm,height=3.5cm]{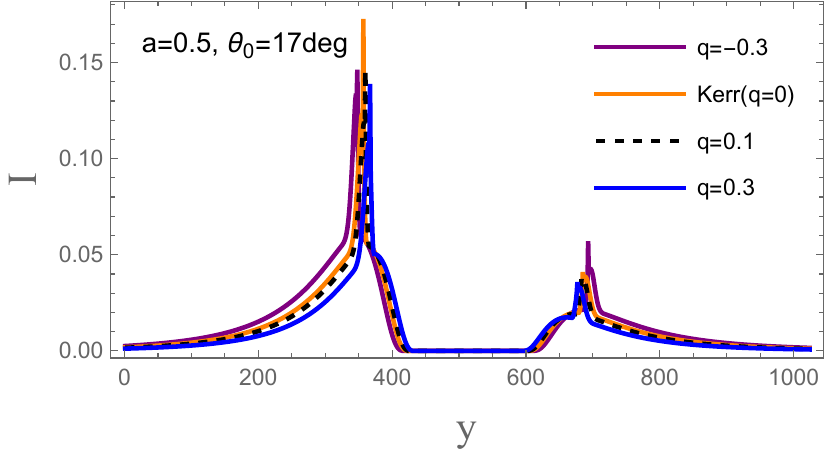}
\vspace{2mm}
\includegraphics[width=4.5cm,height=3.5cm]{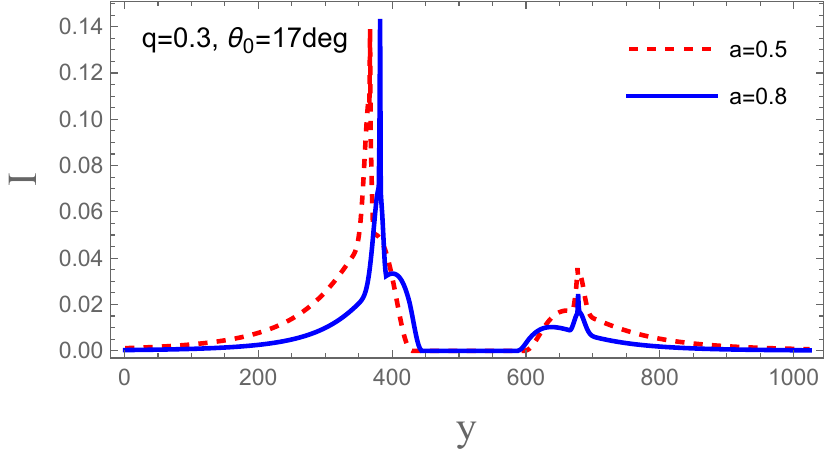}
\vspace{2mm}
\includegraphics[width=4.5cm,height=3.5cm]{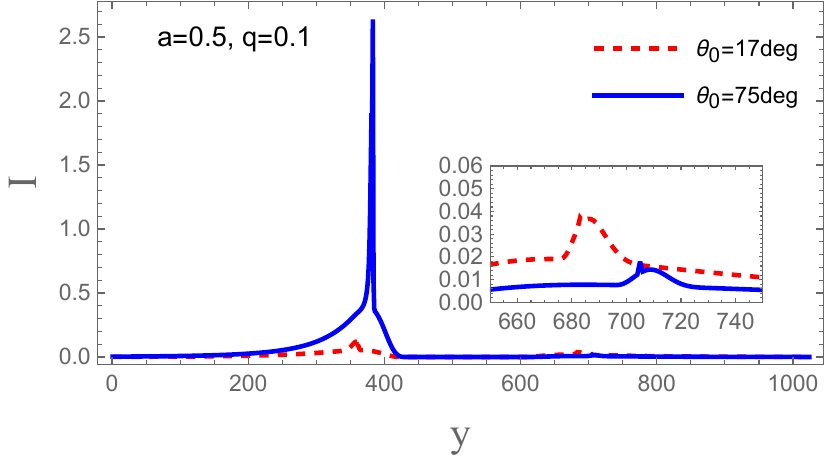}

\caption{Intensity comparison for a prograde accretion disk. The 230~GHz intensity distribution of the braneworld black hole is shown. The first row presents the x-axis slices, while the second row presents the y-axis slices.}
\label{fig:15}
\end{figure}

\begin{figure}[htbp]
\centering
\centering
\includegraphics[width=4.5cm,height=3.5cm]{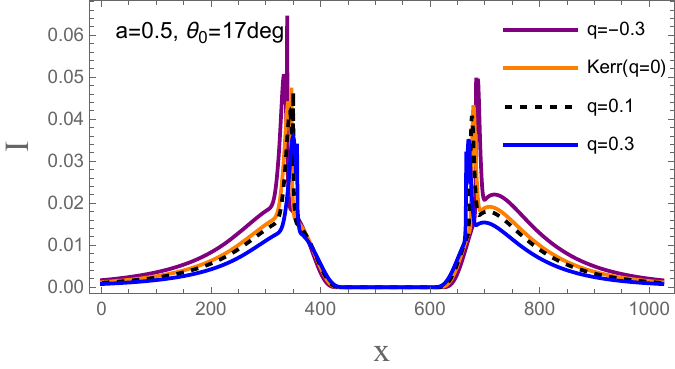}
\vspace{2mm}
\includegraphics[width=4.5cm,height=3.5cm]{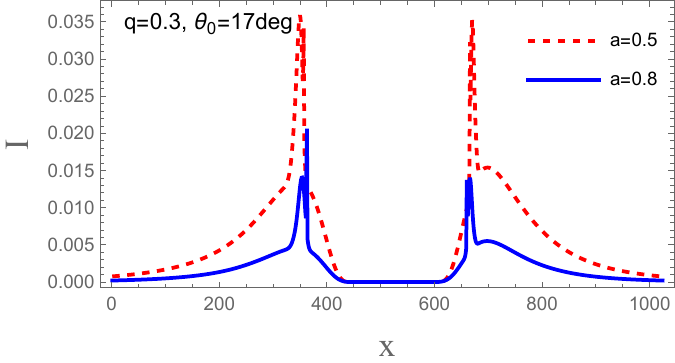}
\vspace{2mm}
\includegraphics[width=4.5cm,height=3.5cm]{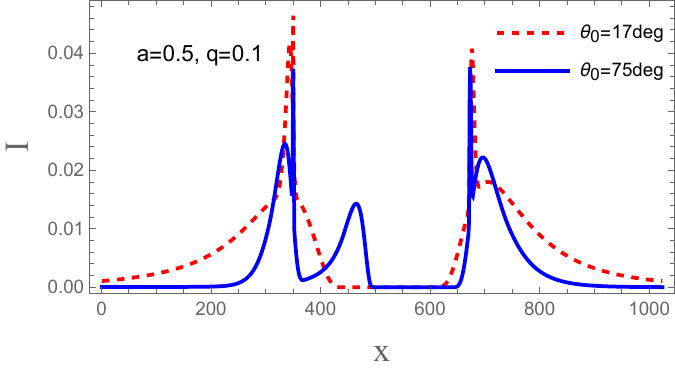}
\vspace{2mm}
\includegraphics[width=4.5cm,height=3.5cm]{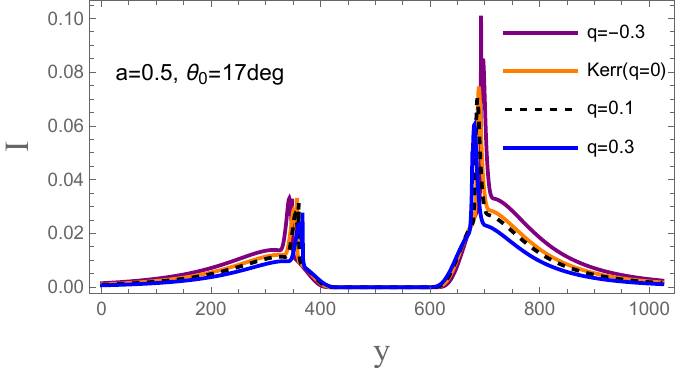}
\vspace{2mm}
\includegraphics[width=4.5cm,height=3.5cm]{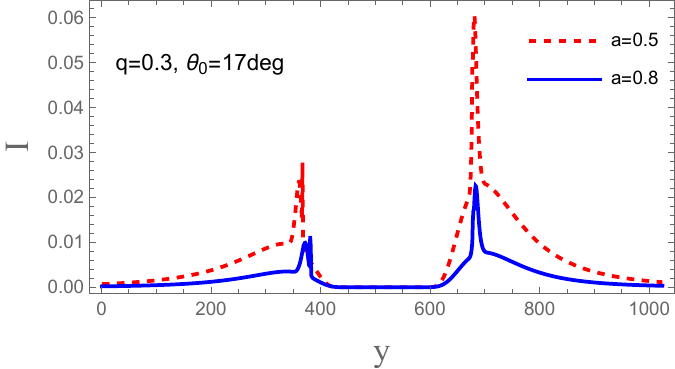}
\vspace{2mm}
\includegraphics[width=4.5cm,height=3.5cm]{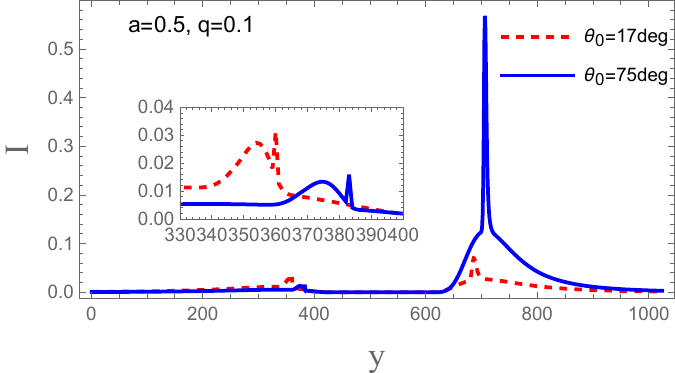}
\vspace{2mm}
\caption{Comparison of the 230~GHz intensity profiles for a retrograde accretion disk around the braneworld black hole. The first and second rows show the intensity slices along the x- and y-axes, respectively.}
\label{fig:16}
\end{figure}

\par
Figures~17 and 18 illustrate the apparent image of a braneworld black hole surrounded by an optically thin accretion disk in both prograde and retrograde accretion scenarios. Regardless of the disk rotation direction, for observer inclinations $\theta < \pi/2$, the morphological distinction between the direct image and the lensed image becomes increasingly pronounced as the inclination angle increases. At an observer inclination of $\theta = 150^\circ$, however, the overall intensity is reduced, which obscures the boundary and makes it substantially more difficult to distinguish the direct emission from the higher-order lensing rings.

Both the tidal charge and the spin parameter affect the shadow of the braneworld black hole, although the spin parameter plays the dominant role. Nevertheless, the inner shadow and the critical curve remain detectable over a wide range of inclination angles, despite variations in the black hole spin and tidal charge, highlighting their fundamental nature in braneworld black hole spacetimes. The limited vertical extent of the accretion disk also leads to a reduction in the central brightness in the simulated images. In addition, with sufficient angular resolution, the Doppler asymmetry induced by prograde and retrograde disk motion can be clearly identified on the image plane. For retrograde accretion, the Doppler signature is significantly stronger on the right-hand side of the image than on the opposite side. At the same time, frame dragging caused by the black hole spin produces an observable suppression of brightness in braneworld black hole images.

\begin{figure}[htbp]
\centering
\includegraphics[width=4.5cm,height=4.5cm]{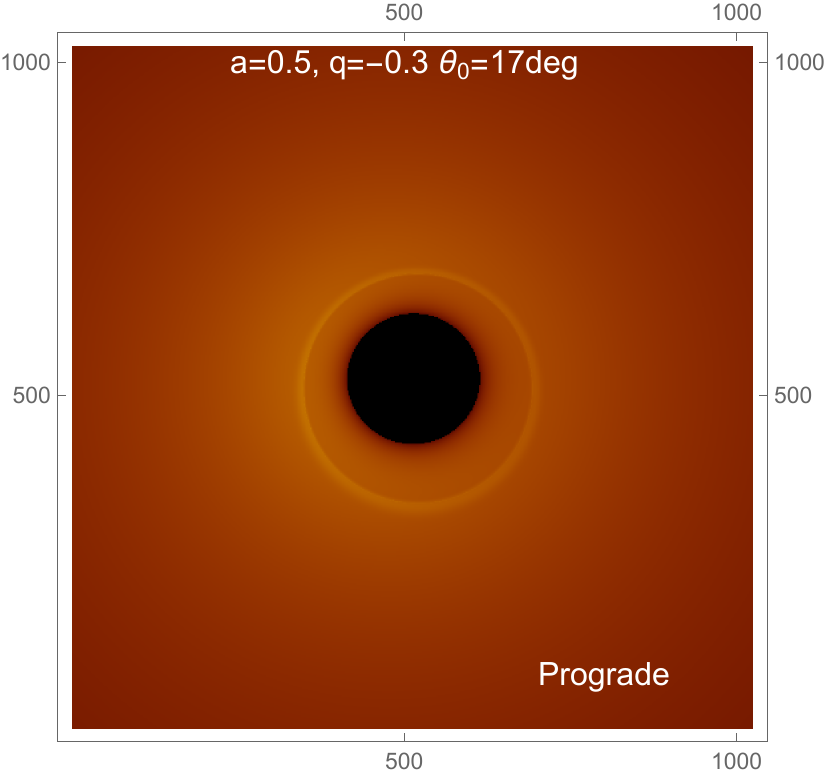}
\vspace{2mm}
\includegraphics[width=4.5cm,height=4.5cm]{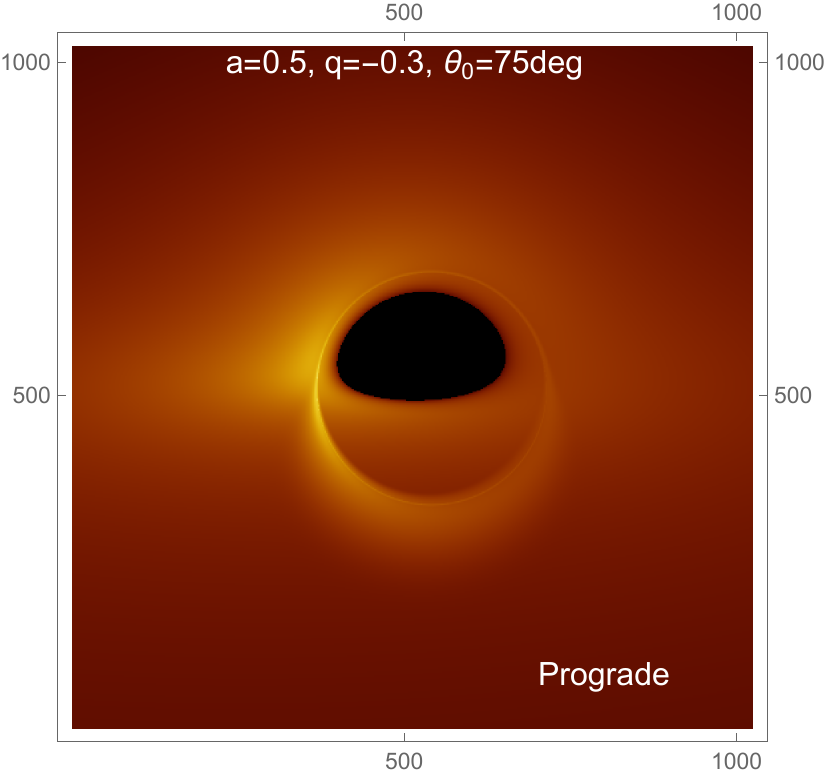}
\vspace{2mm}
\includegraphics[width=4.5cm,height=4.5cm]{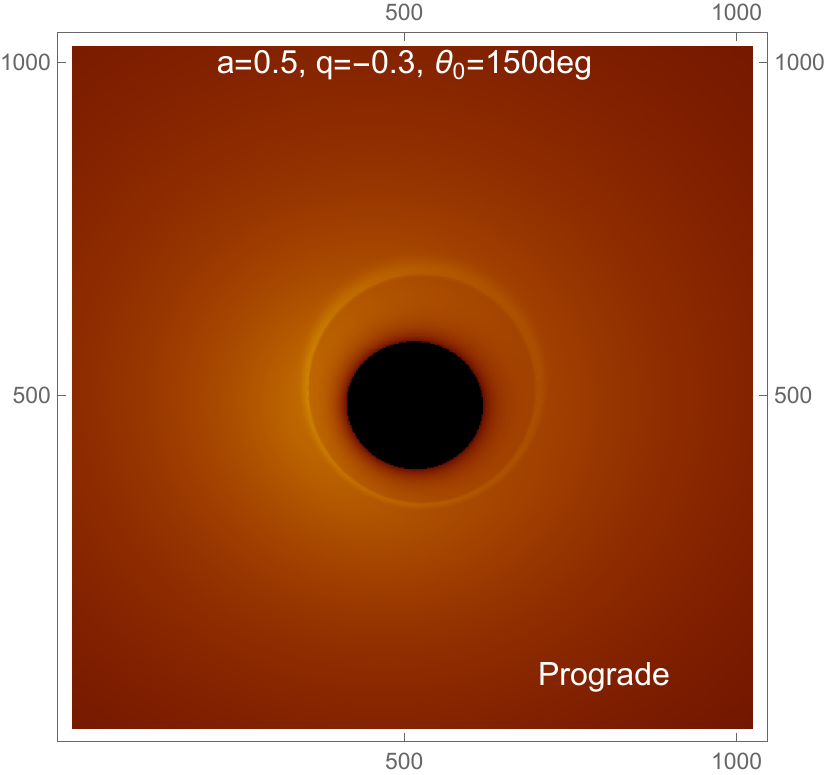}
\vspace{2mm}
\includegraphics[width=.5cm,height=4cm]{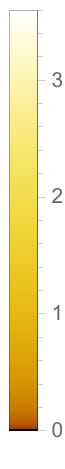}
\vspace{2mm}
\includegraphics[width=4.5cm,height=4.5cm]{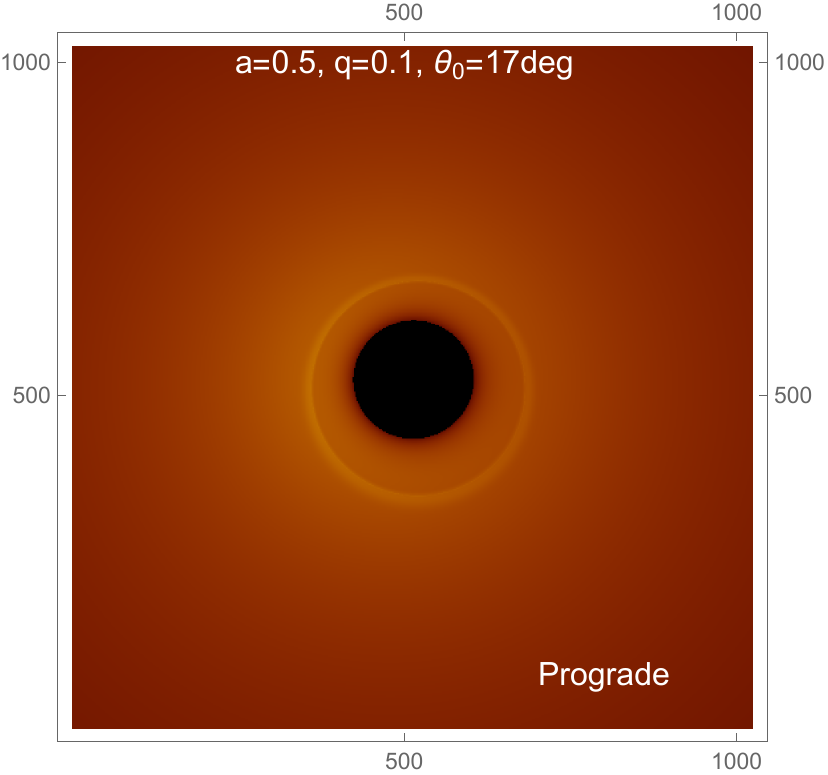}
\vspace{2mm}
\includegraphics[width=4.5cm,height=4.5cm]{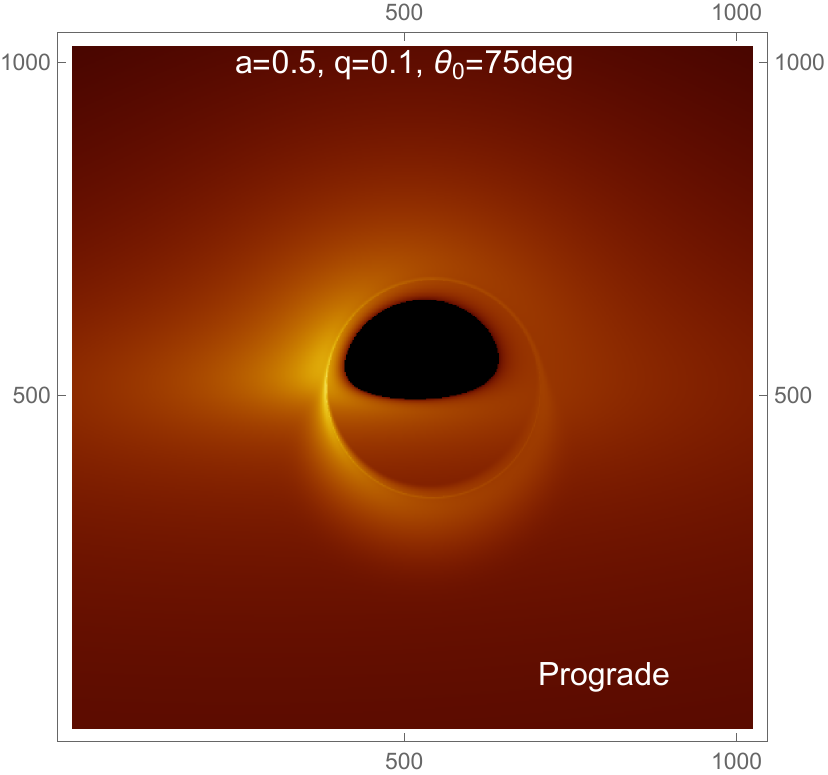}
\vspace{2mm}
\includegraphics[width=4.5cm,height=4.5cm]{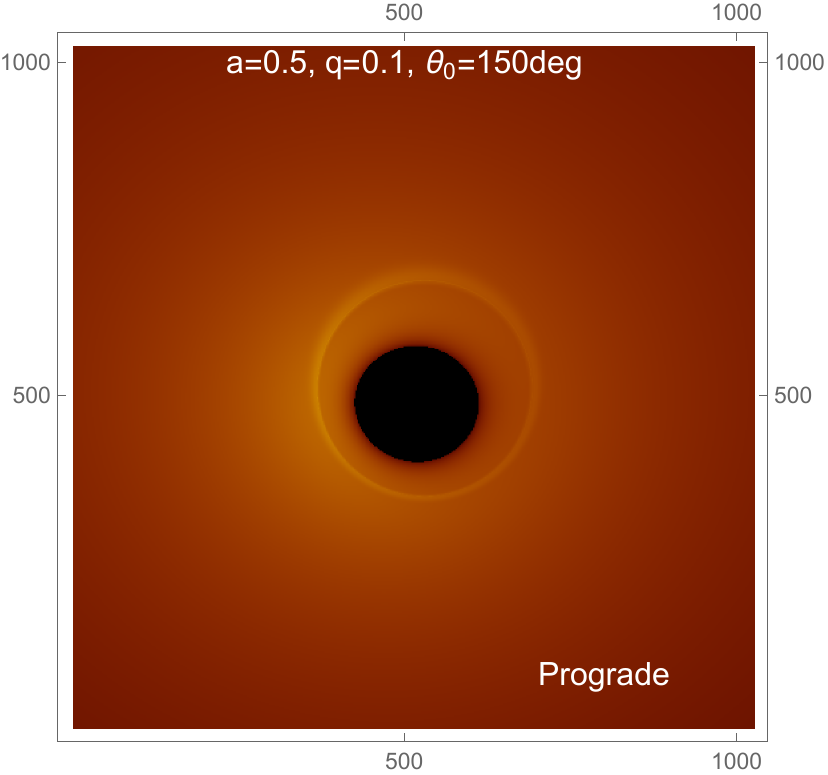}
\vspace{2mm}
\includegraphics[width=.5cm,height=4cm]{gds.pdf}
\vspace{2mm}
\includegraphics[width=4.5cm,height=4.5cm]{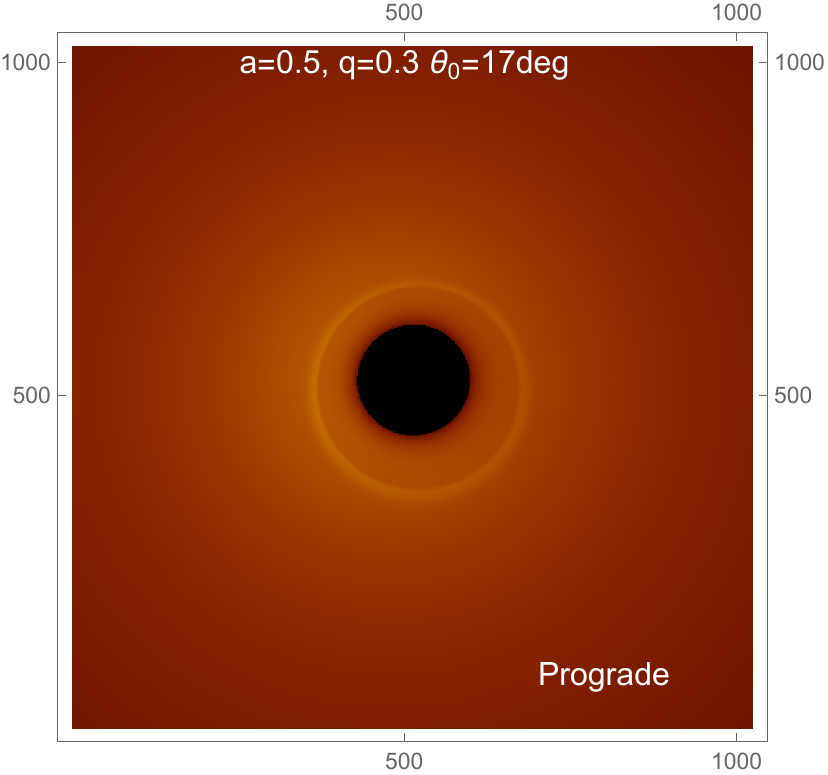}
\vspace{2mm}
\includegraphics[width=4.5cm,height=4.5cm]{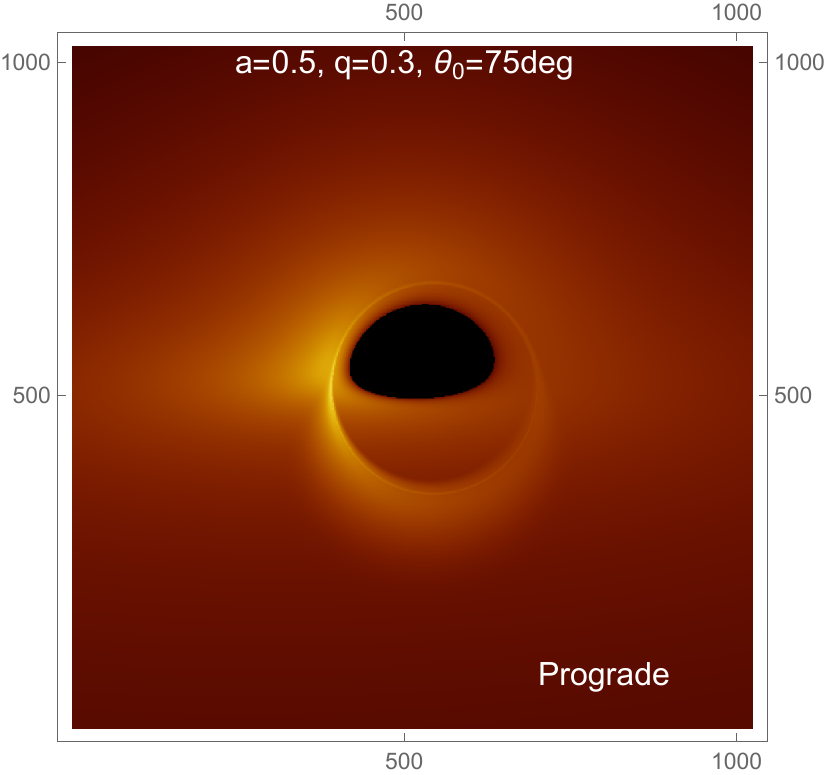}
\vspace{2mm}
\includegraphics[width=4.5cm,height=4.5cm]{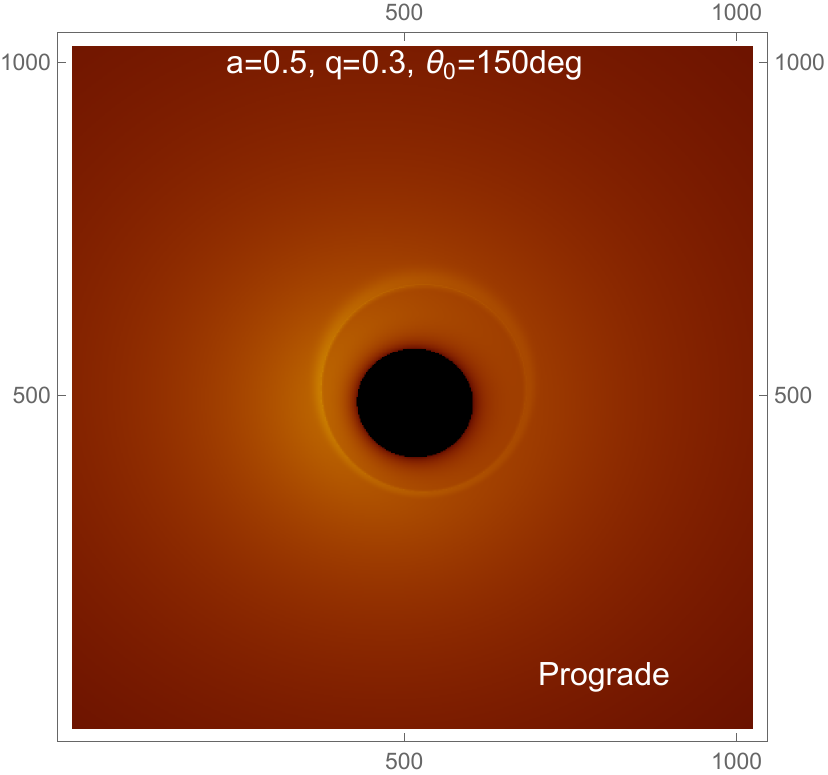}
\vspace{2mm}
\includegraphics[width=.5cm,height=4cm]{gds.pdf}
\vspace{2mm}
\includegraphics[width=4.5cm,height=4.5cm]{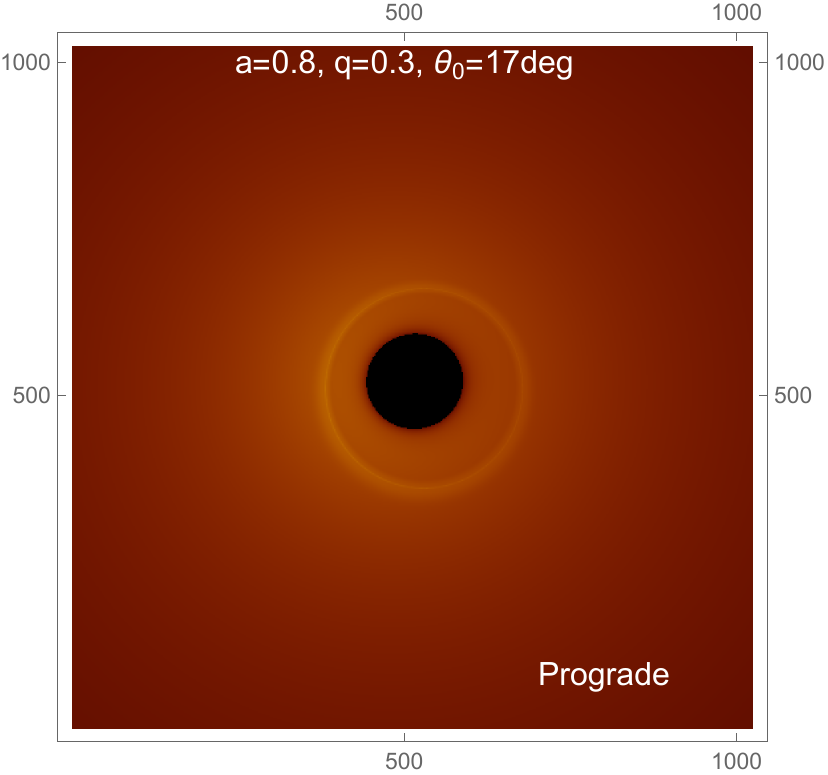}
\vspace{2mm}
\includegraphics[width=4.5cm,height=4.5cm]{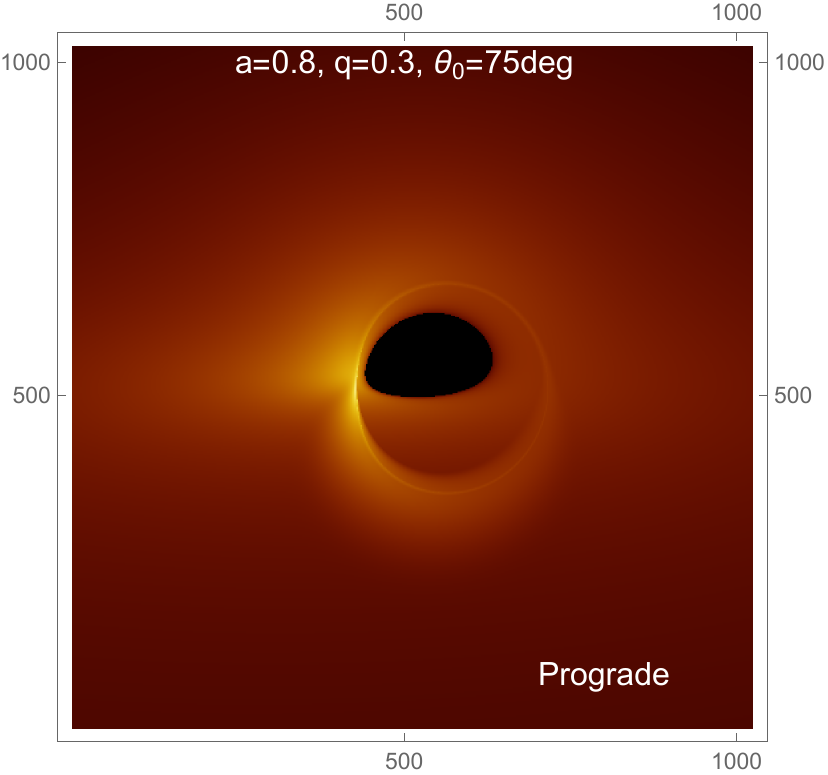}
\vspace{2mm}
\includegraphics[width=4.5cm,height=4.5cm]{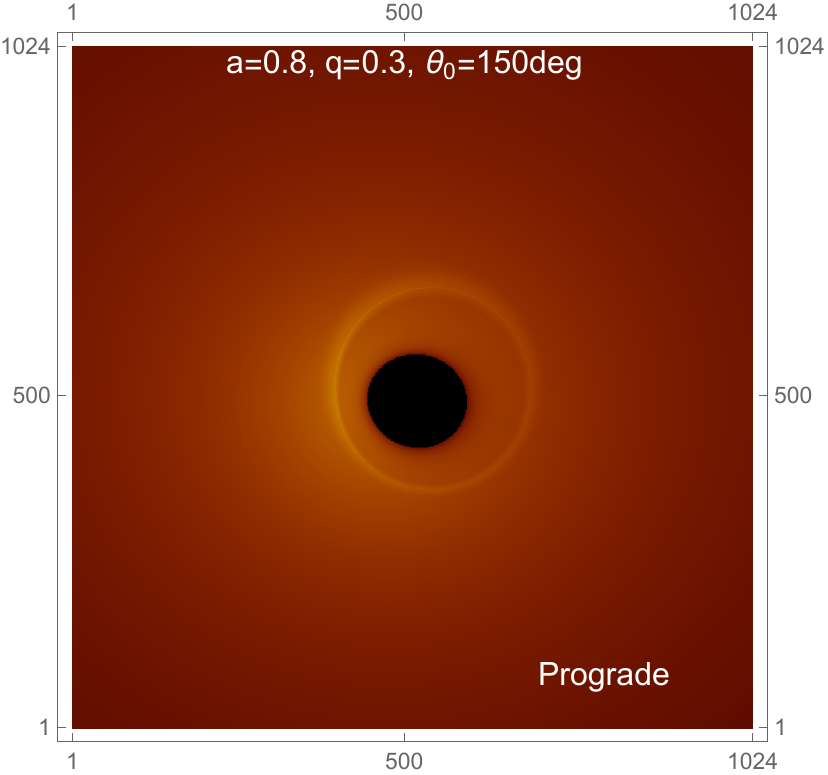}
\vspace{2mm}
\includegraphics[width=.5cm,height=4cm]{gds.pdf}
\vspace{2mm}

\caption{Intensity maps of the accretion disk in the prograde state at 230~GHz. The panels show the 230~GHz intensity distribution of the accretion disk for different black hole parameter sets and observer inclination angles. The four parameter sets considered are $a=0.5$, $q=-0.3$; $a=0.5$, $q=0.1$; $a=0.5$, $q=0.3$; and $a=0.8$, $q=0.3$. For each case, the intensity distribution is shown at observer inclination angles of $17^\circ$, $75^\circ$, and $150^\circ$. In all panels, the black hole mass is fixed at $M=1$.}
\label{fig:17}
\end{figure}

\begin{figure}[htbp]
\centering
\includegraphics[width=4.5cm,height=4.5cm]{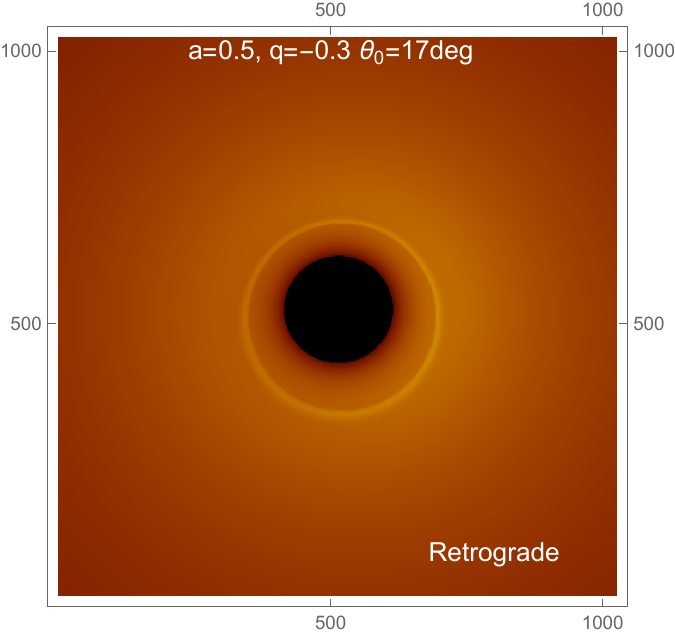}
\vspace{2mm}
\includegraphics[width=4.5cm,height=4.5cm]{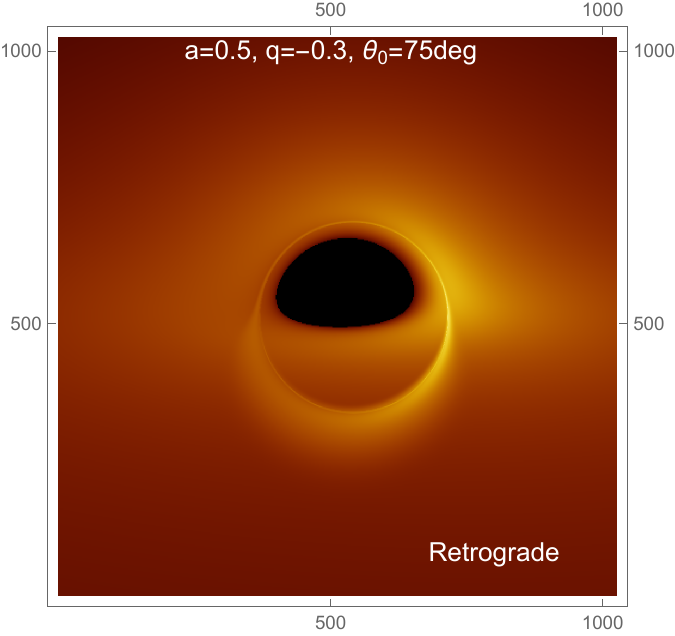}
\vspace{2mm}
\includegraphics[width=4.5cm,height=4.5cm]{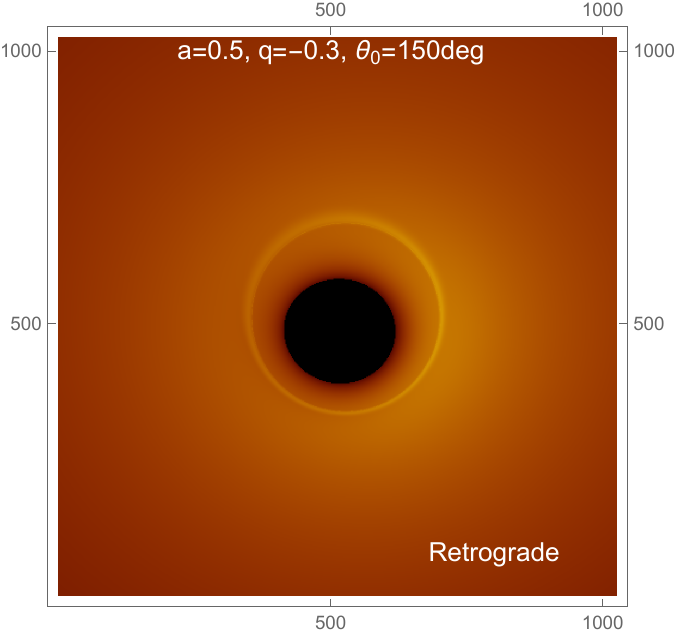}
\vspace{2mm}
\includegraphics[width=.5cm,height=4cm]{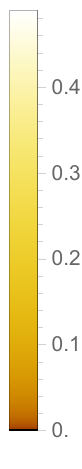}
\vspace{2mm}
\includegraphics[width=4.5cm,height=4.5cm]{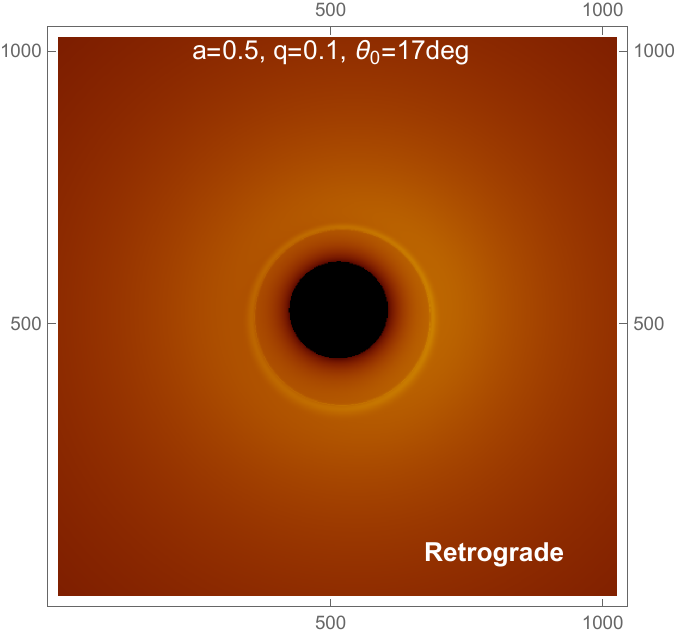}
\vspace{2mm}
\includegraphics[width=4.5cm,height=4.5cm]{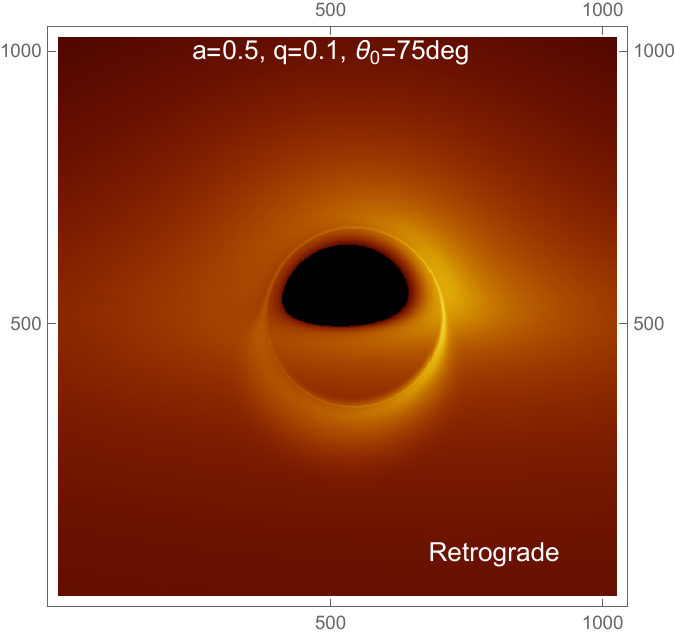}
\vspace{2mm}
\includegraphics[width=4.5cm,height=4.5cm]{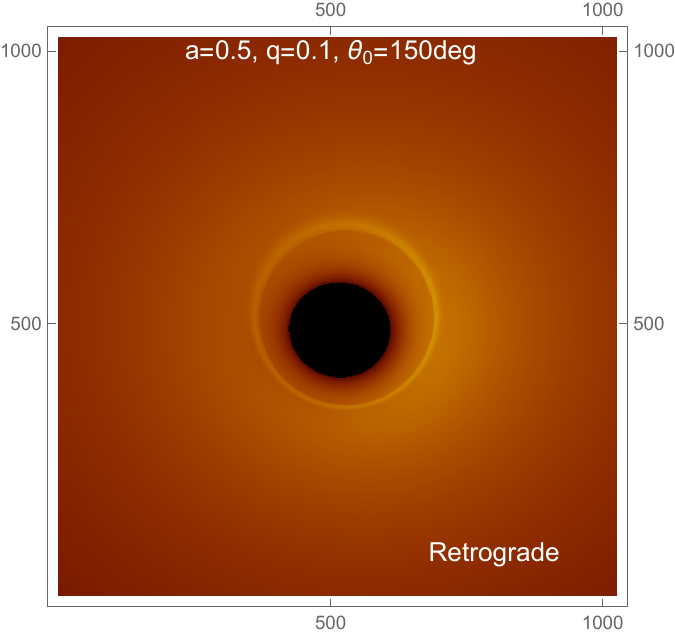}
\vspace{2mm}
\includegraphics[width=.5cm,height=4cm]{gdn.pdf}
\vspace{2mm}
\includegraphics[width=4.5cm,height=4.5cm]{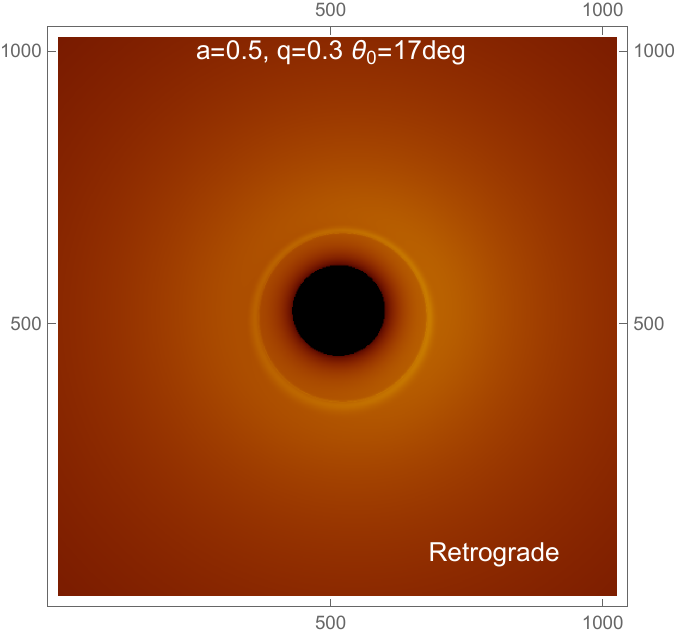}
\vspace{2mm}
\includegraphics[width=4.5cm,height=4.5cm]{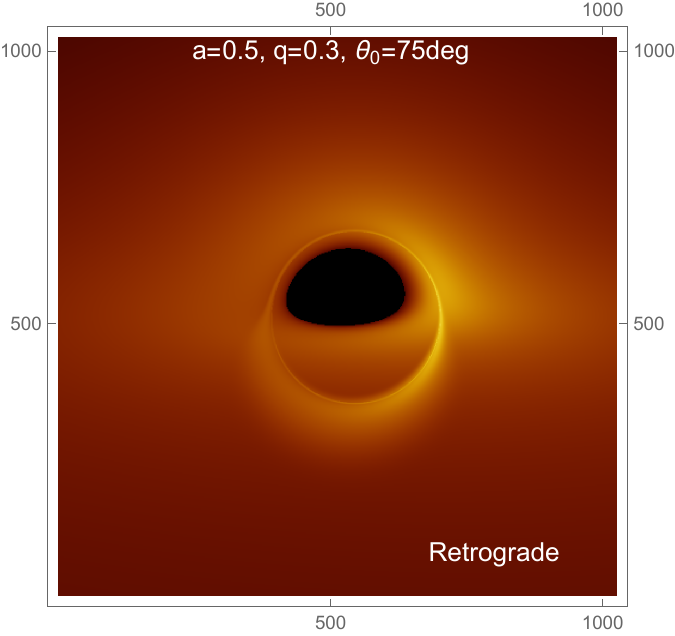}
\vspace{2mm}
\includegraphics[width=4.5cm,height=4.5cm]{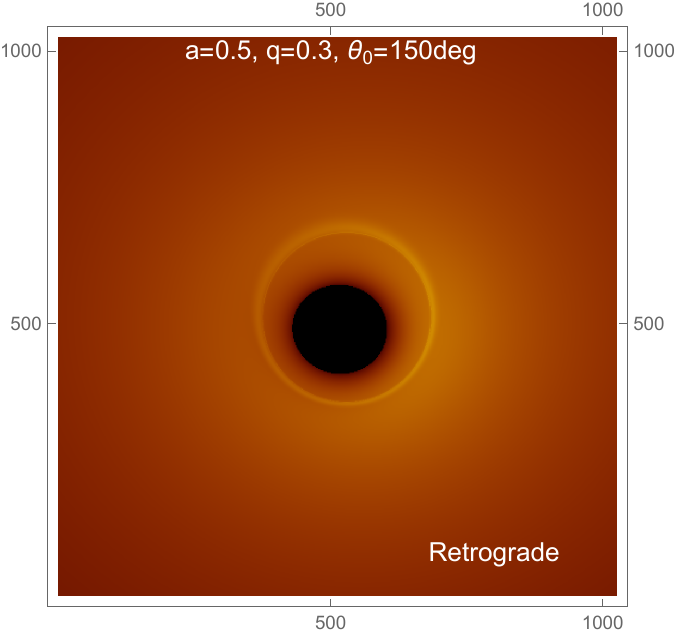}
\vspace{2mm}
\includegraphics[width=.5cm,height=4cm]{gdn.pdf}
\vspace{2mm}
\includegraphics[width=4.5cm,height=4.5cm]{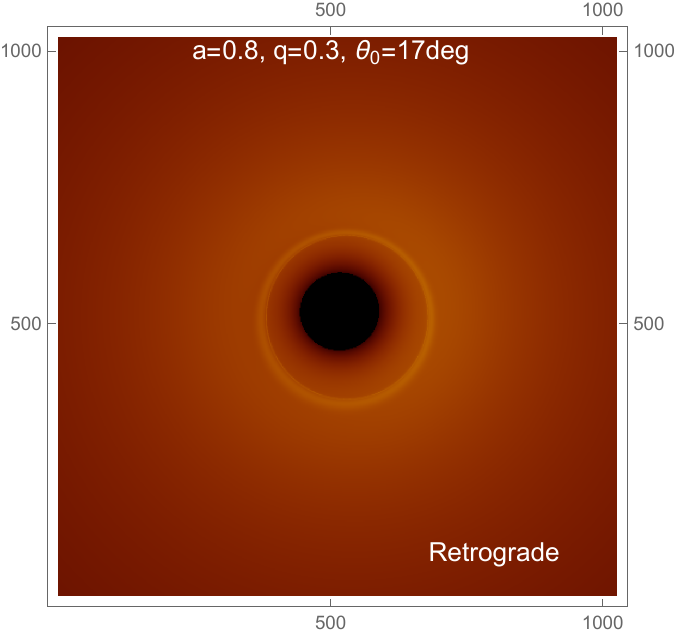}
\vspace{2mm}
\includegraphics[width=4.5cm,height=4.5cm]{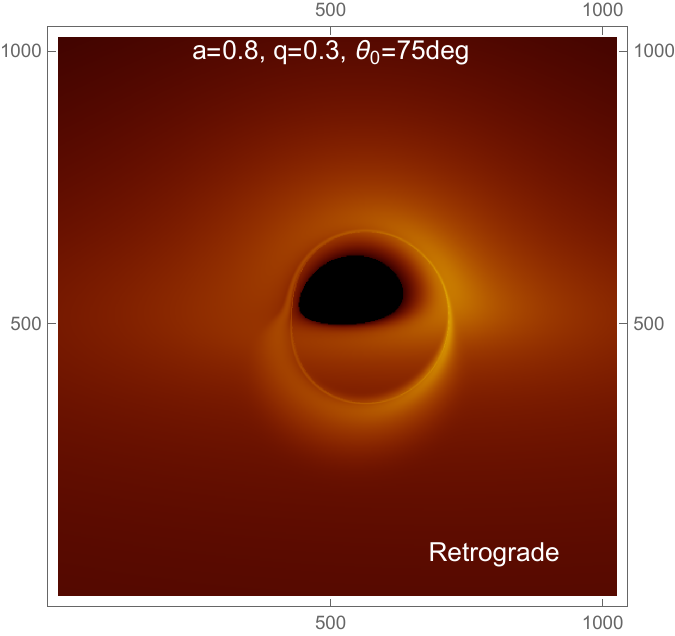}
\vspace{2mm}
\includegraphics[width=4.5cm,height=4.5cm]{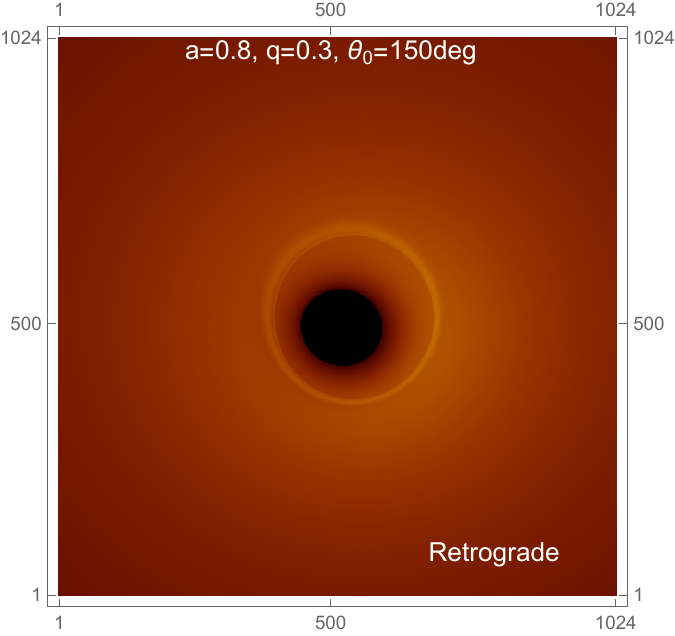}
\vspace{2mm}
\includegraphics[width=.5cm,height=4cm]{gdn.pdf}
\vspace{2mm}

\caption{Intensity maps of the accretion disk in the retrograde state at 230~GHz. The panels show the 230~GHz intensity distribution of the accretion disk for different black hole parameter sets and observer inclination angles. The four parameter sets considered are $a=0.5$, $q=-0.3$; $a=0.5$, $q=0.1$; $a=0.5$, $q=0.3$; and $a=0.8$, $q=0.3$. For each case, the intensity distribution is shown at observer inclination angles of $17^\circ$, $75^\circ$, and $150^\circ$. In all panels, the black hole mass is fixed at $M=1$.}
\label{fig:18}
\end{figure}

\begin{figure}[htbp]
\centering
\includegraphics[width=13cm]{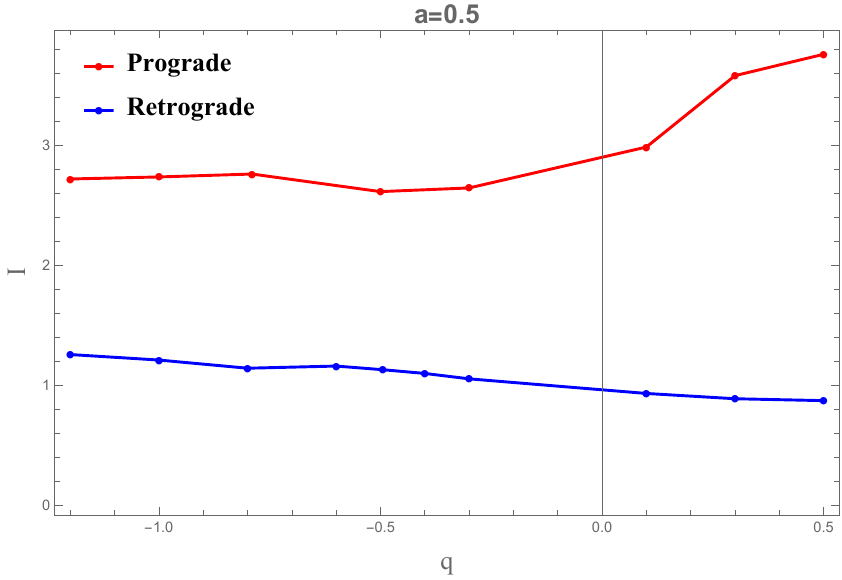}

\vspace{2mm}
\caption{Intensity profiles for a fixed spin parameter $a=0.5$, with the tidal charge $q$ varying over the range $[-1.2,\,0.5]$. In all panels, the black hole mass is fixed at $M=1$.}
\label{fig:19}
\end{figure}

\begin{figure}[htbp]
\centering
\includegraphics[width=13cm]{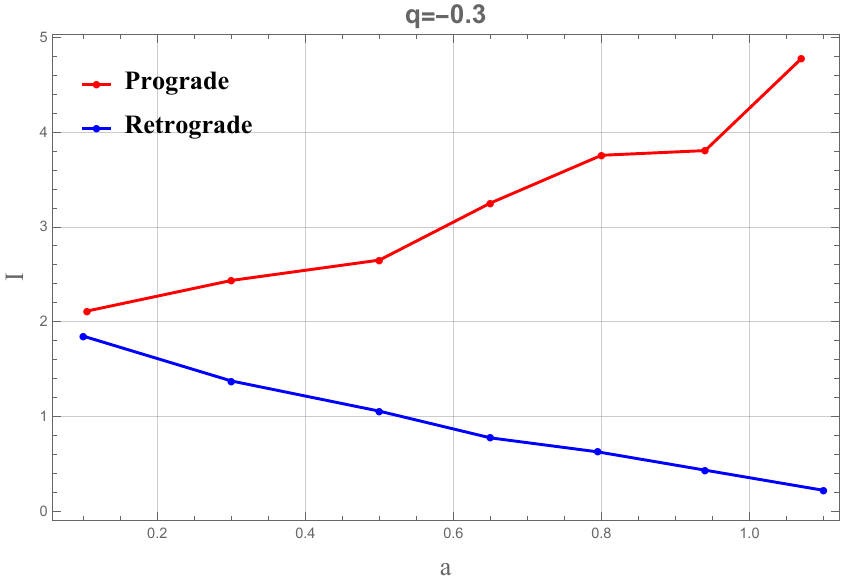}

\vspace{2mm}
\caption{Intensity profiles for a fixed tidal charge $q=-0.3$, with the spin parameter $a$ varying over the range $[0.1,\,1.1]$. In all panels, the black hole mass is fixed at $M=1$.}
\label{fig:20}
\end{figure}

Figures~19 and 20 show how the intensity profiles depend on the tidal charge and spin. At fixed spin, the intensity increases steadily with tidal charge in the prograde configuration but decreases in the retrograde configuration. At fixed tidal charge, larger spin enhances the intensity for prograde accretion and suppresses it for retrograde accretion. Overall, the intensity variation is stronger in the prograde case than in the retrograde case, whether the spin or the tidal charge is varied. Studying negative tidal charges in braneworld spacetime therefore helps characterize how the tidal charge affects the frequency-shift behavior and radiation profiles in the present model.

\begin{table}[h]
\caption{Maximum observed intensity of braneworld black holes with varying parameters. The row with $q=0$ corresponds to the Kerr limit.}
\label{tab2}
\centering
\setlength{\tabcolsep}{4pt}
\begin{tabular}{@{}ccrrrr@{}}
\toprule
$a$ & $q$ & \multicolumn{2}{c}{Prograde} & \multicolumn{2}{c}{Retrograde} \\
\cmidrule(lr){3-4}\cmidrule(lr){5-6}
 & & $\theta_0=75^\circ$ & $\theta_0=150^\circ$ & $\theta_0=75^\circ$ & $\theta_0=150^\circ$ \\
\midrule
0.5 & $-0.3$ & 2.64944 & 0.498447 & 1.05467 & 0.207229 \\
0.5 & 0      & 2.80005 & 0.349915 & 1.28694 & 0.167148 \\
0.5 & 0.1    & 2.98830 & 0.402825 & 0.93276 & 0.152894 \\
0.5 & 0.3    & 3.58403 & 0.376889 & 0.88865 & 0.135587 \\
0.8 & 0.3    & 3.38708 & 0.34138  & 0.26267 & 0.0527602 \\
\bottomrule
\end{tabular}
\end{table}

\par
Table~\ref{tab2} lists the maximum observed intensities for rotating braneworld black holes with different spin parameters $a$, tidal charges $q$, disk rotation directions, and observer inclination angles. The row with $q=0$ corresponds to the Kerr limit. The results show that the prograde disk generally reaches a larger maximum observed intensity than the retrograde disk, especially for $\theta_0=75^\circ$. The maximum intensity also varies with the tidal charge $q$, indicating that the tidal charge can affect the peak specific intensity on the observer's image plane.

\par
As a counterpart to the 230~GHz results, we also present 86~GHz images of the braneworld black hole with an optically thin accretion disk. For this frequency, we use $A=0$ and $B=-\frac{3}{4}$, so that the emissivity profile is~\cite{37}:

\begin{equation}
\label{64}
\log[J_{model}(r)] = -\frac{3}{4}\left[\log\!\left(\frac{r}{r_H}\right)\right]^2.
\end{equation}
Figures~21 and 22 show the 86~GHz intensity profiles of the accretion disk along the x- and y-axes. The main trends are as follows.

Along the x-axis, the emission characteristics of the accretion disk exhibit distinct dependencies on the spin parameter, tidal charge, and observer inclination angle:

\begin{itemize}
    \item \textbf{Impact of the spin parameter:} The disk behavior differs significantly between the two accretion configurations. For prograde accretion, the peak intensity increases with increasing spin, whereas in the retrograde case the peak intensity decreases as the spin increases. Moreover, a larger spin parameter leads to an inward contraction of the central high-intensity region.

    \item \textbf{Impact of the tidal charge:} As the tidal charge increases, the peak intensity gradually decreases in both prograde and retrograde configurations. This attenuation is more pronounced in the retrograde case.

    \item \textbf{Impact of the observer inclination:} The observer inclination angle produces a trend similar to that associated with the tidal charge. Regardless of the disk rotation direction, the peak intensity decreases steadily with increasing observer inclination, while the central emitting region contracts inward.
\end{itemize}

Along the y-axis, the emission characteristics differ markedly from those along the x-axis and exhibit distinct dependencies on the same physical parameters:

\begin{itemize}
    \item \textbf{Impact of the spin parameter:} When the tidal charge and inclination angle are fixed, the effect of the spin parameter depends on the accretion configuration. In the prograde case, the peak intensity decreases with increasing spin. In the retrograde case, this decrease is much stronger and is accompanied by a stronger inward contraction of the central emitting region.

    \item \textbf{Impact of the tidal charge:} Regardless of the disk rotation direction, the peak intensity decreases steadily as the tidal charge increases.

    \item \textbf{Impact of the observer inclination:} The observer inclination angle shows a trend opposite to that found along the x-axis. For both prograde and retrograde configurations, the peak intensity increases significantly with increasing inclination angle, indicating a stronger angular dependence than that observed along the x-axis.
\end{itemize}

Comparing the 86~GHz and 230~GHz synthetic images reveals clear emission anisotropy between the x- and y-axis profiles. Nevertheless, the overall behavior remains consistent for both prograde and retrograde disks, and the 86~GHz images show a much higher total intensity. Across the image plane, the radiation intensity is strongly shaped by gravitational redshift and Doppler modulation. The central brightness depression associated with the inner shadow remains visible, and the critical curve is also preserved. These features suggest that, within the adopted disk model, the inner-shadow-like structure remains visible at both observing frequencies. However, this feature is not unique to braneworld black holes and should be interpreted together with the disk emissivity, observer inclination, and frequency-dependent radiative effects.

\begin{figure}[htbp]
\centering
\includegraphics[width=4.5cm,height=3.5cm]{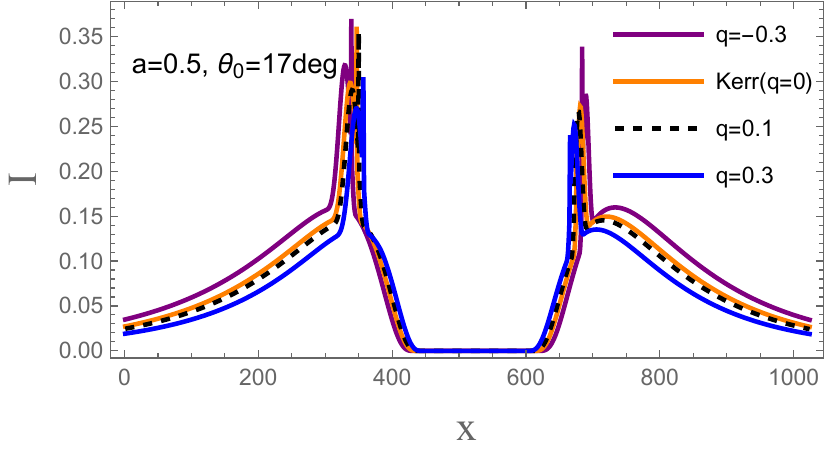}
\vspace{2mm}
\includegraphics[width=4.5cm,height=3.5cm]{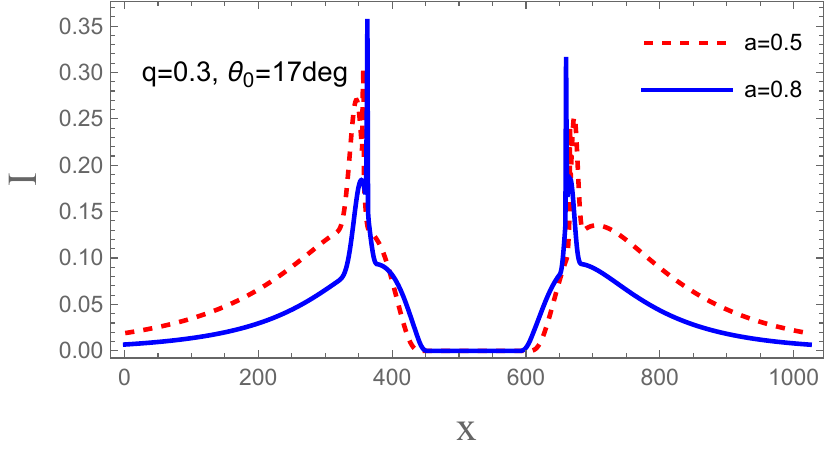}
\vspace{2mm}
\includegraphics[width=4.5cm,height=3.5cm]{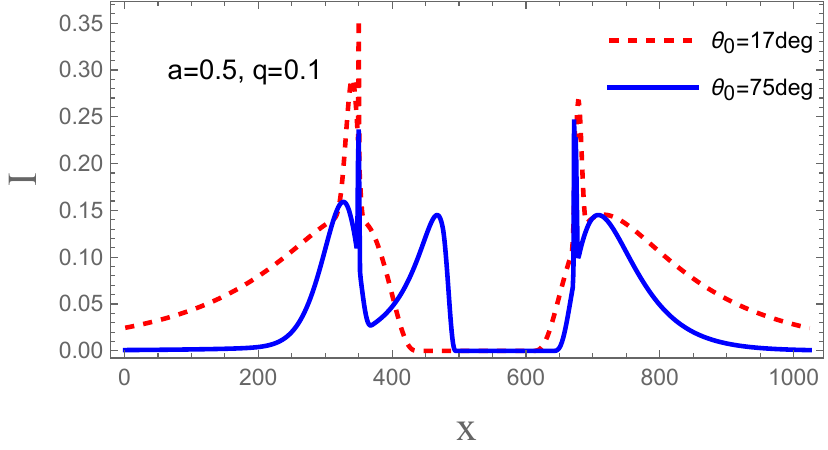}
\vspace{2mm}
\includegraphics[width=4.5cm,height=3.5cm]{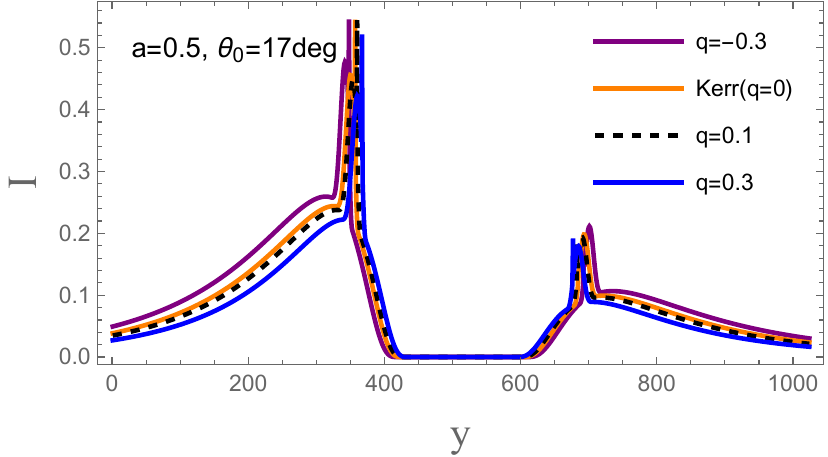}
\vspace{2mm}
\includegraphics[width=4.5cm,height=3.5cm]{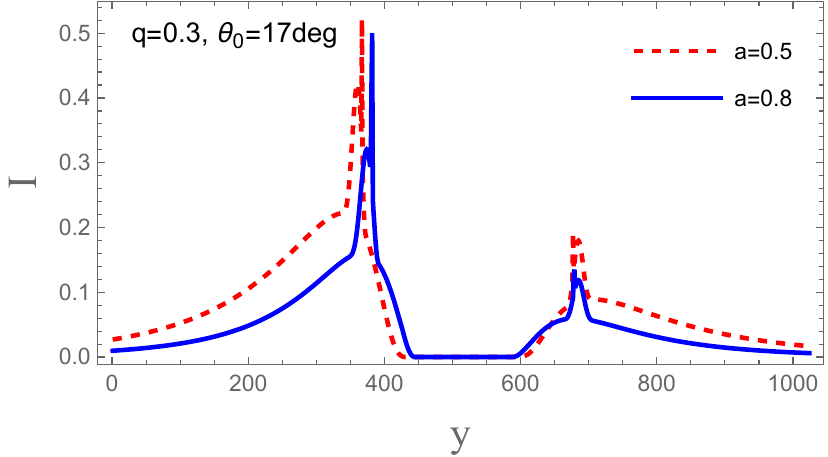}
\vspace{2mm}
\includegraphics[width=4.5cm,height=3.5cm]{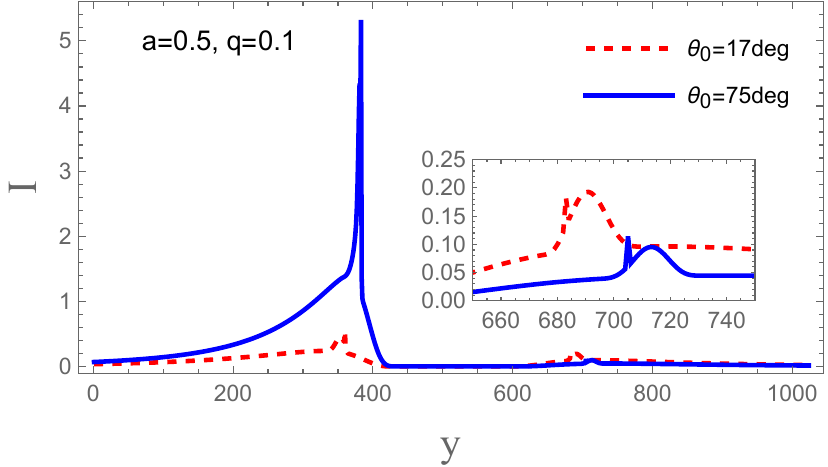}
\caption{Comparison of the 86~GHz intensity profiles for prograde accretion around the braneworld black hole. The first and second rows show the intensity cuts along the x- and y-axes, respectively.}
\label{fig:21}
\end{figure}

\begin{figure}[htbp]
\centering
\includegraphics[width=4.5cm,height=3.5cm]{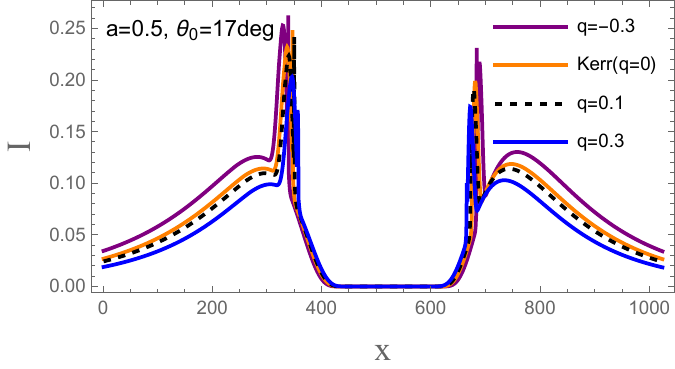}
\vspace{2mm}
\includegraphics[width=4.5cm,height=3.5cm]{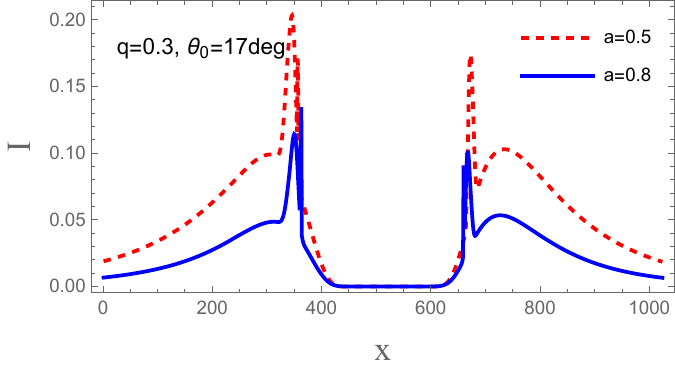}
\vspace{2mm}
\includegraphics[width=4.5cm,height=3.5cm]{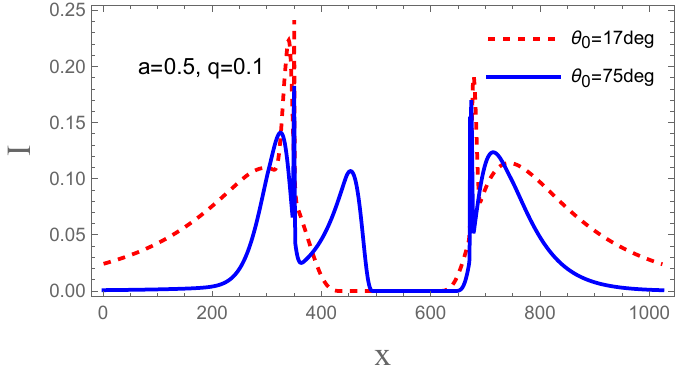}
\vspace{2mm}
\includegraphics[width=4.5cm,height=3.5cm]{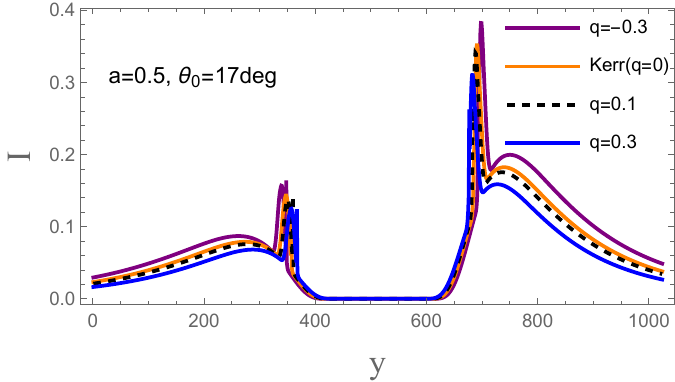}
\vspace{2mm}
\includegraphics[width=4.5cm,height=3.5cm]{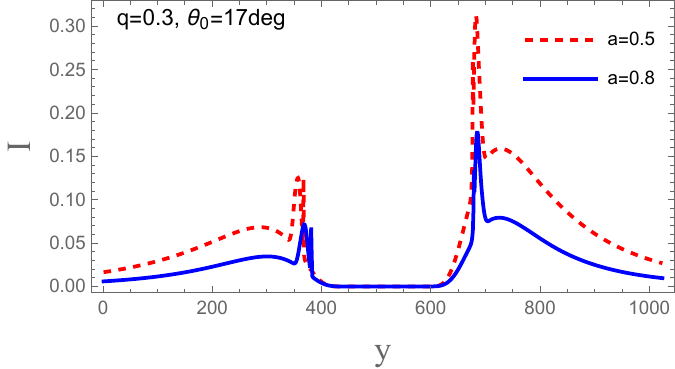}
\vspace{2mm}
\includegraphics[width=4.5cm,height=3.5cm]{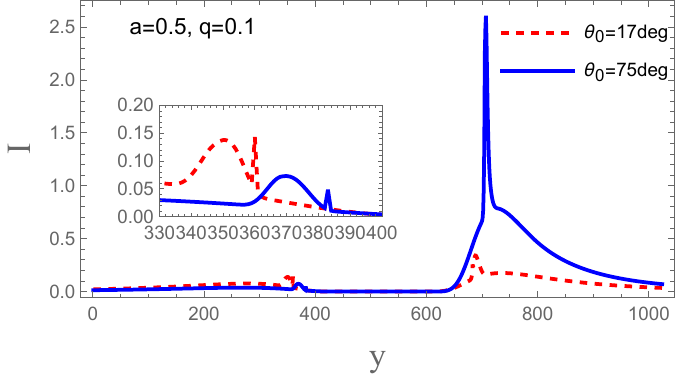}

\caption{Comparison of the 86~GHz intensity profiles for retrograde accretion around the braneworld black hole. The first and second rows show the intensity cuts along the x- and y-axes, respectively.}
\label{fig:22}
\end{figure}

\begin{figure}[htbp]
\centering
\includegraphics[width=4.5cm,height=4.5cm]{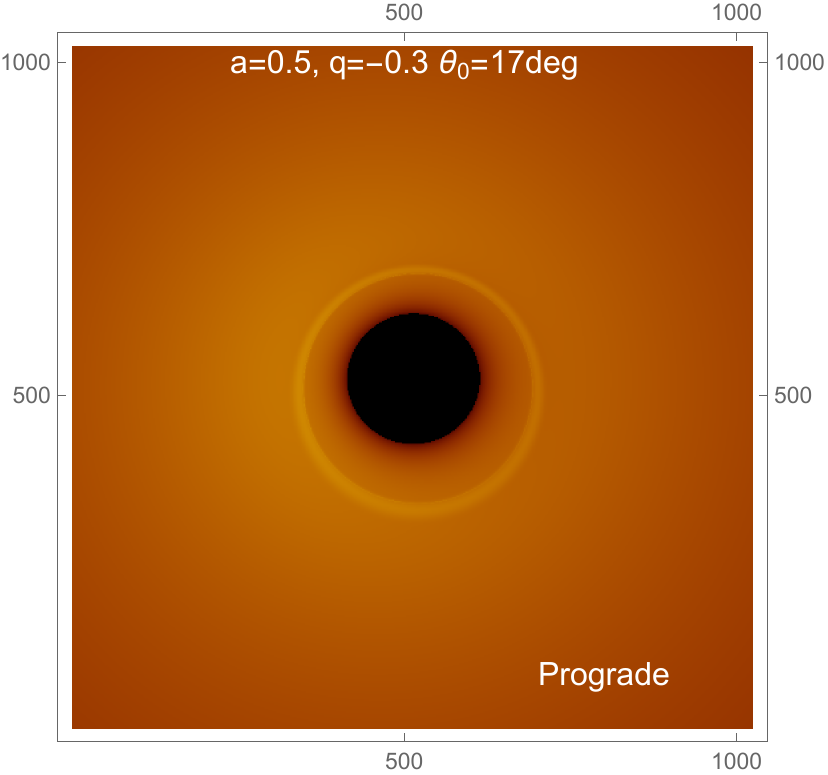}
\vspace{2mm}
\includegraphics[width=4.5cm,height=4.5cm]{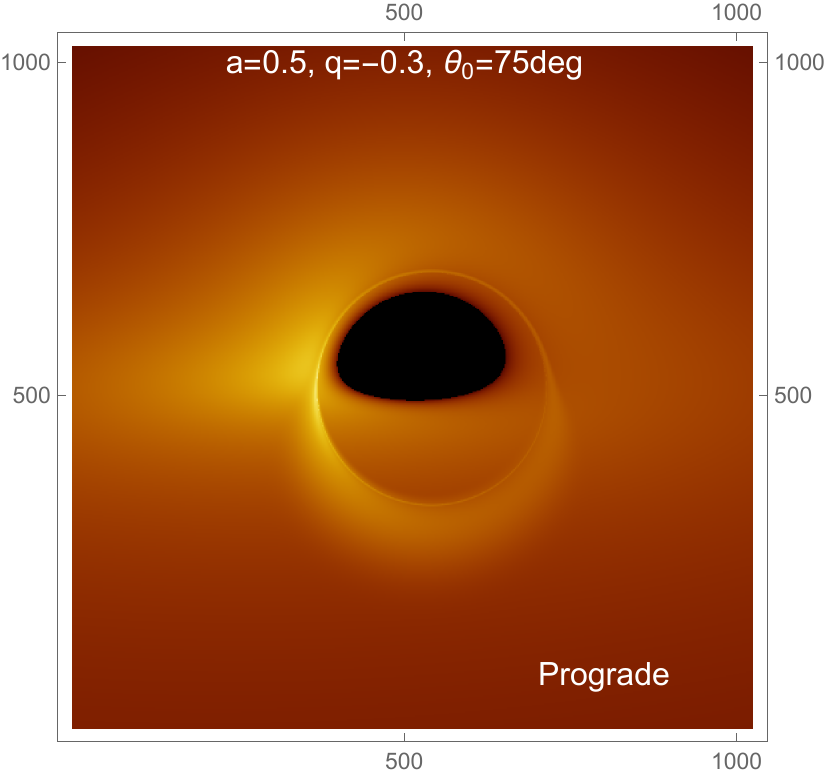}
\vspace{2mm}
\includegraphics[width=4.5cm,height=4.5cm]{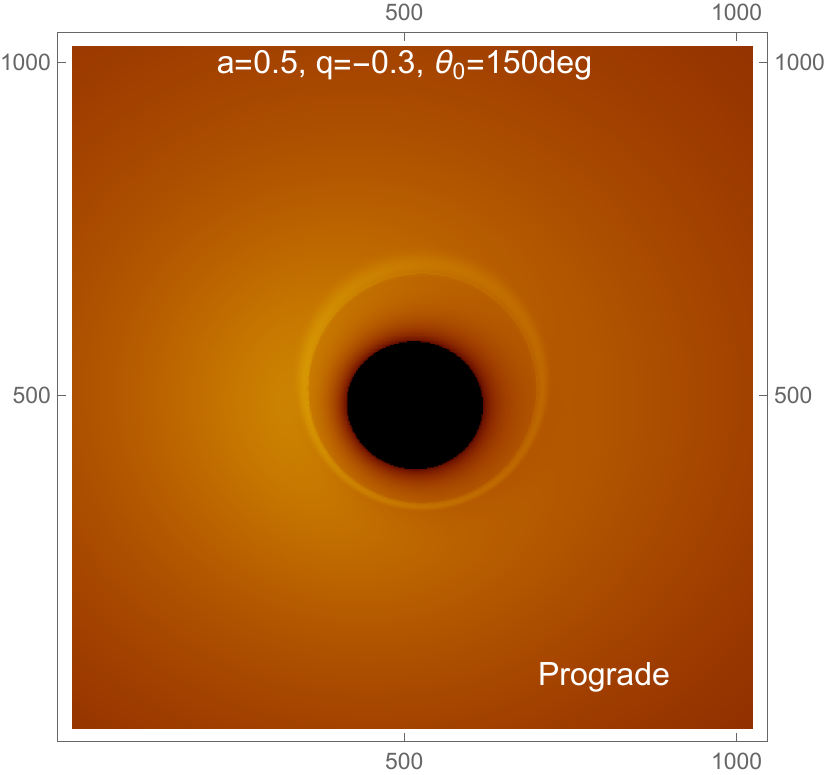}
\vspace{2mm}
\includegraphics[width=.5cm,height=4cm]{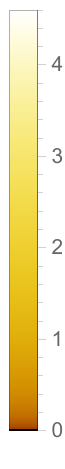}
\vspace{2mm}
\includegraphics[width=4.5cm,height=4.5cm]{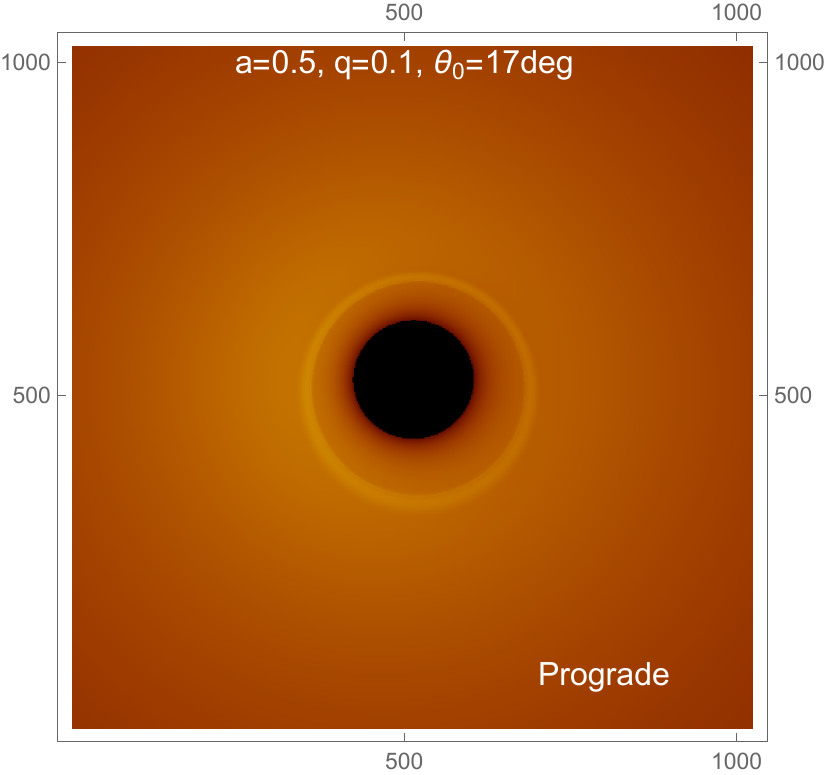}
\vspace{2mm}
\includegraphics[width=4.5cm,height=4.5cm]{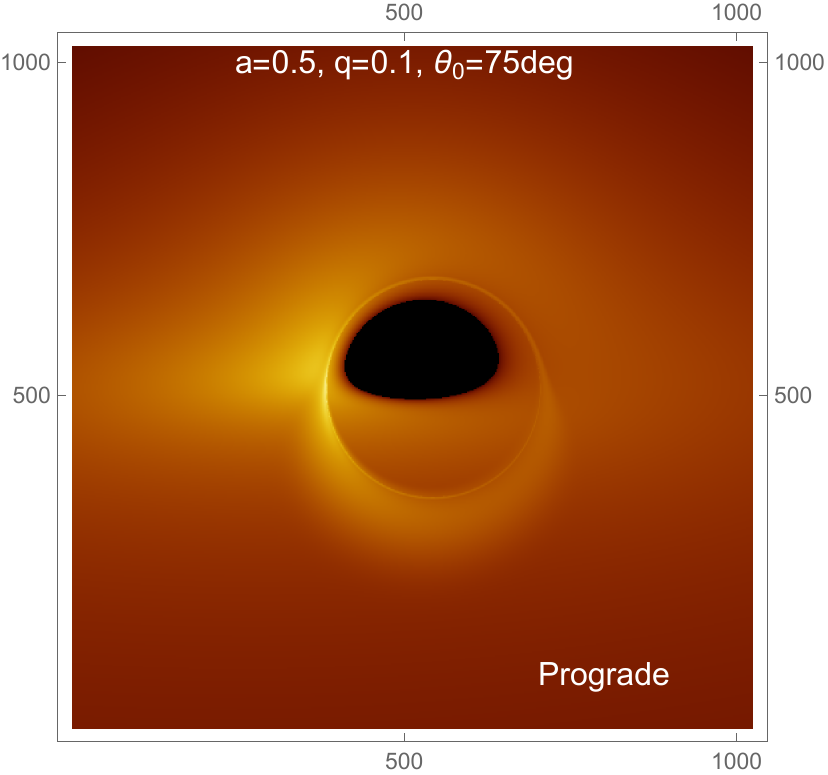}
\vspace{2mm}
\includegraphics[width=4.5cm,height=4.5cm]{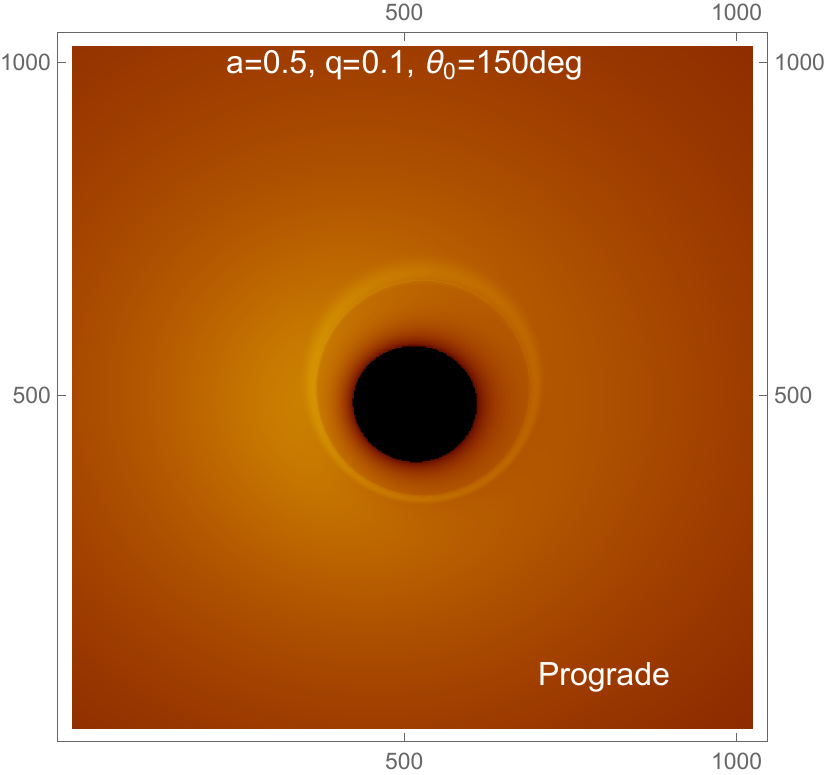}
\vspace{2mm}
\includegraphics[width=.5cm,height=4cm]{8gds.pdf}
\vspace{2mm}
\includegraphics[width=4.5cm,height=4.5cm]{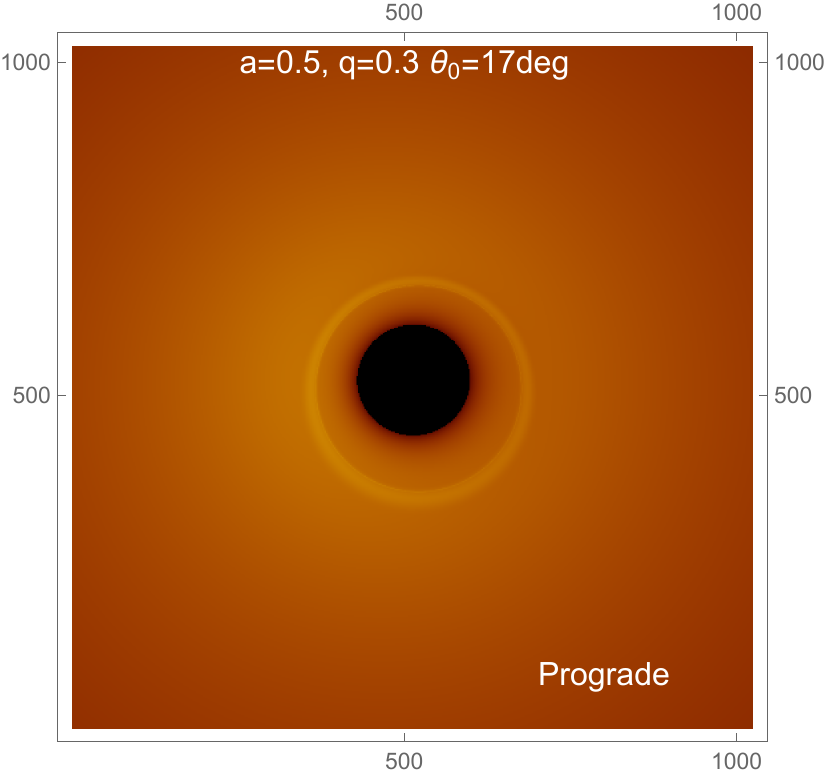}
\vspace{2mm}
\includegraphics[width=4.5cm,height=4.5cm]{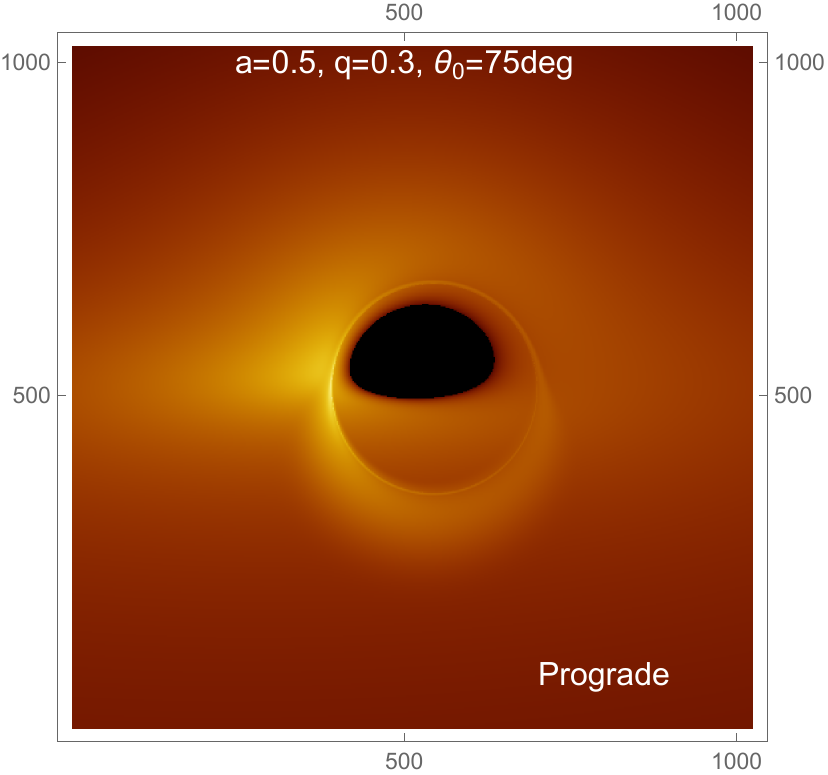}
\vspace{2mm}
\includegraphics[width=4.5cm,height=4.5cm]{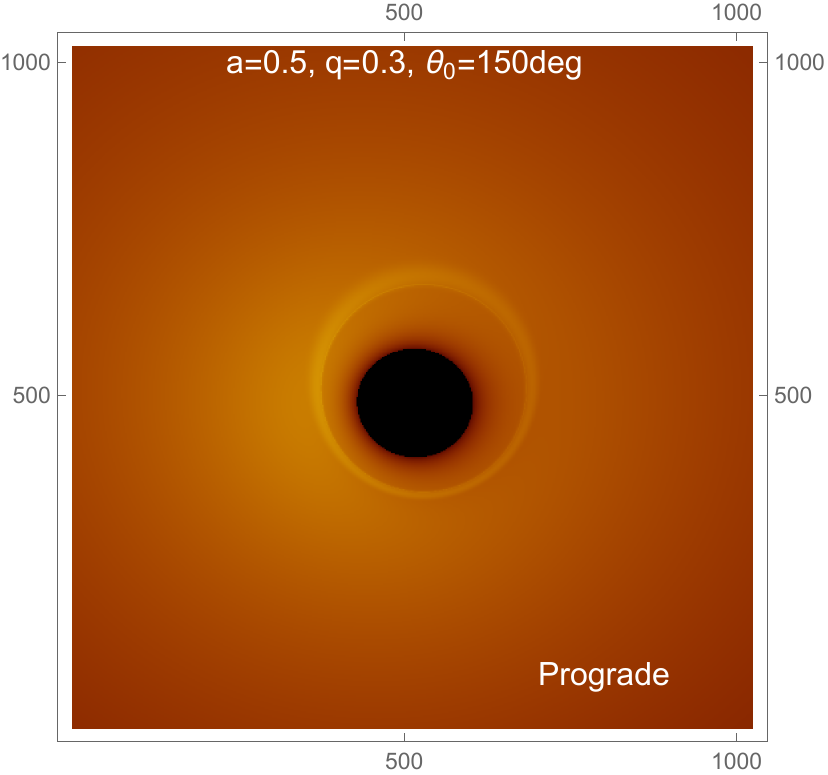}
\vspace{2mm}
\includegraphics[width=.5cm,height=4cm]{8gds.pdf}
\vspace{2mm}
\includegraphics[width=4.5cm,height=4.5cm]{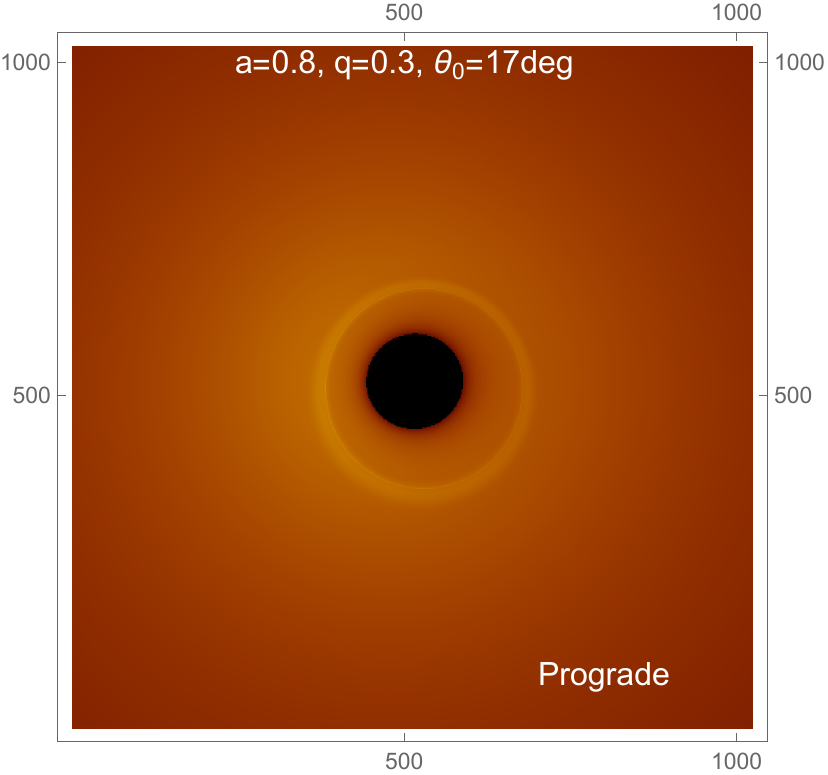}
\vspace{2mm}
\includegraphics[width=4.5cm,height=4.5cm]{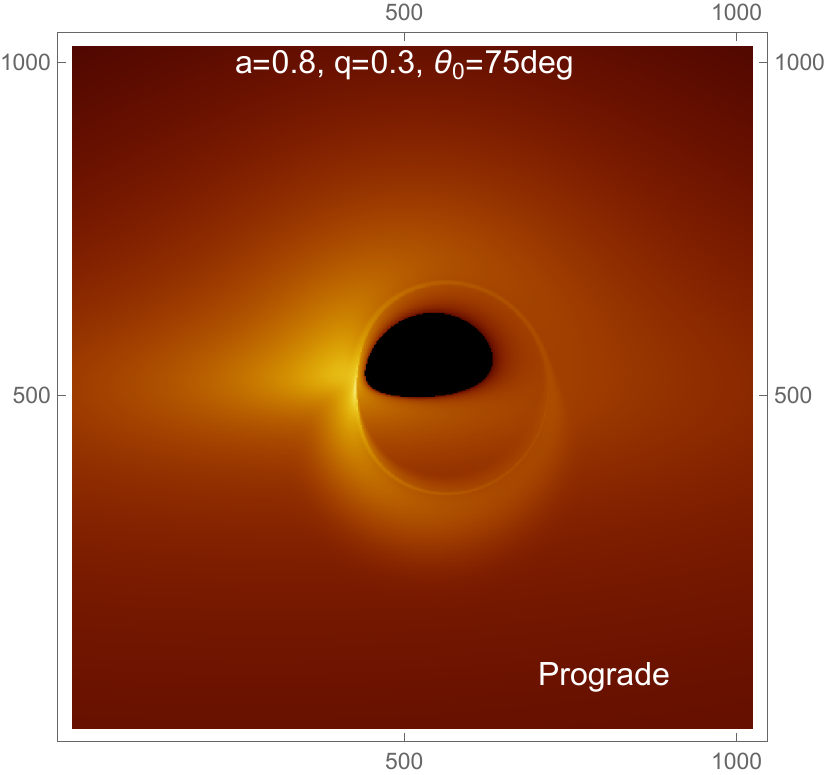}
\vspace{2mm}
\includegraphics[width=4.5cm,height=4.5cm]{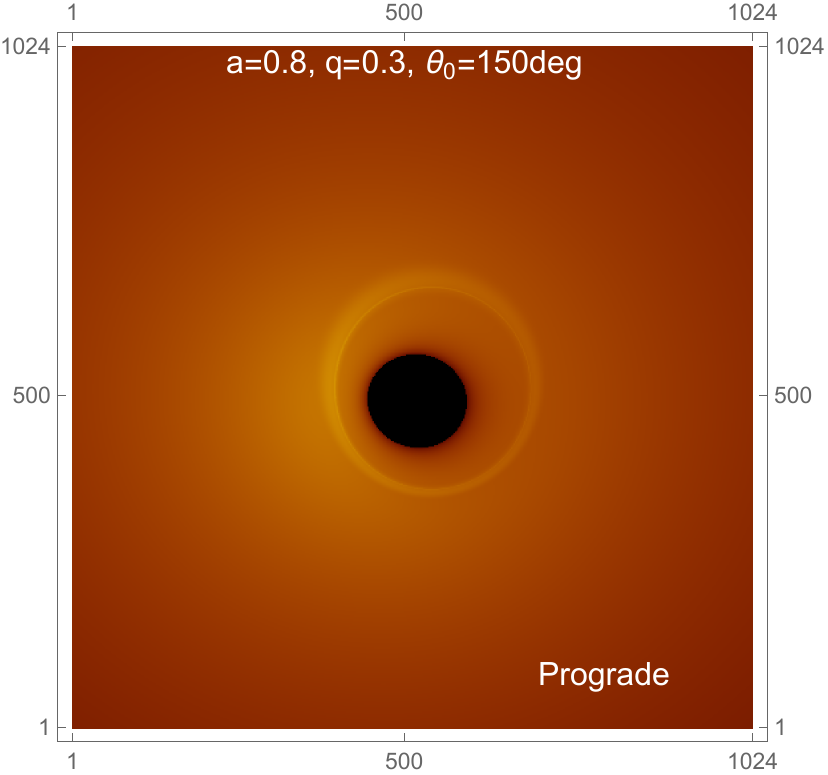}
\vspace{2mm}
\includegraphics[width=.5cm,height=4cm]{8gds.pdf}
\vspace{2mm}

\caption{The 86~GHz intensity maps for a prograde accretion disk are shown. The panels display the 86~GHz intensity distribution for different black hole parameter sets and observer inclination angles. The four parameter sets considered are $a=0.5$, $q=-0.3$; $a=0.5$, $q=0.1$; $a=0.5$, $q=0.3$; and $a=0.8$, $q=0.3$. For each case, the intensity distribution is shown at observer inclination angles of $17^\circ$, $75^\circ$, and $150^\circ$. In all panels, the black hole mass is set to $M=1$.}
\label{fig:23}
\end{figure}

\begin{figure}[htbp]
\centering
\includegraphics[width=4.5cm,height=4.5cm]{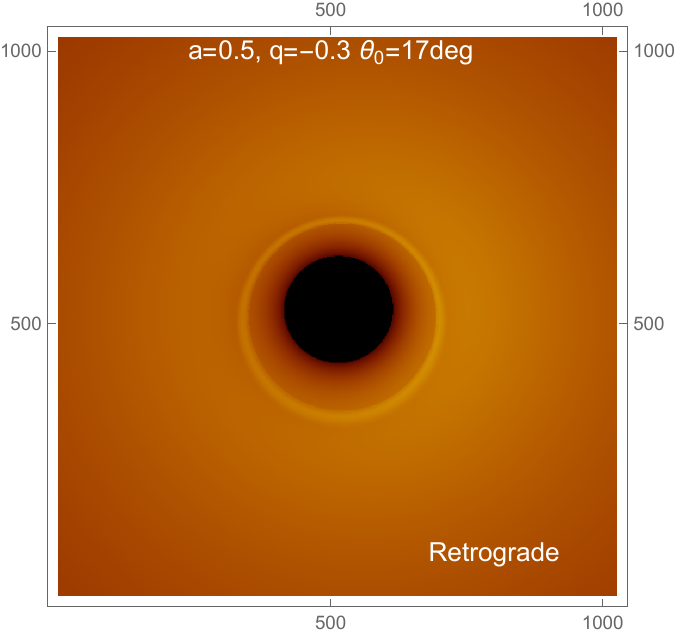}
\vspace{2mm}
\includegraphics[width=4.5cm,height=4.5cm]{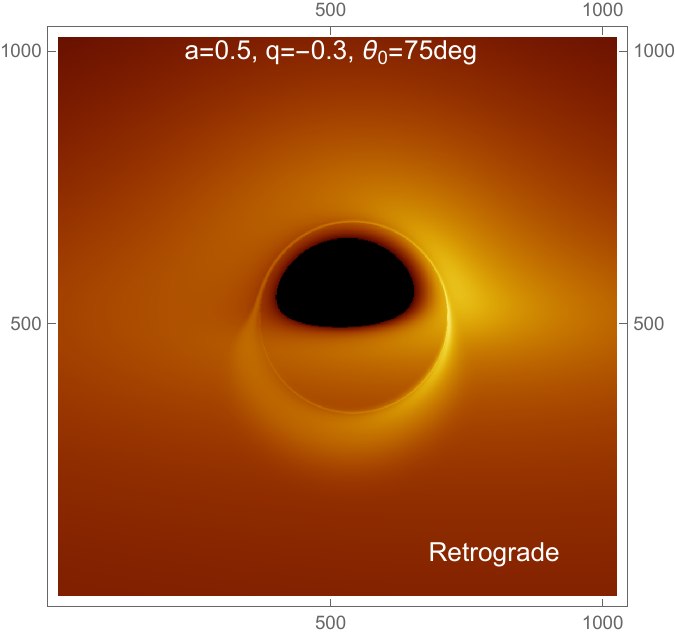}
\vspace{2mm}
\includegraphics[width=4.5cm,height=4.5cm]{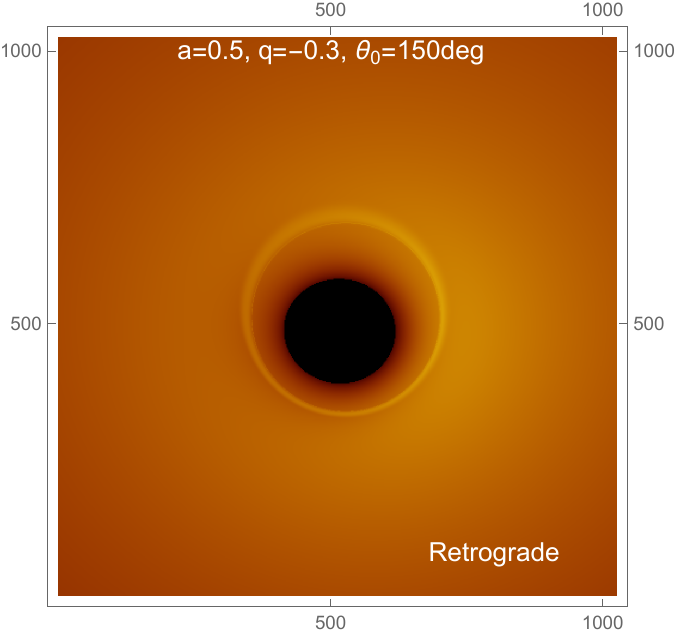}
\vspace{2mm}
\includegraphics[width=.5cm,height=4cm]{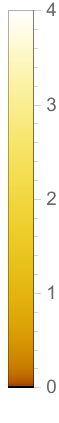}
\vspace{2mm}
\includegraphics[width=4.5cm,height=4.5cm]{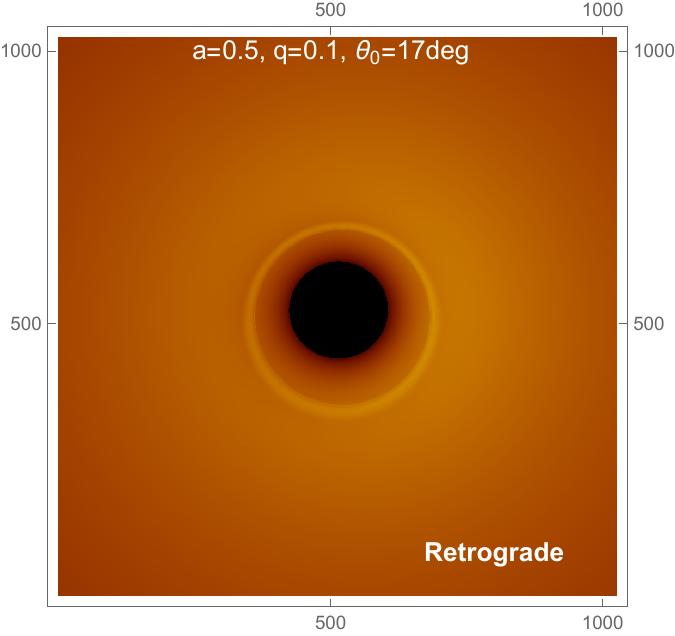}
\vspace{2mm}
\includegraphics[width=4.5cm,height=4.5cm]{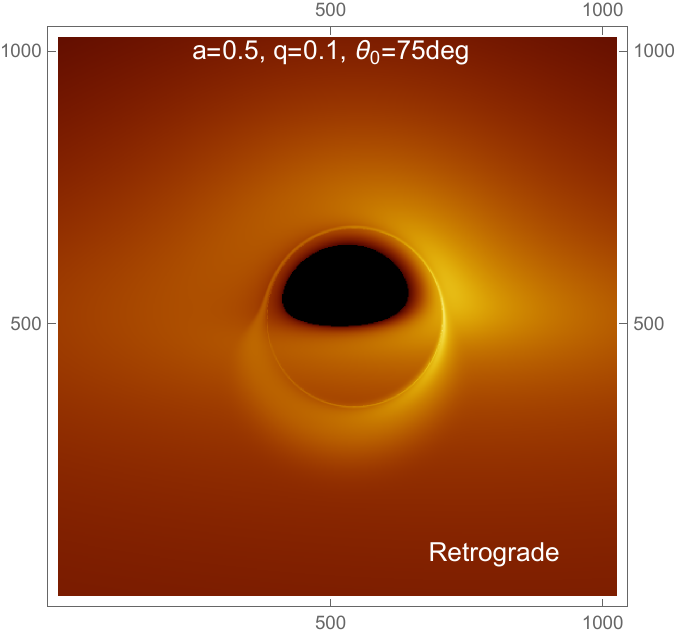}
\vspace{2mm}
\includegraphics[width=4.5cm,height=4.5cm]{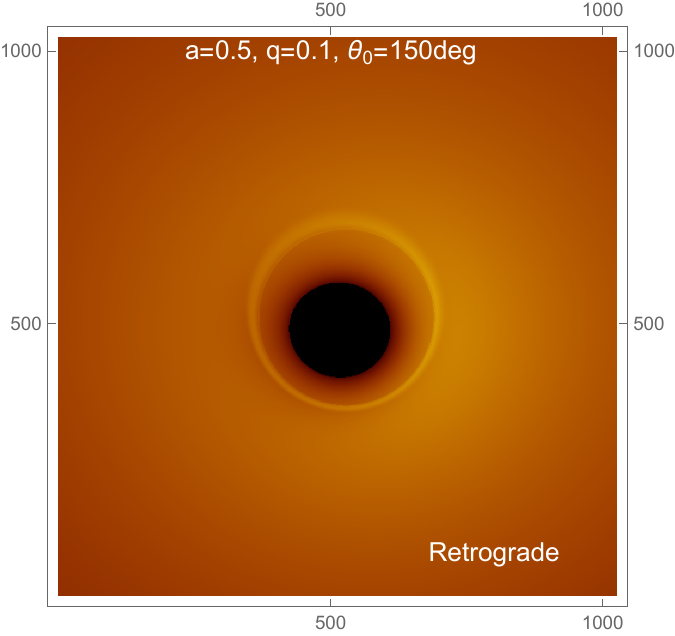}
\vspace{2mm}
\includegraphics[width=.5cm,height=4cm]{8gdn.pdf}
\vspace{2mm}
\includegraphics[width=4.5cm,height=4.5cm]{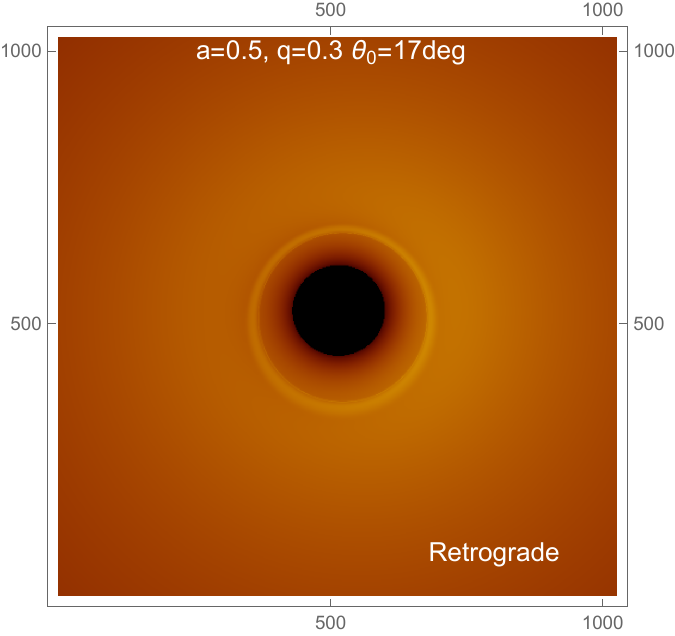}
\vspace{2mm}
\includegraphics[width=4.5cm,height=4.5cm]{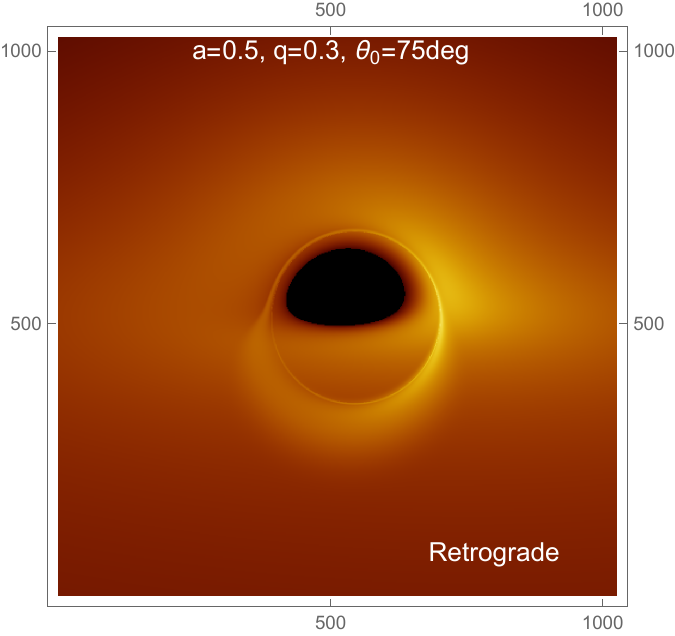}
\vspace{2mm}
\includegraphics[width=4.5cm,height=4.5cm]{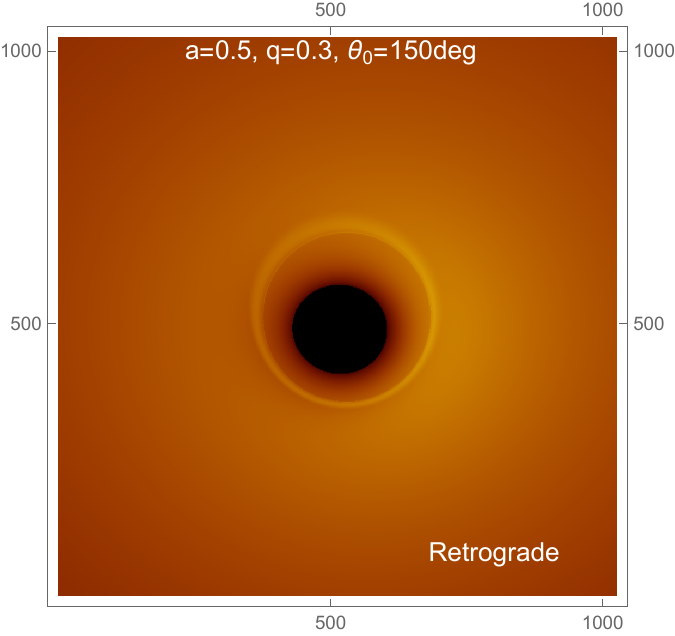}
\vspace{2mm}
\includegraphics[width=.5cm,height=4cm]{8gdn.pdf}
\vspace{2mm}
\includegraphics[width=4.5cm,height=4.5cm]{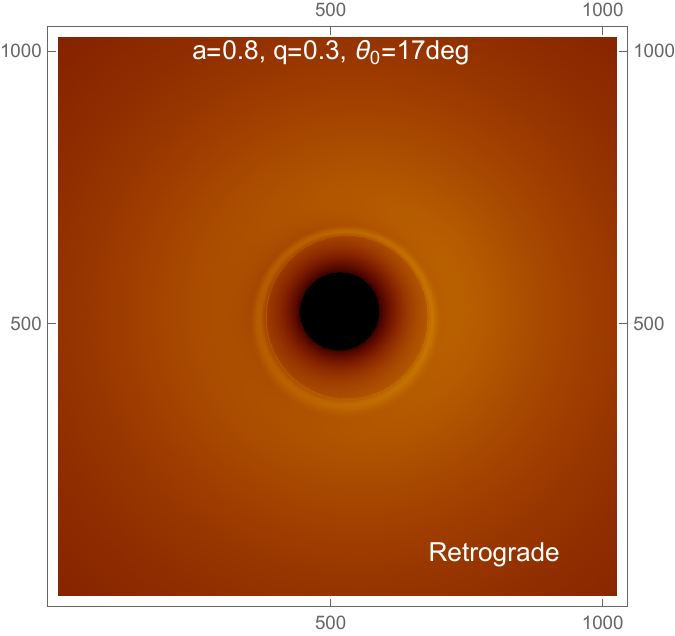}
\vspace{2mm}
\includegraphics[width=4.5cm,height=4.5cm]{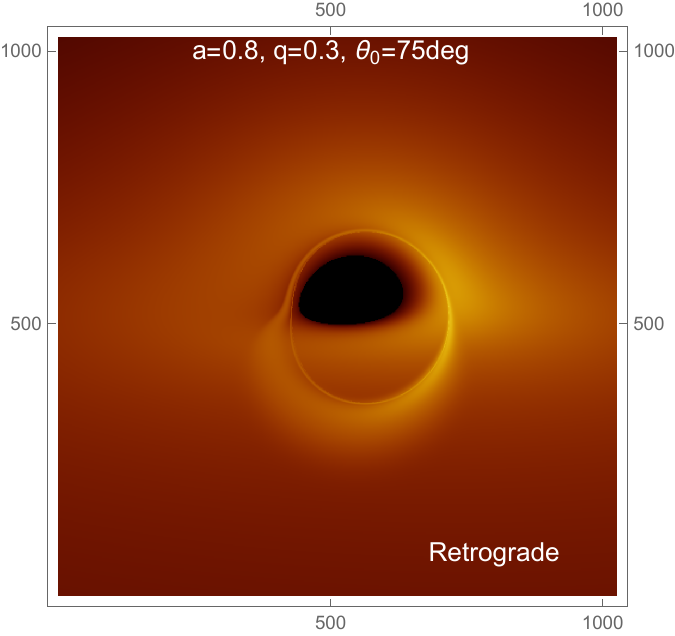}
\vspace{2mm}
\includegraphics[width=4.5cm,height=4.5cm]{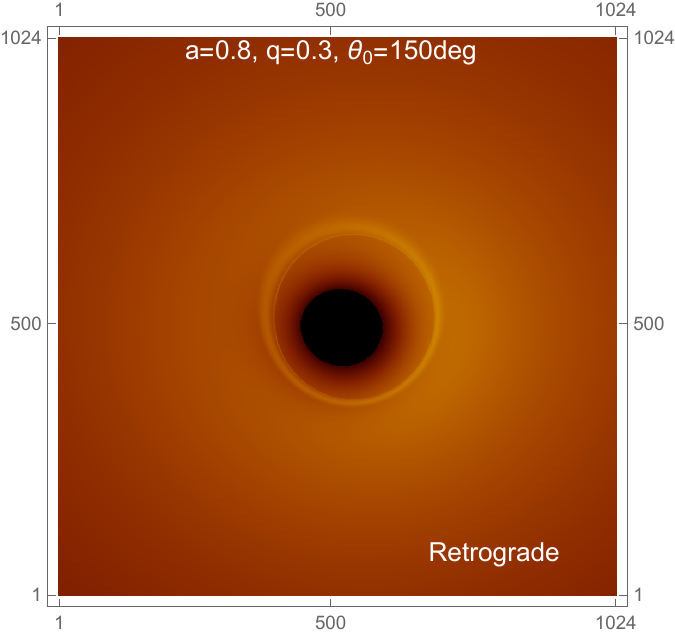}
\vspace{2mm}
\includegraphics[width=.5cm,height=4cm]{8gdn.pdf}
\vspace{2mm}
\caption{The 86~GHz intensity maps for a retrograde accretion disk are shown. The panels display the 86~GHz intensity distribution for different black hole parameter sets and observer inclination angles. The four parameter sets considered are $a=0.5$, $q=-0.3$; $a=0.5$, $q=0.1$; $a=0.5$, $q=0.3$; and $a=0.8$, $q=0.3$. For each case, the intensity distribution is shown at observer inclination angles of $17^\circ$, $75^\circ$, and $150^\circ$. In all panels, the black hole mass is set to $M=1$.}
\label{fig:24}
\end{figure}

\section{Conclusion}
\label{sec:4}

\par
In this work, we carried out a systematic study of the observable properties of a braneworld black hole surrounded by an optically thin accretion disk. Combining ray tracing with a detailed analysis of photon trajectories, we investigated how the black hole image depends on the tidal charge and spin. The critical curve and inner shadow were determined through a quartic equation obtained from the elliptic-integral description of photon motion. The results indicate that both $a$ and $q$ affect the deformation of the inner shadow, with the spin parameter producing the stronger distortion.

\par
We also found that the observed intensity and frequency-shift distributions depend strongly on the observer inclination. This dependence is especially clear near the ISCO, where the viewing angle largely determines the frequency-shift pattern. The frequency-shift asymmetry is generally more pronounced for prograde disks than for retrograde disks. In the prograde case, the left side of the image plane is blueshifted and the right side is redshifted, whereas the pattern is reversed for retrograde accretion.

\par
We further compared the synthetic intensity distributions at 230~GHz and 86~GHz. The 86~GHz images show both higher peak intensity and larger total intensity than the corresponding 230~GHz images. Nevertheless, the main spacetime-related structures, including the inner shadow and the critical curve, remain visible at both frequencies. This frequency dependence helps clarify the relation between disk emission and photon-ring geometry and may provide useful theoretical guidance for future high-resolution black hole imaging studies.

\par
Overall, our results show that the geometry of the accretion disk is crucial for determining the apparent image of a braneworld black hole, especially the photon ring and the inner shadow. Clear image asymmetries appear in both prograde and retrograde accretion cases because of the combined influence of spin, tidal charge, and viewing angle. These results provide a theoretical basis for future studies of high-resolution black hole images and may help characterize how the tidal charge modifies the apparent interaction between rotating braneworld black holes and their surrounding accretion flows.

\appendix
\section{Derivation of the Radial and Angular Integrals}
\label{Appendix A}
In this appendix, we provide the detailed steps leading to Eq.~(15-17). The corresponding radial and angular integrals are defined as

\begin{equation}
I_{t} = -\int\limits_{r_{s}}^{r_{o}} 
        \frac{(2Mr - q)(a^{2} + r^{2} - a\xi) + \Delta(r)\,r^{2}}
             {\pm_{r}\,\Delta(r)\,\sqrt{\mathcal{R}(r)}}\,dr,
\end{equation}

\begin{equation}
I_{\tau} = -\int\limits_{r_{s}}^{r_{o}} 
           \frac{dr}{\pm_{r}\,\sqrt{\mathcal{R}(r)}},
\end{equation}

\begin{equation}
I_{\phi} = -\int\limits_{r_{s}}^{r_{o}} 
           \frac{2aMr - a^{2}\xi - aq}
                {\pm_{r}\,\Delta(r)\,\sqrt{\mathcal{R}(r)}}\,dr,
\end{equation}

\begin{equation}
G_{t} = -\int\limits_{\theta_{s}}^{\theta_{o}} 
        \frac{\cos^{2}\theta}{\pm_{\theta}\,\sqrt{\mathcal{B}(\theta)}}\,d\theta,
\end{equation}

\begin{equation}
G_{\theta} = -\int\limits_{\theta_{s}}^{\theta_{o}} 
             \frac{d\theta}{\pm_{\theta}\,\sqrt{\Theta(\theta)}},
\end{equation}

\begin{equation}
G_{\phi} = -\int\limits_{\theta_{s}}^{\theta_{o}} 
           \frac{\csc^{2}\theta}{\pm_{\theta}\,\sqrt{\Theta(\theta)}}\,d\theta.
\end{equation}

{

\section{Effective Potential and Radial Four-Velocity of Massive Particles}
\label{app:effective_potential}

For the massive particles in the accretion disk, we denote the conserved
specific energy and angular momentum by
\begin{equation}
\mathcal{E}=-u_t,\qquad \mathcal{L}=u_\phi ,
\end{equation}
where $u^\mu$ is the four-velocity of the disk matter. These quantities
refer to timelike particles in the disk and should be distinguished from
the photon constants of motion introduced in the previous section.

The normalization condition for massive particles is
\begin{equation}
u^\mu u_\mu=-1 .
\end{equation}
Equivalently, it can be written as
\begin{equation}
g^{\mu\nu}u_\mu u_\nu=-1 .
\end{equation}
For equatorial motion, $\theta=\pi/2$ and $u^\theta=0$, so we have
\begin{equation}
g^{tt}u_t^2+2g^{t\phi}u_tu_\phi
+g^{\phi\phi}u_\phi^2+g^{rr}u_r^2=-1 .
\end{equation}
Substituting $u_t=-\mathcal{E}$ and $u_\phi=\mathcal{L}$, one obtains
\begin{equation}
g^{tt}\mathcal{E}^2-2g^{t\phi}\mathcal{E}\mathcal{L}
+g^{\phi\phi}\mathcal{L}^2+g^{rr}u_r^2=-1 .
\end{equation}
Therefore,
\begin{equation}
g^{rr}u_r^2
=
-\left(
1+g^{tt}\mathcal{E}^2
-2g^{t\phi}\mathcal{E}\mathcal{L}
+g^{\phi\phi}\mathcal{L}^2
\right).
\end{equation}
We then define the effective potential as
\begin{equation}
V(r,\mathcal{E},\mathcal{L})=
\left(
1+g^{tt}\mathcal{E}^2
-2g^{t\phi}\mathcal{E}\mathcal{L}
+g^{\phi\phi}\mathcal{L}^2
\right)_{\theta=\pi/2}.
\end{equation}
Thus,
\begin{equation}
g^{rr}u_r^2=-V(r,\mathcal{E},\mathcal{L}).
\end{equation}
Since the metric has no $r$--$t$ or $r$--$\phi$ cross terms, one has
$g^{rr}=1/g_{rr}$ and $u^r=g^{rr}u_r$. Hence,
\begin{equation}
(u^r)^2
=
-g^{rr}V(r,\mathcal{E},\mathcal{L})
=
-\frac{V(r,\mathcal{E},\mathcal{L})}{g_{rr}} .
\end{equation}
For the inward plunging branch, we take the negative sign and obtain
\begin{equation}
u^r
=
-\sqrt{
-\frac{V(r,\mathcal{E},\mathcal{L})}{g_{rr}}
}.
\end{equation}
The minus sign in front of the square root denotes inward radial motion.
The allowed radial motion satisfies $V(r,\mathcal{E},\mathcal{L})\leq 0$,
so that the quantity inside the square root is non-negative.

\section{Four-Velocity of the Plunging Flow inside the ISCO}

In this appendix, we derive the four-velocity of the plunging accretion
flow inside the innermost stable circular orbit (ISCO) of a rotating
braneworld black hole. For a massive particle, the conserved specific
energy and conserved specific angular momentum are defined by \cite{38}
\begin{equation}
\mathcal{E}=-u_t,
\qquad
\mathcal{L}=u_\phi.
\end{equation}

On the equatorial plane, $\theta=\pi/2$, the metric function is
\begin{equation}
\Delta=r^2-2Mr+a^2+q.
\end{equation}

Throughout this appendix, the upper and lower signs correspond to
prograde and retrograde motion, respectively. For a fixed orbital
branch, the same choice of sign must be used consistently in all
expressions.

The ISCO radius is determined by the marginal-stability condition
\begin{equation}
Mr_{\rm ISCO}^3
-6M^2r_{\rm ISCO}^2
+9Mqr_{\rm ISCO}
-4q^2
+a^2\left(4q-3Mr_{\rm ISCO}\right)
\pm
8a\left(Mr_{\rm ISCO}-q\right)^{3/2}
=0.
\end{equation}
The physically relevant root is chosen outside the outer event horizon
and must satisfy
\begin{equation}
Mr_{\rm ISCO}-q\geq0,
\end{equation}
together with
\begin{equation}
r_{\rm ISCO}^2
-3Mr_{\rm ISCO}
+2q
\pm
2a\sqrt{Mr_{\rm ISCO}-q}
>0.
\end{equation}

Inside the ISCO, stable circular motion is no longer possible. Following
the Cunningham prescription, we assume that the plunging matter retains
the conserved specific energy and angular momentum of the circular orbit
at the ISCO:
\begin{equation}
\mathcal{E}=\mathcal{E}_{\rm ISCO},
\qquad
\mathcal{L}=\mathcal{L}_{\rm ISCO}.
\end{equation}
These quantities are explicitly given by
\begin{equation}
\mathcal{E}_{\rm ISCO}
=
\frac{
r_{\rm ISCO}^2-2Mr_{\rm ISCO}+q
\pm a\sqrt{Mr_{\rm ISCO}-q}
}{
r_{\rm ISCO}
\sqrt{
r_{\rm ISCO}^2-3Mr_{\rm ISCO}+2q
\pm2a\sqrt{Mr_{\rm ISCO}-q}
}
},
\end{equation}
and
\begin{equation}
\mathcal{L}_{\rm ISCO}
=
\frac{
\pm\sqrt{Mr_{\rm ISCO}-q}
\left(r_{\rm ISCO}^2+a^2\right)
-a\left(2Mr_{\rm ISCO}-q\right)
}{
r_{\rm ISCO}
\sqrt{
r_{\rm ISCO}^2-3Mr_{\rm ISCO}+2q
\pm2a\sqrt{Mr_{\rm ISCO}-q}
}
}.
\end{equation}

For equatorial timelike geodesics, the radial equation can be written as
\begin{equation}
r^4\left(u^r\right)^2
=
\left[
\mathcal{E}_{\rm ISCO}\left(r^2+a^2\right)
-a\mathcal{L}_{\rm ISCO}
\right]^2
-
\Delta
\left[
r^2+
\left(
\mathcal{L}_{\rm ISCO}
-a\mathcal{E}_{\rm ISCO}
\right)^2
\right].
\end{equation}
The inward plunging branch is selected by taking $u^r<0$.

Substituting the explicit ISCO energy and angular momentum and using the
marginal-stability condition, the radial component reduces to
\begin{align}
u^r
={}&
-\frac{
\left(r_{\rm ISCO}-r\right)^{3/2}
}{
r^2r_{\rm ISCO}
\sqrt{
r_{\rm ISCO}^2
-3Mr_{\rm ISCO}
+2q
\pm2a\sqrt{Mr_{\rm ISCO}-q}
}
}
\nonumber\\
&\times
\Bigg\{
Mrr_{\rm ISCO}
\left[
r_{\rm ISCO}^2
-3Mr_{\rm ISCO}
+2q
\pm2a\sqrt{Mr_{\rm ISCO}-q}
\right]
\nonumber\\
&\qquad
-
\left(
a\mp\sqrt{Mr_{\rm ISCO}-q}
\right)^2
\left[
r\left(Mr_{\rm ISCO}-q\right)
+qr_{\rm ISCO}
\right]
\Bigg\}^{1/2}.
\end{align}
This expression satisfies
\begin{equation}
\left.u^r_{\rm in}\right|_{r=r_{\rm ISCO}}=0,
\end{equation}
and the negative sign describes inward motion for
$r<r_{\rm ISCO}$.

The covariant temporal and azimuthal components are
\begin{equation}
u_t
=
-\frac{
r_{\rm ISCO}^2
-2Mr_{\rm ISCO}
+q
\pm a\sqrt{Mr_{\rm ISCO}-q}
}{
r_{\rm ISCO}
\sqrt{
r_{\rm ISCO}^2
-3Mr_{\rm ISCO}
+2q
\pm2a\sqrt{Mr_{\rm ISCO}-q}
}
},
\end{equation}
and
\begin{equation}
u_\phi
=
\frac{
\pm\sqrt{Mr_{\rm ISCO}-q}
\left(r_{\rm ISCO}^2+a^2\right)
-a\left(2Mr_{\rm ISCO}-q\right)
}{
r_{\rm ISCO}
\sqrt{
r_{\rm ISCO}^2
-3Mr_{\rm ISCO}
+2q
\pm2a\sqrt{Mr_{\rm ISCO}-q}
}
}.
\end{equation}
The covariant radial component follows from
\begin{equation}
u_r
=
\frac{r^2}{\Delta}u^r_{\rm in}.
\end{equation}

Raising the indices and substituting the ISCO quantities gives
\begin{align}
u^t
={}&
\frac{1}{
r^2\Delta\,r_{\rm ISCO}
\sqrt{
r_{\rm ISCO}^2
-3Mr_{\rm ISCO}
+2q
\pm2a\sqrt{Mr_{\rm ISCO}-q}
}
}
\nonumber\\
&\times
\Bigg\{
\left[
\left(r^2+a^2\right)^2-a^2\Delta
\right]
\left[
r_{\rm ISCO}^2
-2Mr_{\rm ISCO}
+q
\pm a\sqrt{Mr_{\rm ISCO}-q}
\right]
\nonumber\\
&\qquad
-a\left(2Mr-q\right)
\Big[
\pm\sqrt{Mr_{\rm ISCO}-q}
\left(r_{\rm ISCO}^2+a^2\right)
-a\left(2Mr_{\rm ISCO}-q\right)
\Big]
\Bigg\},
\end{align}
and
\begin{align}
u^\phi
={}&
\frac{1}{
r^2\Delta\,r_{\rm ISCO}
\sqrt{
r_{\rm ISCO}^2
-3Mr_{\rm ISCO}
+2q
\pm2a\sqrt{Mr_{\rm ISCO}-q}
}
}
\nonumber\\
&\times
\Bigg\{
a\left(2Mr-q\right)
\left[
r_{\rm ISCO}^2
-2Mr_{\rm ISCO}
+q
\pm a\sqrt{Mr_{\rm ISCO}-q}
\right]
\nonumber\\
&\qquad
+
\left(
r^2-2Mr+q
\right)
\Big[
\pm\sqrt{Mr_{\rm ISCO}-q}
\left(r_{\rm ISCO}^2+a^2\right)
-a\left(2Mr_{\rm ISCO}-q\right)
\Big]
\Bigg\}.
\end{align}

Therefore, the contravariant four-velocity of the plunging flow is
\begin{equation}
u^\mu
=
\left(
u^t,
u^r,
0,
u^\phi
\right).
\end{equation}}

\end{document}